\documentclass[a4paper,12pt,final]{article}
\usepackage{cite}
\usepackage{amsmath,amssymb,amscd}
\usepackage{graphics,graphicx}
\usepackage{pstricks,pst-node,pst-tree}
\usepackage{indentfirst}

\begin{document}

\title{ On the geometry pervading One Particle States }
\author{  Antonio D\'{\i}az Miranda \\(Universidad Aut\'onoma de Madrid (Retired Professor))\thanks{email diazmir@outlook.es}}

\date{}

\maketitle
\tableofcontents
\vspace{2cm}

\begin{abstract}
In this paper, a way is  given to obtain explicitly the representations of the Poincar\'e group as can be prescribed by Geometric Quantization. Thus one obtains  some forms of the Space of Quantum States of the different relativistic free particles, and I give explicitly these spaces and the corresponding operators for the usually accepted as realistic physical particles. The general description of the massless particles I obtain, is given in terms of solutions of Penrose equations. In the case of Photon, I also give other descriptions, one in terms of the Electromagnetic Field. Since the results are derived from Geometric Quantization, they are related to certain Contact and Symplectic manifols, that I study in detail. The symplectic manifold must be interpreted, according with Souriau, as the  Movement Space  of the corresponding classical particle, and that leads to propose one of the spaces I use as the State Space of the corresponding classical particle. These spaces are also described in each case.
\end{abstract}

\section{Introduction.}

The Wave Equations of relativistic elementary particles,
Klein-Gordon, Dirac, Weil etc.., have been originally
derived each one in a independent way. An unification was
the discovery of the relation of these equations with the
representations of Poincar\'e group (inhomogeneous Lorenz
group). The classification of the representations of
Poincar\'e group made by Wigner \cite{Wig} with important
contributions of Majorana,  Dirac and Proca \cite{maj,dir,pro}
leads to a group theoretical study of Wave Equations by
Bargmann and Wigner \cite{bar-wig}.

On the other hand, in Kirillov-Kostant-Souriau theory (Geometric Quantization),
a description of Quantum Systems is given in terms of
elements of the dual of the Lie algebra of the Lie group
under consideration. Of course, this way of seing Quantum
Mechanics is not completely independent of the preceding 
one, since it has its origin in a method to obtain
representations \cite{kir,aus-kos}. 

The  correspondence  of  Quantum  States  in  the sense
of
Geometric  Quantization  with  Wave  Functions  in the Quantum
Mechanical sense is  not clear in  all cases.   In Souriau's
book \cite{soustr}, a very general way to do the passage
is
given, but it doesn't work in all cases. 

                   In  this  paper  we  give  a geometrical
construction that  stablishes a  one to  one correspondence
from Quantum States in the sense of Geometric  Quantization
to Wave Functions in the Quantum Mechanical sense, that  is
valid for all  kinds of relativistic  elementary particles.
  The idea
is as follows.

    In Geometric Quantization, one
  begins  with  a  regular  contact  manifold  or its
associated hermitian line bundle.  Quantum states ( in  the
sense of  Geometric Quantification  ) can  be considered as
being the  collection of  those sections  of the  hermitian
line  bundle  which  satisfies  "  Planck's condition" (\it
cf.\rm \ Souriau's book).  In this paper we  see
that
 these sections are in a
one to
one  correspondence  with  the  (unrestricted)  sections of
another hermitian line bundle.  Thus, this fibre bundle  is
a  good  setting  to  describe  the quantum processes under
consideration.  The main idea to pass from this description
to  the  usual  one  in  terms  of  Wave  Functions, can be
intuitively explained as follows.   Quantum States in  G.Q.
attribute an ``amplitude of probability'' to each  movement
of the particle.  To obtain the corresponding wave function
one  must  proceed  as  follows:    for  each  event,   the
corresponding  amplitude  of  probability  is  obtained  by
taking all movements passing  through the given event,  and
then ``adding up'' ( in a suitable sense) the corresponding
amplitudes  of  probability.    Of  course  the  concept of
``movement passing  through an  event'' is  only obvious in
the case of the ordinary massive spinless particle, and  it
is defined in section~\ref{clasical}.

The Contact Manifold under consideration, fibers on a Symplectic Manifold that, following Souriau, must be considered as being composed by the "movements of the corresponding classical particle". Then, with the constructions made in this paper,  it becomes clear what space must be considered as composed by the "states of the corresponding classical particle", for each  kind of relativistic particle. 

The spaces of Wave Functions and many relevant isomorphic vector spaces we obtain, are the spaces of the representations of Poincar\'e group that I describe in section \ref{representaciones}.

The preceding constructions are made in a explicit way for the relativistic particles usually considered as having  physical sense. A section is devoted to massive particles and other to massless particles.

In section \ref{massive}  a explicit application of the preceeding constructions is done for   massive particles. One
obtains solutions of Klein-Gordon and Dirac equations,  and
also  a  description  of  the  wave  functions  for massive
particles  of  higher  spin.  

 Massless particles are studied in section \ref{waveNula}.  For  massless
particles  of  spin  1/2  one  obtains  solutions  of  Weyl
equations  and  for  general  spin  this  method leads in a
natural way  to the  description of  massless particles  by
means of solutions of Penrose wave equations \cite{pen}. This  is related to the  fact that the  Contact  and Symplectic Manifolds corresponding to Massless Particles are expresed in a natural way in Twistor Space and its Projective Space, as explained in section \ref{contactnula}. 

In the particular case of Photon, in section \ref{alternphoton},I give also other forms for the Wave Functions. One of them is in terms of the Electromagnetic Field,and are  similar to the Wave Functions proposed by  Bialynicki-Birula.

In section \ref{SS} I describe for each particle the concrete space that must be considered to be the "Space of states of the corresponding classical particle".

\section{Notation\label{Notation} }

All differentiable manifolds appearing in this paper are assumed to
be $C^{\infty}$, finite dimensional, Hausdorff and second countable.

The set composed by the differentiable vector fields on a differential
manifold, $M$, is denoted by ${\cal D}(M)$. The set of differential
$k$-forms on $M$ is denoted by $\Omega^{k}(M)$.

Let $X\in{\cal D}(M),\ \omega\in\Omega^{k}(M)$. We denote by $i(X)\omega$
the interior product of X by $\omega$ and by $L(X)\omega$ the Lie
derivative of $\omega$ with respect to $X$.

Let $G$ be a Lie group. The Lie algebra of $G$ is the set consisting of the left invariant
vector fields on $G$, provided with its canonical structure of Lie
algebra. It is denoted in this paper by $\underline{G}$. I denote by $\underline{G}^{\ast}$
the dual of $\underline{G}$. $\underline{G}^{\ast}$ is canonically
identified with the set composed by the left invariant $1$-forms
on $G$. By identification of each element of $\underline{G}$ or
$\underline{G}^{\ast}$ with its value at the neutral element,
$e$,\ the lie algebra of $G$ is identified to the tangent space at $e$,\  
and $\underline{G}^{\ast}$ to its dual.

There exist a map, the exponential map,\ $Exp: G \rightarrow \underline{G}$, such that: if $X\in \underline{G}$,\  the integral curve of $X$ with initial value $e$ is  $t\in {\rightarrow}  Exp \,tX$.\ This integral curve  is 
a one parameter subgroup of $G$.

The Lie algebra of the circle Lie group, $S^1$, is identified to $R$ in such a way that for all $ t\in R,\ Exp\,t=e^{2 \pi i t}$.

If we have an homomorphism of $G$ in $S^1$, its differential is a linear map from the Lie algebra of $G$ into the Lie algebra of
 $S^1$,\ which is $R$. Thus the differential of an homomorphism of $G$\  into $S^1,$\ can be considered as an element of $\underline{G}^{\ast}$.

In the same way, the Lie algebra of the Lie group $R$ is identified to $R$ in such a way that the exponential map becomes the identity. Thus, also in this case, the differential of  an homomorphism of $G$ into $R$, can be considered as an element of $\underline{G}^{\ast}$.

The coadjoint representation is the homomorphism $Ad^{\ast}:g\in G\to Ad_{g}^{\ast}\in$\ Aut
$(\underline{G}^{\ast})$, given by $(Ad_{g}^{\ast}(\alpha))(X)=\alpha(Ad_{g^{-1}}(X))$
for all $\alpha\in\underline{G}^{\ast},X\in\underline{G}$, where
$Ad_{g^{-1}}$ is the differential of the automorphism of $G$ that
sends each $h\in G$ to $g^{-1}hg.$

Let $M$ be a differentiable manifold and $G$ a Lie group acting
on $M$ on the left (resp. on the right). Given an element, $X$,
of $\underline{G}$, we denote by $X_{M}$ the vector field on $M$
whose flow is given by the diffeomorphisms associated by the action
to $\{$Exp$(-tX):t\in{\bf R}\}$ (resp. $\{$Exp$tX:t\in{\bf R}\}$).
$X_{M}$ is called the infinitesimal generator of the action
associated to $X$.

A principal fibre bundle having $M$ as total space, $B$ as base
and $G$ as structural group, will be denoted by $M(B,G)$.

In all that  concerns  fibre  bundles,  we  use  the  notation of \cite{kn}.

\part{General Method}

\section{A Particular Classical State Space. }\label{clasical}

In this section I motivate the physical interpretation of the geometrical constructions that are to be made in this paper. I do not enter in the details of the computations. 
Most part of this section is an easy consecuence of Souriau's book \cite{ soustr}, which has inspirated   this paper.

                   Let  us  consider  a  classical  \bf  relativistic  free  particle without
spin \rm  in  Minkowski
space-time, with rest mass $ m\ne 0$.

                  Space-time  is interpreted  as
being  an  abstract  four  dimensional  manifold,  M,  each
inertial  observer,  R,  providing  us  with  a  \it global
chart\rm, $\phi_{R}=(x_{R}^1,...,x_{R}^{4})$.   Changes  of
these   global   charts   are   given   by  transformations
corresponding   to   elements   of   the  Poincar\'e  (i.e.
inhomogeneous  Lorenz)  group, ${\cal P}$.\  We  consider  a family of
inertial  frames  such  that  changes  are  all  given   by
ortochronous   proper   Poincar\'e   transformations,  i.e.
elements of the connected component of the identity, ${\cal
P}_{+}^{\uparrow}$.

 A geometrical  object in  \mbox{$\bf R^4
\rm $} invariant under the action on \mbox{$\bf R^4 \rm $},
provides us with a well  defined object in M ,  whose local
expression is the same for all inertial frames.  An example
is the Minkowski metric, $g$. When provided with this metric, M becomes Minkowski space.

The objects defined in  this
way  would  be  well  defined  even  if the charts were not
global.

                   Charts  $\phi_R  $  give  rise  in   the
canonical way to  charts of TM,  $\dot {\phi}_R=(x_R^i,\dot
x_R^i)$,  where  $\dot  x_R^i(v)=v(x_R^i)$, for all $ v \in TM.$    The   charts
$\overline {\phi}_R=(x_R^i,P_R^i) $,\ $P_R^i=m\,\dot x_R^i$,\
are  more  natural  in  Physics.    Changes of these global
charts are  given by  the transformations  corresponding to
elements of the Poincar\'e group , in its canonical  action
on $\bf R^8 \rm\ $$i.e.$\ the action given by
$$
(L,C)*\left(\begin{array}{c}z^1\\ \vdots \\z^8\end{array}\right)=\left(
\begin{array}{l} L\left(\begin{array}{c}z^1\\\vdots \\z^4\end{array}\right)+C\\ L\left(\begin{array}{c}z^5\\\vdots \\z^8\end{array}\right)
\end{array}\right)
$$

                   Geometrical objects  on
{\mbox{$\bf  R^8  \rm$}}\  invariant  under  this  action, give  us well  defined geometrical
objects on  TM, whose  local expressions  are equal  in the
different    inertial    frames.        For   example,   if
$(z^1,\ldots,z^8)$\ is the  canonical coordinate system  of
{\mbox{$\bf    R^8    \rm$}}\    ,    the   1-form:
$$
{\mbox{$\omega$}}_0    \equiv\    z^8    dz^4-\sum_{i=1}^3\
z^{i+4}dz^i
$$
the vector field:
$$
X_0 \equiv\  \frac1m\ \sum_{i=1}^4\ z^{i+4}
  \frac{\partial}{\partial z^i}
$$
and the submanifold, {\mbox{${\cal E}_0$}}, given by:
$$
z^8  \equiv\ \sqrt{m^2+(z^5)^2+(z^6)^2+(z^7)^2}
$$
are invariant by the action on {\mbox{$\bf R^8 \rm$}}.

Thus the following 1-form, vectorfield, and submanifold are well defined
on TM:
 $$
  \widetilde{\mbox{$\omega$}} = P_R^4\,dx_R^4 -\sum_{i=1}^3\ P_R^i\,dx_R^i
$$
$$
\widetilde{X}={  \frac1m}\ {\sum_{i=1}^4}P_R^i\,
\frac{\partial}{\partial x_R^i}
$$

\begin{eqnarray*}
{\cal E} & = & \left \{\, v\in TM: P_R^4(v)=\left
( m^2+\sum_{i=1}^3
(P_R^i(v))^2\right )^{\frac 12} \right \} \\
 & = & \left \{ \,v\in TM:  g(v,v)=1,\dot x_R^4(v) > 0 \right \}
\end{eqnarray*}

where R is an arbitrary inertial frame.

                   The restriction of $ \widetilde{\mbox{$\omega$}} $
to {\mbox{$\cal E,$}}\ $\omega,$\  is a  contact form. $\widetilde X$\  is  tangent to ${\cal E}$. The restriction of $\widetilde X$\  to {\mbox{$\cal E,$}}\ $X$, is the unique
vectorfield such that $$ i_X {\mbox{$\omega$}} = m,\ i_Xd
{\mbox{$\omega$}} = 0. $$

                   Each  inertial  observer  associates  to
each  element,  v,  of  {\mbox{$\cal  E$}}\  eight numbers,
$\overline\phi_R(v)=(x_R^1(v),  ...,\  x_R^4(v),  P_R^1(v),
...,\ P_R^4(v))$ which are interpreted as giving  position,
time and momentum-energy ( we take c=1).  Thus {\mbox{$\cal
E$}}\ must be interpreted as being {\bf state space}.

                   The movements of the free particle under
consideration    are    curves    in   {\mbox{$\cal   E,$}}
{\mbox{$\gamma$}} , having,  for each inertial  observer, a
constant four velocity $i.e.$
  $$
\dot\phi_R\circ\gamma(s)=(
a^i+s(\dot x_R^i)_0,  (\dot x_R^i)_0)  $$ where  the $(\dot
x_R^i)_0$  are  constant,   $(\dot  x_R^4)_0  >   0,  (\dot
x_R^4)_0^2-\sum_{i=1}^3 (\dot x_R^i)_0^2  = 1.$\ 

Since  the
preceeding relation is equivalent to
$$
\overline\phi_R\circ\gamma(s) = ( a^i+s
\frac{{(P_R^i)}_0}m,
{(P_R^i)}_0),
$$
where ${(P_R^i)}_0=m(\dot x_R^i)_0,$\ one sees that
the movements are the integral curves of X.

The parameter $s$\ coincides, since we take c=1, with proper time (of the trajectory in space time).

The  action on {\mbox{$\bf    R^8    \rm$}}\   gives, by means of any
 of the inertial observers, $R,$\  an action on M by means of
$$
(L,C)\odot v=(\overline\phi_R)^{-1}((L,C)*(\overline\phi_R(v))),
$$
for all $v\in {\cal E},\ (L,C)\in {\cal
P}_{+}^{\uparrow}$.

This action obviously preserves {\mbox{$\cal   E,$}} and 
$ {\mbox{$\omega$}} $  in such a way that    $({\mbox{${\cal E}$}},
{\mbox{$\omega$}})$  is a  homogeneous contact  manifold
for  the  given  action, but different inertial observers lead to different actions.

                   Now,  we  shall  describe ${\mbox{${\cal
E}_0$}}$ in a  different way, thus  obtaining a picture  of
state space more suitable  for its generalisation to  other
kind of particles.

                   Let $Y_8$ be the infinitesimal generator
of the action on  {\mbox{$\bf R^8 \rm$}} associated  to the
element Y of the Lie algebra of ${\cal  P}_{+}^{\uparrow},\
\underline{\cal P}$.

                   Since  the action on {\mbox{$\bf R^8 \rm$}} preserves
${\mbox{$\omega$}}_0$,\ we have     $$L_{Y_8} {\mbox{$\omega$}}_0=0,$$ so that 
$$i_{Y_8}\,d    {\mbox{$\omega$}}_0=-d\,(i_{Y_8}{\mbox{$\omega$}}_0)=-d({\mbox{$\omega$}}_0(Y_8)).$$ 

Thus we  define  a  map,
  $ {\mu}_0$, called the  \bf momentum  map,\ \rm from {\mbox{$\bf R^8 \rm$}} into 
	$\underline{\cal P}^*$,\ by means of $$  {\mu_0}(z)\cdot Y=-(  {\mbox{$\omega$}}
_0(Y_8))(z),
$$
for all $Y \in \underline{\cal P}, z \in
\bf R^8 \rm.$

                   Let $\{ Y_{\alpha}^i,  Y_
\beta^i,  Y_{\gamma}^i, Y_\delta:
\ i=1, 2,  3 \}$  be the basis  of $\underline{\cal P}$,
composed by  the usual  generators of  Lorenz rotations and
space-time translations.

It can be proved that 
$$
\mu_0=z^8\,Y_
\delta^*   -  \sum_{i=1}^3\  \{[(z^1,   z^2,
z^3)\times(z^5, z^6, z^7)]^i\, Y_\alpha^{i*}
+ 
$$
$$
+ (  z^4z^{i+4}\,-z^8z^i)\,Y_\beta^{i*} +
z^{i+4}\,    Y_{\gamma}^{i*}\,\}
$$
where   $\{   Y_
\alpha^{1*},...,  Y_\delta
^*\}$  is  the  dual  basis  of $\{ Y_\alpha
^1,..., Y_\delta\}$.

           The map  $\mu_0$  is  equivariant  for  the given
action  on  {\mbox{$\bf  R^8  \rm$}}  and coadjoint action.
Thus,  since  $(0,  ...,0,m)\in  {\mbox{${\cal E}_0$}}$ and
${\mu_0}(0, ...,0,m)=m\, Y_\delta^*,\ {\mu_0}({\mbox{${\cal
E}_0$}})$is the coadjoint orbit of $m\,Y_\delta
^*$.

We also have ${\mu_0^*}\,  {\mbox{$\Omega$}} =d  {\mbox{$\omega$}} _o$, where
 $  {\mbox{$\Omega$}} $ is
the Kirillov symplectic form on the coadjoint orbit.

For each {\mbox{$\alpha$}} in the coadjoint orbit,
$\mu_0^{-1}\,\{{\mbox{$\alpha$}}\}$ is the image of
a integral curve of $X_0$, that we know to be local expresion of movements. Then $\mu_0$ gives a one to one
correspondence
between points of the coadjoint orbit and movements of the particle under consideration.

Now we consider the map
$$
f:\ (a,b)\in {\mbox{${\cal E}_0$}}\longrightarrow (a, \mu_0\,(a,b))\in {{\bf
R^4}\rm}\times \underline{\cal P}^*.
$$

                   By  using   the  above   expression  for
$\mu_0$, one sees that $f$  is an imbedding.  Also,  $f$ is
equivariant  when  one   considers  the  given   action  in
${\mbox{${\cal  E}_0$}}$  and  the  ``product action'' in $
{{\bf R^4}\rm}\times \underline{\cal P}^*$ given by
$$
(L,C)*(a,  {\mbox{$\alpha$} } )=(La+C,Ad_{(L,C)}^*\,
{\mbox{$\alpha$} } ),\ \ \ \ \forall
(L,C)\in
{\cal P}_{+}^{\uparrow}, (a,   {\mbox{$\alpha$} } )
\in f({\mbox{${\cal E}_0$}}).
$$

Observe that,
as a  consequence, $f({\mbox{${\cal E}_0$}})$ is
an orbit of that action.

Each inertial frame, R, enables us to ``see'' {\bf state space},
$\cal E$, by
means of $f\circ\overline\phi_R$, as being $f({\mbox{${\cal E}_0$}})\subset
{ {\bf
R}^4}\times \underline{\cal P}^*$. Changes of inertial frames are now
given by transformations coming from the product action.

Since $f\circ
\overline\phi_R(e)=(x_R^1(e),...,x_R^4(e),\ \mu_0(\overline\phi
_R(e)))$, if we denote $$\mu_R  \equiv \mu_0\circ\overline\phi_R$$ we have
$$f\circ
\overline\phi_R =(\phi_R, \mu_R),$$ and 
 $\mu_R$ stablishes a one to one correspondence of
trajectories of X ($i.e.$  movements of the particle
parametrized by proper
time) with points of the coadjoint orbit. Thus, coadjoint orbit is to be
identified to {\bf movement space} for all inertial frames, althought the identification is different for each $R$.

The picture of a state, $e\in \cal E$, obtained by R is
now
$(\phi_R(e),
\mu_R(e)) \   i.e.$ an event, $\phi_R(e)$, and a movement,
$\mu_R(e)$,
``passing through'' the event.

Each $Y\in\underline{\cal P}$ defines a function on $
\underline{\cal P}^*$, denoted by the same symbol. Thus
$Y\circ\pi_2\circ f\circ \overline\phi_R$ is a function on
$\cal E$ ($i.e.$
a {\bf dynamical variable}) where $\pi_2$ is the canonical projection of
$ { {\bf R}^4}\times \underline{\cal P}^*$ onto $\underline{\cal
P}^*$.

We have:
\begin{eqnarray*}
- {Y_{\gamma}^i}\circ\mu_R &=
&P_R^i      \\
  {Y_\delta}\circ\mu_R
&=& P_R^4   \\
- {Y_{\alpha}}^i\circ\mu_R
&=&({\overrightarrow x_R} \times \overrightarrow P_R)^i \\
 {Y_\beta }^i\circ\mu_R
&=&(P_R^4\overrightarrow x_R -x_R^4\overrightarrow P_R)^i
\end{eqnarray*}
where $i=1,2,3,\overrightarrow x_R= (x_R^1, x_R^2,x_R^3),$\
and $
\overrightarrow P_R= (P_R^1,P_R^2,P_R^3)$.

Thus the function on ${ {\bf R}^4}\times \underline{\cal P}^*$
given by $P=(P^1,P^2,P^3,P^4)         \equiv (- {Y_{\gamma}^1}\circ\pi_2,
- {Y_{\gamma}^2}\circ\pi_2, - {Y_{\gamma}^3}\circ\pi_2,
 {Y_\delta}\circ\pi_2)$,can be considered as an abstraction of
{\bf linear momentum}: each inertial observer obtains
$f\circ\overline\phi_R(e)$
as the picture of
$e\in {\mbox{$\cal E$}}$, \ and the linear momentum he
measures is $P(f\circ\overline\phi_R(e))$.

                   In  the  same  way  the  components   of
$\overrightarrow   l=   (   -   Y_\alpha
^1\circ\pi_2, - {Y_\alpha^2}\circ\pi_2  ,
- {Y_\alpha^3}\circ\pi_2)$   and
$\overrightarrow g= ( Y_\beta^1\circ\pi_2,
Y_\beta^2\circ\pi_2     , Y_
\beta^3\circ\pi_2)$  must  be interpreted as
being  the  components   of  (relativistic)  {\bf   angular
momentum}.

If we denote  ${\cal P}_{+}^{\uparrow}$ by G, we can
summarize what has been said as follows

\vspace{1cm}

\fbox{ 
\begin{minipage}[t]{4.7in} 
                   State space of  our free particle  is such
that  each  inertial  observer  establishes a
diffeomorphism
from it to an orbit  of G in ${ {\bf R}^4}\times  \underline
G^*,$\ for the canonical action  .  Changes  of  inertial  observer  are  given by the
transformations given by the same action.

                   An inertial observer,  R, thus sees  any
state as a  pair, the first  component is an  event and the
second a movement containig the event.  The values at  that
point of $P,\overrightarrow  l,\overrightarrow g$, are  the
values  measured  by  R  of {\bf  linear momentum}  and   {\bf angular
momentum}.
\end{minipage}
}

\vspace{1cm}

In this paper I accept that this is also valid for all relativistic free particles, with or without mass or spin. More precisely,  it is assumed that
the possible state space of the relativistic free particles
are the orbits of G in ${ {\bf R}^4}\times \underline G^*$,
where now G is the  universal (two fold) covering group  of
 ${\cal P}_{+}^{\uparrow}$.    The  preceeding  interpretations   of
movements, states, linear and angular momentum are also considered as valid.

                   In  what follows,   we  assume  that  an
inertial observer, R,  has been fixed.  

If state  space is
the orbit $\cal O$ and $( x, \alpha)\in \cal O$,\ we have seen how 
  $\alpha$ \ can be considered as a movement. In fact, this movement is composed by the elements of  $\cal O$\ of the form  $( y, \alpha).$\ The set composed by such $y$\ , $M_{\alpha},$\ compose the ordinary portrait of the movement $\alpha$\ in ${\bf R}^4,$ obtained by the inertial observer R. Obviously
	\begin{equation}\label{puntpormovim}
	M_{\alpha}=\{ g*x:  g\in G_\alpha \}
	\end{equation}
where  $G_\alpha$  is  the
isotropy subgroup at $\alpha$\ of the coadjoint
representation.  

This way of looking at states also enables
us to determine all movements containing a given event: if $( x, \alpha)\in \cal O$,\ the movements containing the event $x$\ compose the set
\begin{equation}\label{movimporpunt}
N_x=\{ Ad^*_g\alpha:g\in G_x\}
\end{equation}
where $G_x$\ is the isotropy group at $x$\ of the action of $G$\ on ${\bf R}^4.$

\parindent=1cm
\parskip=5mm

\section{Universal covering group of the Poincar\'e Group. }
\label{sec-grupo}

It is a well known fact that the universal covering group of
Poincar\'e
group is a semidirect product of $\bf {SL}(2,\bf {C})$
\rm by a four dimensional
real vector space. In this section, I recall some general
facts about this group and stablishes the notation.

Let $(x^1,x^2,x^3,x^4)$ be the canonical coordinates in
$\bf
{R}^4$, I
the $2\times 2$ unit matrix and ${\sigma}_1, {\sigma}_2, {\sigma}_3$
 the Pauli matrices, $i.e.$
  $$
\left (\begin{array}{cc} 0 & 1\\
1 & 0
\end{array}
\right ),\
\left (\begin{array}{cc}
0 & -i\\
i & 0
\end{array}
\right ), \,\,
\left (\begin{array}{cc}
1 & 0\\
0 & -1
\end{array}
\right )
$$
respectively. A generic point of $\bf {R}^4$ will be denoted
$x=(x^1,x^2,x^3,x^4)$.

We define an isomorphism $h$, from $\bf {R}^4$ onto the
real vector
espace, $\bf {H}(2)$, of the hermitian $2\times 2$
matrices, by means of
$$
h(x)=x^4I+\sum _{i=1}^{3}x^i{\sigma}_i
=\left (\begin{array}{cc}
x^4+x^3 , & x^1-ix^2\\
x^1+ix^2 , &x^4-x^3
\end{array}
\right )
$$

We have $Det \, h(x)=<x,x>_{ m}$, where $<\, ,\, >_m $ is
 Minkowski pseudo-scalar product
$$
<x,y>_m=x^4y^4-\sum _{i=1}^3 x^iy^i.
$$
 If $x=(x^1,x^2,x^3,x^4)$\ we denote $\vec x=(x^1,x^2,x^3)$\ and $h(\vec x,x^4)=h(x).$

The following formulae are useful for many computations along this paper
\begin{equation}\label{conmherm}
	[h(\vec k,k_4),h(\vec x,x^4)]\stackrel {def}=h(\vec k,k_4)h(\vec x,x^4)-h(\vec x,x^4)h(\vec k,k_4)=2ih(\vec k \times \vec x,0),
\end{equation}
where $\times$\ means ordinary vector product,
\begin{eqnarray*}\label{anticomhertm}
\{h(\vec k,k_4),h(\vec x,x^4)\}& \stackrel {def}=&h(\vec k,k_4)h(\vec x,x^4)+h(\vec x,x^4)h(\vec k,k_4)=\\ &=& 2h(k_4\vec x +x^4 \vec k,k_4x^4+\langle \vec k,\vec x \rangle),	
\end{eqnarray*}
where $\langle \, . \, ,\, . \, \rangle$\ is the usual scalar product in $\mathbb{R}^3,$
\begin{eqnarray}\label{casicom}
h(\vec k,k_4)\varepsilon \overline{ h(\vec x,x^4)} \varepsilon-h(\vec x,x^4)\varepsilon \overline{ h(\vec k,k_4)} \varepsilon=\\\nonumber
=2[h(k_4 \vec x-x^4 \vec k,0)+ih(\vec k \times \vec x,0)]
\end{eqnarray}
where the bar means complex conjugation, and we denote by ${\varepsilon}$  the
matrix $i\sigma_2,\ i.e.$
$$
{\varepsilon}=\left (\begin{array}{cc}                   
0 & 1\\
-1 & 0
\end{array}
\right ).
$$

Notice that $${ }^t\negthinspace A\,{\varepsilon}\,A=(Det \,
A)\,{\varepsilon},$$
so that
\begin{equation}\label{Aepsi}
 A\,{\varepsilon}={\varepsilon}\,({
}^t\negthinspace A)^{-1} 
\end{equation}
if  $A\in \bf {SL}(2,\bf {C})$.

 Also we have
$$
\frac {1}{2}Tr(h(x){\varepsilon}\overline
{h(y)}{\varepsilon})=-<x,y>_m
$$
for all $x, y\in \bf {R}^4.$

We define an action on the left of the Lie group $\bf {SL}(2,\bf {C})$ on
the abelian Lie group $\bf {H}(2)$ by means of
$$
A*H=AHA^*
$$
for all $A\in \bf {SL}(2,\bf {C})$, $H\in\bf {H}(2)$, where $A^*$ is the
transposed of the complex conjugate of $A$. To this action by
automorphisms of $\bf {H}(2)$ then corresponds a semidirect product,
$\bf {SL}(2,\bf {C})\oplus \bf {H}(2)$, whose group law is given by
$$
(A,H)*(B,K)=(AB,AKA^*+H)
$$

The identity element is $(I,0)$ and $(A,H)^{-1}=(A^{-1},
-A^{-1}HA^{*^{-1}})$.

This semidirect product acts on the left on $\bf {R}^4$\
by means of
$$
(A,H)* x=h^{-1}(Ah(x)A^*+H).
$$
Poincar\'e group, ${\cal {P}}$, is identified to the closed subgroup of
$GL(5; \bf {R})$ composed by the matrices
$$
\left (\begin{array}{cc}
L & C\\
0 & 1
\end{array}
\right )
$$
where $C\in \bf {R}^4$ and $L\in {\cal {O}}(3,1)$ (such a
matrix is denoted in
the following simply by $(L,C)$).

For all $(A,H)\in \bf {SL}\oplus \bf {H}(2)$ (where
$\bf {SL}$ stands for
$\bf {SL}(2;\bf {C})$), there exists a unique $(L,C)\in
{\cal {P}}$ such
that $(A,H)*x=Lx+C$ for all $x\in \bf {R}^4$.

The map, $\rho$,  from $\bf {SL}\oplus \bf {H}(2)$ into ${\cal {P}}$ defined
by sending such
a $(A,H)$ to the corresponding $(L,C)$, is a homomorphism of Lie
groups, whose kernel consists of $(I,0)$ and
$(-I,0)$. In fact, $\rho$ is
a two fold covering map
of the identity component in ${\cal {P}}$,\ ${\cal {P}}_+^{\uparrow }$.\
Since $\bf {SL}$ and $\bf {H}(2)$ are connected and simply
connected it follows that
$\bf {SL} \oplus \bf {H}(2)$ is the universal covering group of ${\cal                    
{P}}_+^{\uparrow }$.  

The differential of   $\rho$ is an isomorphism from the Lie algebra of  $\bf {SL}\oplus \bf {H}(2)$ onto the Lie algebra of 
 ${\cal {P}}$.

The standard method to handle semidirect products enable us to identify
the Lie algebra of $\bf {SL}\oplus \bf {H}(2)$ with $\bf {sl(2,{\bf C})}\times \bf
{H}(2)$, the Lie bracket being
$$
[(a,k),(a',k')]=([a,a'], ak'+ k'a^*-(a'k+ka'^*)).
$$

If $(a,h)\in \bf {sl(2,{\bf C})}\times \bf{H}(2),\ t\in {\bf R},$\ we have
\
\begin{equation}\label{exp}
Exp[t(a,h)]	=\left(e^{ta},\int_0^t e^{sa}\,h\, e^{sa^*}\,ds \right).
\end{equation}

In this paper, we use the basis of
$\bf {sl}\times \bf
{H}(2)$ \ corresponding by $d\rho$ to the basis of ${\cal                    
{P}}$ associated to { linear momentum} and { angular momentum} in section \ref{clasical} . This basis is 
composed by the following elements

\begin{eqnarray}\label{limoang}
P^k  &=&(0,-\sigma_k),\ \ \ k=1,2,3,   \nonumber \\
P^4  &=&(0,\sigma_4)=(0,I),   \nonumber \\
l^k  &=&(i\frac {\sigma_k}2,0),    \\
g^k  &=&(\frac {\sigma_k}2,0).  \nonumber
\end{eqnarray}

An element, $X$,\ of $\bf {sl(2,{\bf C})}\times \bf {H}(2)$\ defines a (linear) function on \\ $(\bf {sl(2,{\bf C})}\times \bf
{H}(2))^*.$\ Its restriction to {\bf movement space} ( a coadjoint orbit) must be considered as a {\bf  dynamical variable}. But also, if we have a representation on a vector space (resp. an action on a manifold) of $\bf {SL}\oplus \bf {H}(2),$\ $X$ gives rise to a infinitesimal generator of the representation (resp. the action), {\it i.e.},  an \bf endomorphism \rm of the vector space (resp. a \bf vector field \rm on the manifold). 

We define vector valued functions on the dual of the Lie algebra as follows
\begin{eqnarray*}
P&=&(P^1,\,P^2,\,P^3,\,P^4)  \\
\overrightarrow{l}&=&(l^1,\,l^2,\,l^3)\\
\overrightarrow{g}&=&(g^1,\,g^2,\,g^3)
\end{eqnarray*}
what will be considered as being {\bf linear momentum} and  {\bf angular momentum}.

We define a non degenerate scalar product in $\bf {sl}\times
\bf {H}(2)$ by means of
\begin{eqnarray}
<(a,k),(b,l)> & =&2Re Tr (\frac{1}{4}k{\varepsilon}\overline
{l}{\varepsilon}-ab)=  \nonumber \\
& =&\frac {1}{2}Tr(k{\varepsilon}\overline
{l}{\varepsilon})-2 Re Tr\, ab.  \nonumber
\end{eqnarray}

This scalar product defines in the standard way an isomorphism from the
Lie algebra of $\bf {SL}\oplus \bf {H}(2)$ onto its dual. The image   of $(a,k)\in
\bf {sl}\times \bf {H}(2)$   by this isomorphism,  will be denoted by $\{a,k\} \in \left(\bf {sl}\times \bf {H}(2)\right)^{*}$, and is given by
$$
\{a,k\}\left((b,m)\right)=<(a,k),(b,m)>.
$$

With this notation, the values of $P,\ \overrightarrow{l}$ {\rm and} $\overrightarrow{g}$\ at $\{a,k\}$\ are, when 
 written in terms of its hermitian form 
\begin{eqnarray}\label{vardin}
h(P( \{ a,\ k \}))&=&-k, \\
h(\vec l(\{ a,\ k \}),\ 0)&=& i\,(a^*-a), \\
h(\vec g(\{ a,\ k \}),\ 0)&=& -(a+a^*).
 \end{eqnarray}
so that
\begin{equation}\label{paralgP}
	\{a,k\}=\{-\frac{1}{2}h(\vec{g}(\{ a,\ k \}),0)+\frac{i}{2}h(\vec{l}(\{ a,\ k \}),0),-h(P( \{ a,\ k \}))\}.
\end{equation}
\emph{i.e.} $-(1/2)h(\vec{g}(\{ a,\ k \}),0)$\ is the hermitian real part of $a$\ and the matrix $(1/2)h(\vec{l}(\{ a,\ k \}),0)$\ its hermitian imaginary part.

A straightforward computation, leads to  the following formula for the coadjoint
representation 
\begin{equation}\label{coadjunta}
Ad^*_{(A,H)}\{a,k\}=\{AaA^{-1}+\frac {1}{4}(AkA^*{\varepsilon}\overline
{H}{\varepsilon}-H{\varepsilon}\overline {AkA^*}{\varepsilon}),
AkA^*\}. 
\end{equation}

To end this section, we define  other  functions on the
dual of the  Lie  algebra  of  $\bf  {SL}\oplus  \bf  {H}(2).$\

One of these is $\vert P \vert $,\  defined by
$$
\left| P \right| ( \{a, k\})= Det\,\left( h(P( \{ a,\ k \})) \right)=Det\,\left( k \right),
$$
 whose physical meaning is \bf {mass square}\rm. 

Then,
$$
\left| P \right| ( Ad^*_{(A,H)}\{a, k\})=Det (-AkA^*)=Det(k),
$$
so that the value of $\vert P \vert $,\ is constant along any coadjoint orbit.

The other is defined in terms of the Pauli-Lubanski fourvector, given by
\begin{eqnarray}
W&=&(\overrightarrow{ W},W^4)\\
	\overrightarrow{ W}&=&P^4 \vec l+\vec P \times \vec g,\label{P-L3}\\
	W^4&=&\langle \vec P,\vec l \rangle.\label{P-L4}
\end{eqnarray}

 Using \eqref{paralgP}, \eqref{conmherm} and \eqref{anticomhertm}, the hermitian form of $W$\ is found to be
\begin{equation}\label{P-L1}
h(W(  \{a, k\}))= i(a\ k\ -k\ a^*).	
\end{equation}

         One can prove that
$$h(W(Ad^*_{(A,H)}\{a,\ k\}))=A\ h( W(\{a,\ k\}))\ A^*,$$ so that
the function  $$\vert W \vert(  \{a, k\})= Det(h(W( \{a, k\}))),$$ is also constant along each coadjoint orbit.

\parindent=1cm
\parskip=5mm

\section{Quantizable forms}\label{quantizable}

                  In this section I recall some of the geometric constructions done in \cite{souquant, adm182,adm282,admacad,adm95}. 

 I shall give a way  to construct homogeneous contact manifolds that fibers  on the coadjoint orbits of Lie groups. This is not possible on an arbitrary orbit, but only on the so called \bf quantizable\rm \ ones.

The idea is to consider the quantizable coadjoint orbits of $\bf {SL}(2;{\bf C}) \oplus \bf {H}(2)$ as the classical movement space of  relativistic free particles. The corresponding homogeneous contact manifolds are  geometrical objects enabling us to construct the Wave Functions, and representations of this group.

                   Let $G$ be a Lie group and  $\alpha $\  an element of the dual of the Lie algebra of $G$, $\underline
G^*$, where we consider the \bf coadjoint action.\rm

We say that , $\alpha $ is \bf quantizable\rm \  if there exists a surjective
homomorphism, $C_\alpha$,\  from the  isotropy subgroup  at
{$\alpha$},\   {$G_\alpha$}   ,   onto   the  unit  circle,
{\mbox{${\bf   S}^1   \rm   $}},   whose   differential  is
{$\alpha$} (cf. section \ref{Notation} where the way in which the differential is identified to an element of $\underline
G^*$, is detailed). 

In a more explicit way, this means that {$\alpha$} is \bf quantizable\rm \ if there exists a surjective
homomorphism, $C_\alpha,$\  from the  isotropy subgroup  at
{$\alpha$},\   {$G_\alpha,$}\  onto   the  unit  circle, such that, for all $X$\ in the Lie algebra of {$G_\alpha$} and t  in \bf R\rm  \[
C_{\alpha}(Exp\,tX)=e^{2 \pi i t {\alpha}(X)}.
\]

 If a form is quantizable, all the elements of its coadjoint orbit are quantizable. These orbits are called
 \bf quantizable orbits \rm.

The form {$\alpha$}  is said  to be  \bf R-quantizable  \rm if
there exists a  surjective homomorphism  from
{$G_\alpha$}  onto the usual additive Lie group of reals, \bf  R   \rm,  whose  differential  is
{$\alpha$}. 

In other words, {$\alpha$}  is   \bf R-quantizable  \rm if
there exists a  surjective homomorphism  from
{$G_\alpha$}  onto  \bf  R  \rm  ,\ $H_{\alpha}$,\  such that, for all $X$\  in the Lie algebra of {$G_\alpha$}, and $t$\   in \bf R\rm   \[
H_{\alpha}(Exp\,tX)=t\,{\alpha}(X).
\]

To such a $H_{\alpha}$, one can associate an homomorphism, $C_{\alpha}$, onto ${\bf S}^1$ \ by means of \[
C_{\alpha}:g \in {G_\alpha}  \rightarrow      e^{2 \pi i H_{\alpha}(g)} \in  {\mbox{${\bf   S}^1   \rm   $}}.
\]
The differential of $C_{\alpha}$ is $\alpha$. Thus when $\alpha  $ is \bf  R-quantizable \rm ,\ 
it is \bf quantizable \rm.

All elements of a coadjoint orbit containing a  R-quantizable  form, are  R-quantizable. In this case, the orbit is  said to be a \bf R-quantizable  orbit\rm.

 In \cite{adm95} a  slightly
more general concept of  quantizability is used, but  it is
unnecesary for the purposes of the present paper.

                   In what follows we assume that  $\alpha$
is  quantizable  and  $C_\alpha$\  is  a  homomorphism from
{$G_\alpha$} \bf onto \rm the unit  circle, whose  differential is
{$\alpha$}.   We  identify  the  coadjoint  orbit, ${\cal O}_{\alpha},$\   with
$G/G_{\alpha}$ by means of the diffeomorphism
$$
g G_{\alpha} \in \frac{G}{G_{\alpha}} \rightarrow Ad^*_g \alpha \in {\cal O}_{\alpha}.
$$

                   We define an action of ${\bf S}^1$  on
$G/Ker\,C_\alpha$  by   means  of
\begin{equation}\label{bundle}
  (g\,Ker\,C   _\alpha
)*s=g\,h\,Ker\,C_\alpha
\end{equation}
 where  $h$  is  any  element  of
$G_{\mbox{$   \alpha   $}}$   such   that  $C_\alpha(h)=s$.

Actually $(G/Ker\,C_\alpha)$ $(G/G_{\mbox{$\alpha$}},\bf  S
\rm ^1)$\ is a principal  fibre bundle , the bundle  action
 is given by \eqref{bundle}  and the  bundle projection, by the
canonical map, 
$$
\pi: g{KerC_\alpha}  \longrightarrow  g {G_\alpha}
$$

This action of ${\bf S}^1$,\ commutes with the canonical action of $G$ on \\ $G/Ker\,C_\alpha$.

We denote by $\pi^c$\ and $\pi^s$\ the canonical maps

\begin{eqnarray}
\pi^c&:& g\in G \longrightarrow g {KerC_\alpha} \in \frac{G}{{KerC_\alpha}}\\
\pi^s &:&  g\in G \longrightarrow g {G_\alpha} \in \frac{G}{{G_\alpha}}.
\end{eqnarray}

There exist an unique 1-form, $\Omega$,\    on the homogeneous space 
$G/Ker\,C_\alpha$\ such that 
$$
(\pi^c)^*\Omega=\alpha.
$$

The 1-form $\Omega$\ is a contact form, invariant under the action of $G.$

                   Let  $Z(\Omega)$   be  the   vectorfield
defined by  $$ 
i_{Z  (\Omega)} \Omega=  1,\ \ i_{Z(\Omega)}
\,d\Omega=0. 
$$

 All the integral curves of $Z(\Omega)$ have
the same period. If         we         denote          by
${T(\Omega)}$ this period, then
$\Omega/{T(\Omega)}$  is  a   connexion
form. 

 Since the structural group is abelian, the curvature
form is  $d\Omega/{{T(\Omega)}} $. 

 There
exist an  unique 2-form, $\omega,$\   on  $G/G_{\alpha}$,\  such that
$$
\pi_*\omega=\frac{d\Omega}{{T(\Omega)}}.
$$

Thus
\begin{equation}\label{omega}
(\pi^s)^*\omega=\frac{d\alpha}{T(\Omega)}.
  \end{equation}
	
	The
form  $\omega$\ is    an invariant 
 symplectic form, and   its
cohomology class is  integral.

Each  $m\in \underline G,$\ define a function on $\underline G^*$\ so that it does also on the coadjoint orbit. Since we think in the coadjoint orbit as being the movement space of a particle, we can say that m defines a dynamical variable,  $D_m.$
Such a $m\in \underline G,$\ also define a infinitesimal generator of the canonical action of $G$ on $G/Ker\,C_\alpha$\ , denoted by $X_m^c,$\ and  a infinitesimal generator of the canonical action of $G$ on $G/G_\alpha$\ , denoted by $X_m^s.$

Since $\Omega$\ is left invariant by the action of $G,$\ we have
$$
L_{X_m^c}\Omega=0,
$$
so that the relation $L_X=d\,i_X+i_X\,d$ leads to
$$
i_{X_m^c}\,d\Omega=-d[\Omega(X_m^c)].
$$

A computation gives
$$\begin{gathered}
(\Omega(X_m^c) \circ \pi^c)(g)= Ad^*_g \alpha_0 (-m)=-D_m(Ad^*_g \alpha_0)=\\=- D_m(g\, G_\alpha)=-D_m\circ \pi (g\, Ker\,C_\alpha),
\end{gathered}
$$
so that
\begin{equation}\label{funcabaj}
\Omega(X_m^c)=-D_m\circ \pi.
\end{equation}

Thus \eqref{omega}, \eqref{funcabaj}, and the fact that
$$
\pi_* X_m^c=X_m^s,
$$
lead to
\begin{equation}\label{hamilt}
i_{X_m^s}\omega=\frac{1}{T\left( \Omega  \right)}d D_m.
\end{equation}

Since the flow of the vector field $X_m^s$ (resp. $X_m^c$) preserves $\omega$ 
(resp. $\Omega$), it is an infinitesimal automorphism of the symplectic (resp. contact) structure, \emph{i.e.} a locally hamiltonian vector field. Equation \eqref{hamilt} tell us that in fact 
$X_m^s$ is  globally hamiltonian and $ D_m/{T\left( \Omega  \right)}$ is the corresponding hamiltonian.

As usual in the Theory of Connections,  a map, $f$,\  into $G/{KerC_\alpha}$\ is called \emph{horizontal}, if $f^{*}\Omega=0.$
Let $f_0$\ be  a map into $G/{G_\alpha}$.\ An \emph{horizontal lift} of $f_0$ is an horizontal map into $G/{KerC_\alpha}$,\ such that $\pi \circ f =f_0$.

                   The  horizontal  lift  of  curves can be
described  as  follows.     Given  a  curve,  $\gamma$   in
$G/G_{\mbox{$\alpha$}}$, the horizontal lift of $\gamma$ to
$g\,KerC_\alpha$ is
 \begin{equation}
 \widetilde \gamma\left( t  \right)=
\left(  \overline  \gamma  \left(  t  \right)   KerC_\alpha
\right) * e^{-2 \pi i \int _{\overline \gamma |_{[0,\  t]}}
\alpha}    \label{horiz}
 \end{equation}
  where  $\overline  \gamma$  is  any lifting of
$\gamma$ to G, such that $\overline \gamma(0)=g$, and the
vertical bar means restriction.

                   Associated  to   this  principal   fibre
bundle and the canonical action of ${\bf S}^1$
on ${\mathbb C}$, one can consider the 1-
dimensional vector bundle whose total space is
 $$
\left( \frac{G}{KerC_\alpha}\right) \times_{{\bf S}^1}  {\mathbb C}.
$$
 Let us recall its definition.

Consider in $\left( G/KerC_\alpha \right) \times
 {\mathbb C}$,\ an action of ${\bf S}^1$ defined by
$$
 (g\, KerC_\alpha , t)*z=\left(\left(g\, KerC_\alpha\right)*z ,z^{-1} t\right),
$$
for all $z\in {\bf S}^1$.

The elements of $\left( G/
KerC_\alpha
\right)
 \times_{{\bf S}^1}
  {\mathbb C}$\ are the orbits of this action.

Let us denote by $[g\, KerC_\alpha , t]$ the orbit of $(g\, KerC_\alpha , t).$\  We have 
\begin{eqnarray}
\nonumber [g\, KerC_\alpha , t]&=&\left\{ \left(\left(g\,KerC_\alpha\right)*s, s^{-1} t\right):\ s\in S^{\rm 1}\right\} = \\  &=& 
\nonumber       \left\{ \left(gh\,KerC_\alpha, C_\alpha\left(h^{\rm -1}\right) t\right):\ h\in G_\alpha\right\}
\end{eqnarray}

 We define a map, $\overline{\pi}$,\ from 
$$ \frac{G}{KerC_\alpha}   \times_{{\bf  S }^1}   {\mathbb C}$$
 onto the coadjoint orbit by means of
 \[\overline{\pi}([g\, KerC_\alpha , t])=g\,G_\alpha\].

Let $m\in G/G_\alpha$, and $g\in G$  one of its representatives. Since the action of 
{${\bf  S}^1$} on $\pi^{-1}(m)$ is transitive, we have
$$
{\overline \pi}^{-1}(m)=\{[g\,Ker\,C_{\alpha},t]:t\in \bf C\rm\}. 
$$

Thus, for each $g$ we obtain a bijection of ${\overline \pi}^{-1}(m)$ onto $\bf C$.

 We consider in ${\overline \pi}^{-1}(m)$ the structure of hermitian one dimensional complex vector space such that this bijection is a unitary isomorphism.
 
This structure is independent of the representative $g$. 

If $g$ and $g'$ are representatives of $m$,\ we have for all $t,\ t^\prime,\ a \in \bf C$

 \[ \left[g\,KerC
_\alpha  ,\   t\right]  +   \left[g^\prime\,KerC_\alpha  ,\
t^\prime    \right]=\left[g\,KerC_\alpha    ,\   t+C_\alpha
(g^{-1}\,g^\prime)\,t^\prime\right]\],

\[  a  \cdot
\left[g\,KerC _\alpha ,\ t\right] =\left[g\,KerC_\alpha  ,\
a\,t\right],\]

  \[  
	\langle    \left[g\,KerC
_\alpha
,\    t\right]   ,
\left[g^\prime  \,  KerC
_\alpha
   ,\  t^\prime  \right]   \rangle=
 {\overline t}\,{C_\alpha (g^{-1}g^\prime)\,t^\prime} .\]

With these operations, $\overline\pi$ becomes a complex vector bundle of dimension
 one with a hermitian product in each fiber i.e. a \bf hermitian line bundle.\rm

                The
sections of the hermitian line  bundle are in a one  to one
 correspondence
with the functions on $G/{KerC_\alpha} $, f, such that f((g
${\mbox{$Ker\,C_\alpha$}}    )\,*\,s)=    s^{-1}    $   f(g
${\mbox{$Ker\,C_\alpha$}}  $), for all $s\in {\mbox{$\bf S \rm  ^1$}}$.    These  functions  will be
called from now on \bf pseudotensorial functions \rm . This
correspondence is as  follows. 

 If  f is a  pseudotensorial
function, the corresponding section sends $m \in G/G_\alpha$
to  $\left[  r,f(r)  \right]$  where  r is arbitrary in  $\pi^{-1}(m)$.  

 If $\sigma$  is a given section  of the hermitian
line bundle, the corresponding pseudotensorial function, f,
is defined by  $\sigma (\pi (r))  = \left[ r,f(r)  \right]$
for all $ r \in G/Ker C_\alpha$.

The canonical action of $G$ on $\left( G/
KerC_\alpha\right)$, leads to an action on
 $$
\left( G/
KerC_\alpha
\right)
 \times_{\mbox{$\bf  S  \rm  ^1$}}
 {\mbox{\boldmath {$C$}} \rm},
$$
\ given by 
\[
g*[h\, Ker C_\alpha,t]=[gh\, Ker C_\alpha,t].
\]

This action is well defined, as a consecuence of the fact that the action of {\mbox{$\bf  S  \rm  ^1$}} conmutes with the canonical action of $G$.

In the following, I use the  same construction of a hermitian line bundle, for each principal bundle, $ G/(Ker\,C)(G/H,S),$\ where $C$ is an homomorphism of $H$ into {\mbox{$\bf  S  \rm  ^1$}}, whose image is $S$.

\parindent=1cm
\parskip=5mm

\section{Geometric  Quantum States .}\label{quantum}

                  In this section I recall some definitions and results from \cite{adm96}.

Let $G=SL(2,{\bf C}) \oplus H(2)$ and $\alpha$ a quantizable form of $G$. We  use the notation of  section \ref{quantizable}.

The  sections  of  the  hermitian   line
bundle whose total space is \[  \left( G/
KerC_\alpha
\right)
 \times_{\mbox{$\bf  S  \rm  ^1$}}
 {\mbox{\boldmath {$C$}} \rm} \] are  called  \bf  Prequantum  States  \rm  .
We use the same denomination for the
corresponding pseudotensorial functions.

Now, let us consider the actions of the abelian subgroup $\{I\}\times H(2)$ on $G/Ker\,C_\alpha$\ and $G/G_\alpha$,\ induced by the canonical action.

There exist an unique action of
$\{I\}\times H(2)$
 on $G/Ker C_\alpha$
whose orbits are horizontal and such that $\pi$ becomes
equivariant.

 This action is  called \bf horizontal action \rm and is given by
 \begin{equation}
(I,\ K)*((A,\ H)\,KerC_\alpha)=((A,\ H+K)\,KerC_\alpha)*e^
{-i \pi Tr
\left (
AkA^* \varepsilon \overline K \varepsilon \right )}
\label{achor}
 \end{equation}
for all $ K \in H(2),\ (A,\ H)  \in G $,
where $*$ in the left hand side stands for the new
 action and in the right hand one, corresponds
to the bundle action. k is given by $\alpha = \{a,\ k\}$.

 We define \bf    Quantum   States \rm as being the Prequantum
States that correspond to
pseudotensorial functions left invariant by the
horizontal  action.

Let $\pi_1$ and $\pi_2$ be the canonical projections of
$G$ on ${ SL(2,\bf C)}$ and $ H(2)  $ respectively.
We denote $\pi_1(G_\alpha)$ by  ${\mbox{$(G_\alpha)_{SL}$}}$
and $\pi_2(G_\alpha)$ by ${\mbox{$(G_\alpha)_{H}$}}$.

In section 4 of \cite{adm96 } it is proved that the  map  
$$
(C_{{\mbox{$\alpha$}}})_{SL}                              :
(G_{{\mbox{$\alpha$}}})_{SL} \longmapsto   S^1  ,
$$

defined    by  
\begin{equation}\label{calfasl}
  (C_{{\mbox{$\alpha$}}})_{SL}(g)\   =\
C_{{\mbox{$\alpha$}}}(g,h)\     e^{-i\pi     Tr\left     (k
{\mbox{$\varepsilon$}}  \overline   h{\mbox{$\varepsilon$}}
\right ) }, 
\end{equation}
 for all $(g,h)\in{\mbox{$G_\alpha$}},$\  is
well defined and a homomorphism.

 As a consecuence, we can  define 
\begin{equation}\label{widecalfasl}
\widetilde  C_\alpha              :
(G_\alpha)_{SL}\  \oplus\   H(2)
 \longmapsto    S^1 , 
\end{equation}
 by means of
 \[ \widetilde
{C}_{{\mbox{$\alpha$}}}(g,r)\                            =\
(C_{{\mbox{$\alpha$}}})_{SL}(g)\         e^{i\pi         Tr
\left (
k{\mbox{$\varepsilon$}} \overline r{\mbox{$\varepsilon$}}
\right )}.
\]

$ \widetilde    C_{{\mbox{$\alpha$}}}$ is
 an extension of
$C_\alpha$ to $
(G_{{\mbox{$\alpha$}}})_{SL}\  \oplus\  {{\mbox{$   H(2)
\rm$}}},$
and a homomorphism
. Its differential coincides with the restriction
of $\alpha $ to the Lie algebra of this group.

 The canonical action of $G$ on $G/KerC_\alpha$ maps
horizontal orbits to horizontal orbits, thus defining a
transitive   action on the space of horizontal orbits, ${\cal W}_{\alpha}.$

Let us consider the canonical map
$$
\tau:\frac{G}{KerC_\alpha} \rightarrow {\cal W}_{\alpha}
$$
defined by sending each element of $G/KerC_\alpha$ onto its horizontal orbit.

The
isotropy subgroup at  $\tau(KerC_\alpha)$\ 
is $Ker\widetilde C_\alpha$.\    As   a   consequence, we  can identify ${\cal W}_{\alpha}$ to
$G/Ker\,\widetilde C_{\alpha},$\ by means of the bijective  map
 $$
(A,H)\,Ker\,\widetilde C_{\alpha}\in \frac{G}{Ker\,\widetilde C_{\alpha}}\rightarrow \tau((A,H)Ker\,C_{\alpha}) \in {\cal W}_{\alpha}.
$$

With the definitions given in section \ref{quantizable}, we have seen that 
$$
\frac{G}{Ker\,C_{\alpha}}\left(\frac{G}{G_{\alpha}},{\bf S}^1\right)
$$
becomes a principal fibre bundle.

Similar definitions provides 
$$
\frac{G}{Ker\,\widetilde C_{\alpha}}\left(\frac{G}{(G_{\alpha})_{SL}\oplus H(2)},{\bf S}^1\right)
$$
with a structure of principal fibre bundle.

The bundle projection is the canonical map from ${G}/{Ker\,\widetilde C_{\alpha}}$\ onto \\
${G}/({(G_{\alpha})_{SL}\oplus H(2)}).$
The bundle action is given by
$$
((A,H){Ker\,\widetilde C_{\alpha}})*s=(A,H)(B,K){Ker\,\widetilde C_{\alpha}},
$$
where $(B,K)$\ is such that 
$$
\widetilde C_{\alpha}(B,K)=s.
$$

 But the map 
\begin{equation}\label{identif}
(A,H)((G_\alpha)_{SL} \oplus  H(2))\in \frac{G}{(G_{\mbox{$\alpha$}})_{SL} \oplus  H(2)}\rightarrow A\,(G_{\alpha})_{SL} \in
\frac{SL}{(G_{\alpha})_{SL}}
\end{equation}
is a diffeomorphism.

We identify these homogeneous spaces by means of this map, so that the principal fibre bundle becomes
$$
\frac{G}{Ker\,\widetilde C_{\alpha}}\left(\frac{SL}{(G_{\alpha})_{SL}},{\bf S}^1\right),
$$
where the bundle projection is
$$
\tau_2:(A,H){Ker\,\widetilde C_{\alpha}}\rightarrow A\,(G_{\alpha})_{SL}.
$$

The canonical maps define
 the
homomorphism of principal {\mbox{${\bf S}^1 \rm$}}-bundles given in Figure \ref{diagr1}

\begin{figure}[h]
\begin{center}
$
\begin{psmatrix}
\frac{G}{KerC_\alpha} \ \  \  & \ \  \   \frac{G}{Ker \widetilde C_\alpha}    \\
	\frac{G}{G_\alpha}\ \  \    & \ \  \  \frac{SL}{(G_\alpha)_{SL}}
\psset{arrows=->,nodesep=3pt}
\ncline{1,1}{1,2}^{\iota_1}
\ncline{->}{2,1}{2,2}^{\iota_2}
\ncline{->}{1,1}{2,1}<{\tau_1}
\ncline{->}{1,2}{2,2}>{\tau_2}
\end{psmatrix}
$
\end{center}
\caption{Fibre Bundles for Quantum States}
\label{diagr1}
\end{figure}

\parindent=1cm
\parskip=5mm

Since  Quantum   States  correspond   to
pseudotensorial functions left invariant by the  horizontal
action, 
\\
\framebox{  
\begin{minipage}[t][1.2\height][c]{4.3 in}
              Quantum   States   are the pull back by $\iota_1$\  of \bf unrestricted \rm pseudotensorial functions 
 on $G/Ker\widetilde C_\alpha$.
\end{minipage}
}

\parindent=1cm
\parskip=5mm

\section{Wave Functions}\label{wave}

 Now, let us  associate, to each Quantum State in the sense of section \ref{quantum}, a Wave Function in the ordinary sense of Quantum Mechanics.

From now on, we assume, unless the contrary is explicitly stated, that  the State Space  is  the orbit of $(0,\alpha)$ in $H(2)\times \underline G^*.$  

This choice do not carry very important consequences for the study of the free particle at the quantum level, in the sense that the other choices lead to isomorphic spaces of Quantum States (in all of its forms), and the same Wave Functions. These facts are proved in Remark 5.2 of \cite{adm96}.

The isotropy subgroup at  $(0,\alpha),$\ $G_{(0,\alpha)},$ is composed by the elements of $G_\alpha$ of the form $(A,0)$ \emph{i.e.}
$$
G_{(0,\alpha)}=G_\alpha \cap (SL \oplus \{0\}).
$$

Let $\alpha=\{a,k\}$ and denote
$$
SL_1=\{A\in SL: AaA^{-1}=a\},
$$
$$
SL_2=\{A\in SL: AkA^*=k\}.
$$

Then 
$$
G_{(0,\alpha)}=(SL_1 \cap SL_2)\oplus \{0\},
$$

Notice that $(G_\alpha)_{SL}\subset SL_2$ and $SL_1 \cap SL_2\subset(G_\alpha)_{SL}$ so that 
$$
SL_1 \cap SL_2=(G_\alpha)_{SL}\cap SL_1.
$$

But the map
$$
(A,H)((G_\alpha)_{SL}\cap SL_1)\oplus \{0\} \in \frac{G}{((G_\alpha)_{SL}\cap SL_1)\oplus \{0\}}\rightarrow $$$$  \rightarrow (H,A((G_\alpha)_{SL}\cap SL_1))\in H(2) \times \frac{SL}{(G_\alpha)_{SL}\cap SL_1} 
$$ 
is a diffeomorphism.

Thus, State Space is the image of the injective map
\begin{equation}\label{iii}
i:(H,A(G_\alpha)_{SL}\cap SL_1))\in H(2) \times \frac{SL}{(G_\alpha)_{SL}\cap SL_1} \rightarrow
\end{equation}
$$ 
 \rightarrow (H, Ad^*_{(A,H)}\alpha)\in H(2)\times \underline G^*.
$$

This image need not, a priori, be  a proper submanifold of $H(2)\times \underline G^*,$\ and we consider it provided with the topology and differentiable structure such that $i$ becomes a diffeomorphism. In the following we identify each $i(X)$ with $X,$ so that we can say that
$$
H(2) \times \frac{SL}{(G_\alpha)_{SL}\cap SL_1} 
$$
is State Space.

The canonical map from State Space onto Movement Space, can be generalised 
 to all the  homogeneous  spaces  
appearing in the
commutative diagram of Figure 1. This will be done with the following
 geometrical construction.

						Let $ \cal L $\ be a closed subgroup  of
G, and \[ {\cal S}=\{ s\in SL : (s,0) \in {\cal L} \},
\]
so that \[
{\cal S} \oplus \{ 0\}= {\cal L} \cap		(SL \oplus \{0\}).
\]	
			
Thus we define the map
\[
(H,A	{\cal S})\in {H(2) \times \frac{SL}{\cal S} }  \stackrel{\nu}{\longrightarrow}  (A,H) {\cal L} \in \frac{G}{\cal L}.
\]
						
This map is  well	defined and, if $H$ is a fixed element in $H(2)$, its restriction to \[ \{H\} \times \frac{SL}{\cal S} \] 
is injective.

					Now we apply this   to  the  cases in Figure 1.

						When ${\cal L}=Ker\,C_\alpha$ we have ${\cal S}=Ker(C_\alpha)_{SL} \cap SL_1$ ,\  so that we have a map		
								\[
		(H,A	( Ker(C_\alpha)_{SL} \cap SL_1))\in {H(2) \times \frac{SL}{ Ker(C_\alpha)_{SL} \cap SL_1 }}   \stackrel{\nu_1}{\longrightarrow} 
		\]  
		\[
  \stackrel{\nu_1}{\longrightarrow} (A,H) {Ker\,C_\alpha} \in \frac{G}{Ker\,C_\alpha}.		
	\]					
	
		If 	 ${\cal L}=G_\alpha$ we have ${\cal S}= 	SL_1 \cap SL_2=(G_{\alpha})_{SL} \cap SL_1 $,\ and we thus obtain the map 
$$
(H,A	((G_{\alpha})_{SL} \cap SL_1 ))\in {H(2) \times \frac{SL}{(G_{\alpha})_{SL} \cap SL_1 }}   \stackrel{\nu_2}{\longrightarrow} 
	$$$$
  \stackrel{\nu_2}{\longrightarrow} (A,H) {G_\alpha} \in \frac{G}{G_\alpha}.		
	$$						

When ${\cal L}=Ker{\widetilde C}_\alpha$ we have ${\cal S}=  Ker(C_\alpha)_{SL}, $ \  so that we have a map		
$$
	(H,A	 Ker(C_\alpha)_{SL})\in {H(2) \times \frac{SL}{ Ker(C_\alpha)_{SL}}}   \stackrel{\nu_3}{\longrightarrow} 
		  (A,H) {Ker\,\widetilde C_\alpha} \in \frac{G}{Ker\,\widetilde C_\alpha}.	
$$

	If 	 ${\cal L}=(G_\alpha)_{SL} \oplus H(2)$ we have ${\cal S}= (G_\alpha)_{SL} $,\ and we  thus	have a map 
\[
		{(H,A	(G_\alpha)_{SL}))\in {H(2) \times \frac{SL}{(G_\alpha)_{SL}}} }  \stackrel{\nu_4}{\longrightarrow} 
		\]  
		\[
  \stackrel{\nu_4}{\longrightarrow} (A,H) {((G_\alpha)}_{SL}\oplus H(2)) \in  \frac{G}{(G_\alpha)_{SL}\oplus H(2)} 
	\]
which, using the identification \eqref{identif}, can be written
\[
		{(H,A	(G_\alpha)_{SL}))\in {H(2) \times \frac{SL}{(G_\alpha)_{SL}}} }  \stackrel{\nu_4}{\longrightarrow} 
		\]  
		\[
  \stackrel{\nu_4}{\longrightarrow}		A		(G_\alpha)_{SL}	\in \frac{SL}{(G_\alpha)_{SL}}.
			\]

\begin{figure}
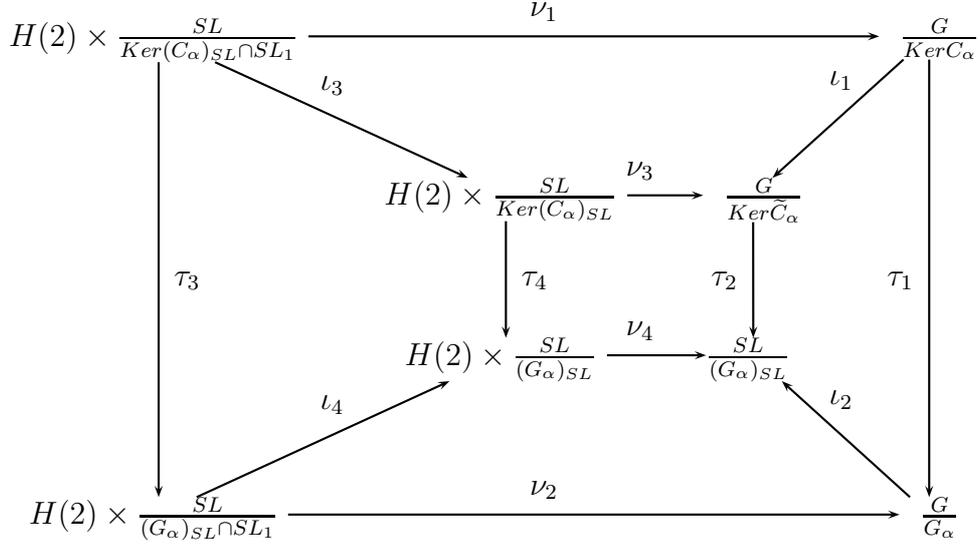

  \centering
$
\begin{psmatrix}[colsep=1cm,rowsep=1.5cm]
H(2) \times \frac{SL}{Ker(C_\alpha)_{SL} \cap SL_1}\ \   & & &\ \   \frac{G}{KerC_\alpha}   \\
   &  H(2) \times \frac{SL}{Ker(C_\alpha)_{SL}} \ \   &\ \    \frac{G}{Ker \widetilde C_\alpha}  &  \\
	  & H(2) \times \frac{SL}{(G_\alpha)_{SL}}\ \    &  \frac{SL}{(G_\alpha)_{SL}}\ \   &  \\
H(2) \times \frac{SL}{ (G_\alpha)_{SL}\cap SL_1}\ \   & & &\ \  		\frac{G}{G_\alpha}
\ncline{->}{1,1}{1,4}^{\nu_1}
\ncline{->}{4,1}{4,4}^{\nu_2}
\ncline{->}{2,2}{2,3}^{\ \ \nu_3}
\ncline{->}{3,2}{3,3}^{\ \ \nu_4}
\ncline{->}{1,1}{4,1}>{\tau_3}
\ncline{->}{2,2}{3,2}>{\tau_4}
\ncline{->}{2,3}{3,3}<{\tau_2}
\ncline{->}{1,4}{4,4}<{\tau_1}
\ncline{->}{1,1}{2,2}^{\iota_3}
\ncline{->}{1,4}{2,3}^{\iota_1}
\ncline{->}{4,1}{3,2}^{\iota_4}
\ncline{->}{4,4}{3,3}^{\iota_2}
\end{psmatrix}
$
\caption{Fibre Bundles for Wave Functions}
\label{diagr2}
\end{figure}

When we denote by $\iota_3,\ \iota_4,\ \tau_3,\ \tau_4,$\ the  canonical maps of homogeneous spaces that appears in Figure \ref{diagr2}, we obtain the  commutative diagram in that Figure. 

The maps $\tau_i$\ are the bundle maps of 
principal  fibre  bundles   whose  structural  groups   are
identified by means of $(C_\alpha)_{SL},$\
$\widetilde C_\alpha$
,\ 
or $C_\alpha $ to subgroups of
$S^1$. 

We already know the bundle actions of ${\bf S}^1$\ on $G/KerC_\alpha$\ and $G/Ker\widetilde C_\alpha.$\ The other are defined in a similar way as follows.

The action for
 the principal bundle corresponding to $\tau_4,$
$$
H(2) \times \frac{SL}{Ker(C_\alpha)_{SL}}\left(H(2) \times \frac{SL}{(G_\alpha)_{SL}},(C_\alpha)_{SL}((G_\alpha)_{SL})\right),
$$
is given by
$$
(H,A{Ker(C_\alpha)_{SL}})*s=(H,AB{Ker(C_\alpha)_{SL}})
$$
if $s\in (C_\alpha)_{SL}((G_\alpha)_{SL}$\ and $B\in (G_\alpha)_{SL}$\ is such that 
$$
(C_\alpha)_{SL}(B)=s.
$$

In the case of the  principal bundle corresponding to $\tau_3,$
$$
H(2) \times \frac{SL}{Ker(C_\alpha)_{SL} \cap SL_1}\left(H(2) \times \frac{SL}{(G_\alpha)_{SL}
\cap SL_1},(C_\alpha)_{SL}((G_\alpha)_{SL}\cap SL_1)\right),
$$
the bundle action is defined in the same way but using the restriction of $(C_\alpha)_{SL}$\ to 
$(G_\alpha)_{SL}\cap SL_1.$

The pairs $(\nu_1,\nu_2),(\nu_3,\nu_4),\ (\iota_1,\iota_2),\ (\iota_3,\iota_4)$\ define
 homomorphisms of principal fibre bundles. In what concerns the structural groups, we have
$$
(C_\alpha)_{SL}((G_\alpha)_{SL}\cap SL_1) \subset (C_\alpha)_{SL}((G_\alpha)_{SL})\subset {\mathbf S}^1
$$
and the homomorphism of structural groups is, in all cases, the canonical injection of the group of the first bundle  into the group of  the second.

As a consecuence of (\ref{vardin}), the {\bf linear momentum},  is given on
the coadjoint orbit of $\alpha$ by $P(Ad^{*}_{(A, H)} \alpha )= -AkA^{*}$,\ so that, with the identification of the coadjoint orbit with $G/G_\alpha$,\ we can write $P((A, H)G_\alpha)=-AkA^{*}$. 

On the other hand, since $  (G_\alpha)_{SL} \subset SL_2,$\ we can  define in $ {SL}/{(G_\alpha)_{SL}} $ a function, $P_0$,\ by means of $P_0(A(G_\alpha)_{SL})=-AkA^{*}$. 

But then, $P_0$ is the projection of $P$ by the canonical map
 \[
(A,\ H)G_\alpha \in \frac{G}{G_\alpha}  \stackrel{\iota_0}{\longrightarrow} A (G_\alpha)_{SL}\in \frac{SL}{(G_\alpha)_{SL}},
\]
and will be denoted in the following simply by $P$ and also called \emph{linear momentum}, thus we can write
\begin{equation}\label{linear momentum}
P(A (G_\alpha)_{SL})=-AkA^{*}
\end{equation} 
where $k$ is given by $\alpha=\{a,k\}$.

Let us denote $(C_\alpha)_{SL}(G_\alpha)_{SL})$\ by $S$. 

In \cite{adm96} I prove that

\vspace{1cm}

\fbox{  
\begin{minipage}[t]{4.7in} 
 The pull back
by $\nu_3$ , maps in  a one  to one  way the  set of quantum
states   (considered   as   pseudotensorial   functions  on
$G/Ker\widetilde C_\alpha$\,   onto the  set composed  by the
   functions      on     $H(2)  \times   (SL/Ker(C_\alpha)_{SL})$ 
		having      the       form
 \begin{equation} \label{preandas}
\phi_f(H,A\,Ker(C_\alpha)_{SL})\,=
\,f(A\,Ker(C_\alpha)_{SL})\,e^{i\pi Tr(P(A (G_\alpha)_{SL})\varepsilon \overline{H}  
\varepsilon)}
 \end{equation}
 where f  is a pseudotensorial function  on
the         principal         fibre         bundle 
 $$      
\frac{SL}{Ker(C_\alpha)_{SL}} \left(\frac{SL}{(G_\alpha)_{SL}},S\right) .
$$
\end{minipage}
}

\vspace{1cm}

In order to be more precise we will use the following definitions.

Let ${\cal C}$\ be  the complex vector space composed by the pseudotensorial functions of the bundle
 \[      
\frac{SL}{\mbox{$Ker\,(C_\alpha)_{SL}$}} \left(\frac{SL}{\mbox{$(G_\alpha)_{SL}$}},S\right),
\]
and $\widetilde{\cal V}$\ the complex vector space composed by the pseudotensorial functions on $G/Ker\widetilde C_\alpha.$

For each $f\in {\cal C}$\ we define a pseudotensorial function on $G/Ker\widetilde C_\alpha,$\ $\Psi_f,$\ by
$$
\Psi_f \circ \nu_3 \left(H,A\,Ker(C_\alpha)_{SL}\right)= \phi_f\left(H,A\,Ker(C_\alpha)_{SL}\right)= 
$$ 
\begin{equation}\label{psif}
  =f(A\,Ker(C_\alpha)_{SL})\,e^{i\pi \,Tr(P(A
\,(G_\alpha)_{SL}     )\,\varepsilon\,     \overline    H\,\varepsilon)}.
 \end{equation}
The map
$$
\Psi:f\in {\cal C} \rightarrow \Psi_f \in \widetilde{\cal V}
$$
is an isomorphism.

Also we define
\begin{equation}
\label{fif}
\Phi_f=\Psi_f \circ \iota_1.
\end{equation}
These functions are pseudotensorial on $G/KerC_\alpha$\ and invariant by the horizontal action,  so that they represent Quantum States in the most primitive sense adopted in this paper.

We denote by  ${\cal V}$ the complex vector space composed by these $\Phi_f,$\ so that the map 
\begin{equation}\label{pseud-contacto}
\Phi:f\in {\cal C} \rightarrow \Phi_f \in {\cal V}
\end{equation}
is an  isomorphism.

Thus, we can consider Quantum States as being elements of ${\cal C}$\ or elements of $\widetilde{\cal V}$\ or elements of ${\cal V}.$

Now we shall regard Quantum States under another form: Wave Functions.

The differentiable manifold
 $$W \stackrel{\mathrm def}{=}( H(2)  \times
(SL/ Ker\,(C_\alpha)_{SL} )  \times_S \,{\bf C}$$  is defined in  the same way as we have defined $$\left( G/
KerC_\alpha
\right)
 \times_{\mbox{$\bf  S  \rm  ^1$}}
 {\mbox{\boldmath {$C$}} \rm}$$ in section \ref{quantizable}.

$W$\  is the total space of      
 the hermitian line bundle associated to the principal fibre bundle 
$$\left(H(2) \times \left(SL/Ker \left(C_\alpha \right)_{SL}
\right)\right)\left(H(2) \times \left(SL/
\left(G_\alpha
\right)_{SL} \right),\,S \right)$$  and the canonical action of $S$ on
{\bf C} .

 The bundle projection is
$$
\eta:[(H,A Ker\,(C_\alpha)_{SL} ), c]_S \in W  \rightarrow (H,A (G_\alpha)_{SL} ) \in 
 H(2)
\times (  SL/(G_\alpha)_{SL}),
$$
where $[(H,A Ker\,(C_\alpha)_{SL} ), c]_S$\ is the orbit of $((H,A Ker\,(C_\alpha)_{SL} ), c)$\ under the action of $S$.

Since for each $f\in {\cal C}$\ the function $\Psi_f \circ \nu_3$\ is pseudotensorial in 
 $$H(2) \times \left(SL/Ker \left(C_\alpha \right)_{SL}\right),$$ it defines a section of $\eta.$\ This section can be considered as another description of the Quantum State given by $f$.

To complete our way towards
Wave Functions, we need to
 ``represent''  Quantum  States   as
functions  with  values  in  a  fixed complex vector space, not in a different vector space for each point in the base space, as does the sections of $\eta$.

                   If    $(C_\alpha)_{SL}(g)=1,\ \forall g\in (G_\alpha)_{SL},$\ , we have $S=\{1\},$\ and $Ker\,(C_\alpha)_{SL}=(G_\alpha)_{SL}$\ so that $W \approx ( H(2)  \times (SL/ (G_\alpha)_{SL} )  \times \,{\bf C},$\ and $\eta$ is the canonical  projection onto the first two factors. The section of $\eta$	corresponding		to the function $\phi_f $						
 in (\ref{preandas}) is 
$$
(H,\,A\,(G_\alpha)_{SL})\, \longrightarrow    (H,\,A\,(G_\alpha)_{SL},\phi_f(\,H,\,A\,\,(G_\alpha)_{SL})).
$$

Thus, if $(C_\alpha)_{SL}={\bf 1 },$\ our task is accomplished by 
$$
\phi_f(\,H,\,A\,Ker\,{\mbox{$(C_\alpha)_{SL}$}})\,=
\,f(\,A\,Ker\,{\mbox{$(C_\alpha)_{SL}$}})\,e^{i\pi \,Tr(P(A
\,(G_\alpha)_{SL}     )\,\varepsilon\,     \overline    H\,
\varepsilon}
$$
 itself as a complex valued function on the base space. In this  case, $\phi_f$ is  called the \bf Prewave Function \rm associated to $f,$\ and denoted by $\psi_f.$

 In the case where
${\mbox{$(C_\alpha)_{SL}$}}$\ is not trivial, our goal will be attained
by imbedding the hermitian  fibre
bundle
in a trivial one.  We do  this in a direct way, but a  more
geometrical view of the method is exposed in remark 5.1 of \cite{adm96}.

A key  concept in our construction of Wave Functions is the following:

A \bf Trivialization \rm of $ C_\alpha $ is a triple $(\rho,\ {
L},\ z_0)$,\ where $ { L}$ \ is a finite dimensional complex vector
space, $z_0\in { L}$ \ and $\rho$\ is a representation of {\mbox{$
SL$}}(2,{\mbox{\bf C}}) in ${ L}$\ such that
\begin{eqnarray} \label{trivialization}
       &1)&\ \rho(A)(z_0)\,=\,(C_\alpha)_{SL}(A)\
z_0,\ \ \ \forall A\in (G_\alpha)_{SL}. \nonumber\\
       &2)&\ {\rm The \ isotropy \ subgroup\  at}\   z_0\   {\rm is}\ Ker\,(C_\alpha)_{SL}. 
\end{eqnarray}

                In  what  follows,  we  assume  that
a trivialization of $C_\alpha$ is given.

The homogeneous space $ { SL}/{\mbox{$Ker\,(C_\alpha)_{SL}$}} $\  is identified to  the orbit of $z_0$\,,\ {$\cal B$}, by means of 
\[
A Ker\,(C_\alpha)_{SL} \in \frac{ SL}{Ker\,(C_\alpha)_{SL}} \longrightarrow  \rho(A)(z_0)\in {\cal B}.
\]

The action of $S$ on $ { SL}/{\mbox{$Ker\,(C_\alpha)_{SL}$}} $\ becomes, with this identification,  multiplication in $L$ of  elements of $S,$\  as complex numbers, by  elements of ${\cal B},$\ as elements of $L$ .
                
The      canonical      map   from $ { SL}/{\mbox{$Ker\,(C_\alpha)_{SL}$}} $\    onto      $     {SL}/{\mbox{$(G_\alpha)_{SL}$}} $\ will be denoted by {  \bf
r}, and is given by \[
r\left(\rho(A)(z_0)\right)=A {\mbox{$(G_\alpha)_{SL}$}} .
\]

Each $f\in {\cal D}$\   thus    becomes a  function on ${\cal  B},$\  homogeneous  of  degree  $-1$\  under ordinary 
multiplication  by  elements  of  $S$. The functions having these characteristics, 
will be called \bf S-homogeneous of degree -1.\ \rm 
 The $S$-homogeneous of degree T
functions are defined in a similar way.
Let us denote by ${\cal C}$\ the complex vector space composed by the S-homogeneous of degree -1 functions on ${\cal B}$. 

Now, $W$ is $( H(2)  \times
{\cal B} ) \times_S \,{\bf C}$,\ and $\eta$\ maps $[(H,z),c]$\ onto $(H,r(z)).$

The sections of $\eta$ corresponding to the functions having the form of $\phi_f$ in (\ref{preandas}) are as follows. 

Let $f$ be a $S$-homogeneous of degree -1 function. The corresponding section, $\sigma,$\   maps $(H,m)$\ to \[ 
\sigma(H,m)=[(H,z),\phi_f(H,z)]  
\]
 where $z$
\ is arbitrary in $r^{-1}(m).$

                   We  define  a  map,  {\mbox{$\chi$}} ,\
from W  into $ H(2) \times ( SL/(G_\alpha)_{SL}) \times\
L  $,   by  sending     $[(H,z ),  c]_S  \in   ( H(2)  \times   
{\cal B})  )  \times_S  \,{\bf  C}$  to  $   (H,\
r(z),  cz)$.

The map {\mbox{$\chi$}} is injective . In fact  the relation \[
\chi([(H,z),c]_S)=\chi([(H^\prime,z^\prime),c^\prime]_S )
\]
is equivalent to \[
(H,r(z),cz)=(H^\prime,r(z^\prime),c^\prime z^\prime)
\]
so that there exist $e^{i\gamma}\in S$\ such that $H^\prime=H,$\ $z^\prime=e^{i\gamma} z,$\ and $c^\prime z^\prime=cz$. Thus    
 \[
[(H^\prime,z^\prime),c^\prime]_S=[(H,ze^{i\gamma} ),ce^{-i\gamma} ]_S=[(H,z),c]_S.
\]

The fiber of $W$ on $(H,m)$ is $\{[(H,z  ),c]: c\in {\bf C}\}$,\ where $z$ is any fixed element in $r^{-1}(m).$\ Its image under {\mbox{$\chi$}} is composed by the $(H,m,y)$\ such that $y$ is in the  one dimensional subspace of $L$ generated by $z.$

We  have thus inmersed our, in general, non trivial bundle, in a trivial one with fiber $L$. This enable us to identify sections of $\eta$ with functions with values in $L,$\ as follows.

The section $\sigma$ of $\eta$, that corresponds to  $f$\ can be identifed with
$$
\chi\circ \sigma(H,m) = \left(H,m,\phi_f(z)z\right),
$$
where $z$ is arbitrary in $r^{-1}(m).$\ 

The right hand side in the preceeding equation is completely determined by its third component:
\begin{equation}\label{prefunc}
\psi_f(H,m)= \phi_f(z)z=f(z)\,e^{i\pi \,Tr(P(m  )\,\varepsilon\,     \overline    H\, \varepsilon)}z.
\end{equation}

The function $\psi_f,$\ with $f$ $S$-homogeneous of degree -1, will be called   {\bf
Prewave Function} associated to $f$.

The complex vector space composed by the Prewave Functions  is denoted by {\cal PW},\ so that the map 
\begin{equation}
\psi:f\in {\cal C}\ \rightarrow   \psi_f\in {\cal PW}
\end{equation}
 is an isomorphism.

The composition of any Prewave Function with $\iota_4$ is a function in State Space that can be considered as giving an amplitude of probability for each state. 

As said in the Introduction, we obtain from this Prewave Function  a Wave Function, defined on space-time points ($i.e.$ hermitian matrices), by adding up, for each $H \in H(2),$ the amplitudes of probability corresponding to the states $(H,\beta)$ where $\beta $ is a movement whose portrait in space-time contains $H$.

According to \eqref{movimporpunt}, we see that these states are the elements of the set
$$
\{(H,Ad^*_{(B,H)}\alpha: B\in SL\},
$$ that is identified by $i$ ($c.f.$ \eqref{iii}) with
$$
\{H\}\times {\frac{SL}{(G_\alpha)_{SL}\cap SL_1}},
$$
whose image by $\iota_4$ is 
$$
\{H\}\times \frac{SL}{(G_\alpha)_{SL}}.
$$

Then     to any  given  Prewave
Function,
$\psi_f$,\  we  associate  a   {\bf   Wave
Function},  $\widetilde{  \psi_f}$,\  as follows 
\begin{equation}\label{WF}
\widetilde     {\psi_f}(H)=\int_{{SL}/(G_\alpha)_{SL}} {\mbox{$\psi$}}_f(H,\,m)\ {\mbox{$\omega$}}_m
\end{equation} 
where $m$ is a generic element in $ {  SL}/({\mbox{$G_\alpha$}})_{SL},$\ and  {\mbox{$\omega$}} \ is  a {\bf  volume element} on $ {  SL}/({\mbox{$G_\alpha$}})_{SL},$\ left invariant by the canonical action of $SL$ on $ {  SL}/({\mbox{$G_\alpha$}})_{SL}.$ This means that, if we denote by $d'_A$ \ the diffeomorphism of $SL/(G_\alpha)_{SL}$ defined by sending $B\,(G_\alpha)_{SL}$\ to $AB\,(G_\alpha)_{SL},$\ then $(d'_A)^*\omega=\omega.$

In the following, we assume that such a invariant volume element is given.

                   This definition of Wave Functions, forces
us to do a restriction on the class of the functions to  be
considered:  it is necessary that the integral exists.

                   In what follows I  consider  only Quantum States corresponding to the functions in ${\cal C}$\ that are 
continuous  with compact  support. The complex vector space composed by these Wave Functions is denoted by ${\cal WF}.$

 Of course, there are other possible 
 conditions that can be imposed on $f$ in order to assure integrability.
Also, if one obtains for some choice a Prehilbert space, one can consider its completed, in order  to have a Hilbert Space. But in   this paper I prefer to maintain the  "continuous with compact support" condition.

The Wave Functions are another form of description of Quantum States, in fact it is the most usual in Quantum Mechanics.

In the following I change the notation in such a way  that, ${\cal C}$\ stands for the complex vector space composed by the  $S$-homogeneous of degree -1 functions on ${\cal B}$\ that also are continuous\ with compact support,
 and ${\cal PW},\ {\cal V},\ \widetilde{\cal V},$\  the corresponding isomorphic spaces.

\newtheorem{coment}{Remark}[section]
\begin{coment}\label{bajar}

\rm Now, let us  assume that we know  a section of the map $r,$  defined an open set,  $D,$\  in ${SL/(G_\alpha)_{SL}}$,
\[
\sigma : D \longrightarrow {\cal B}.
\]

In this case, we can associate a prewave function, and thus a wave function, to each $C^\infty$ function, with compact support contained  in $D,$\ as follows.

Let $f_0$\ be a $C^\infty$ function, with compact support contained  in $D$ . We define a  function on $\cal B$,\ $f$,\  by means of

\[
f(z)= 
\begin{cases}
t^{-1} f_0(r(z))& {\rm if}\  z\in r^{-1}(U)\ {\rm where}\  t \ {\rm is\ such\ that}\ z= t \sigma (r(z))\\   
0&  {\rm if}\    z\notin r^{-1}(U)
\end{cases}
\]

This function  is $S$-homogeneous of degree -1 since, for all $e^{i a}\in S,\ z\in{r^{-1}(U)},$\ we have    
\[
f(e^{i a}z)=t'^{-1} f_0(r(e^{i a}z)),
\]
where $t'$ is such that $$e^{i a}z=t' \sigma(r(e^{i a}z))=t' \sigma (r(z))=t' t^{-1} z,$$
so that
$$
e^{i a}=t' t^{-1}
$$
and
$$
 f(e^{i a}z)=e^{-i a}t^{-1} f_0(r(e^{i a}z))=e^{-i a}t^{-1} f_0(r(z))=e^{-i a}f(z).
$$

If $z\notin{r^{-1}(U)}$,\ $f(e^{i a}z)=0=f(z)=e^{-i a}f(z)$.

The Prewave Function $\psi_f$  then   is
\[
\psi_f(H,\ m)=
\begin{cases}
f_0(m)\,e^{i\pi Tr(P(m)\,\varepsilon \overline H \varepsilon)}\ \sigma(m) & \mathrm{if}\ m\in U  \\
0&   \mathrm{if}\ m\notin U 
\end{cases}
\]
and we have an associated Wave Function, $\tilde \psi_f.$

This expression contains no reference to any homogeneous function and suggest directly the usual form of Wave Functions. 

\end{coment}

\section{Representation of $SL(2,\mathbb{C}) \oplus H(2)$ \ on Quantum States}\label{representaciones}

\subsubsection{ Representation on $S$-homogeneous functions of degree -1 }\label{rephomog}

If $(C_\alpha)_{SL}={\bf 1},$\ , the $S$-homogeneous functions of degree -1 are simply functions on 
 $SL/(G_\alpha)_{SL}.$

In this case we denote by $\cal C$  the complex vector space composed by the continuous functions on  $SL/(G_\alpha)_{SL}$  with compact support.

For all $B\in SL$ we denote by $d^\prime_B$ the canonical diffeomorphism of $SL/(G_\alpha)_{SL}$ given by
$$
d^\prime_B(A (G_\alpha)_{SL})=BA (G_\alpha)_{SL}).
$$

Then, we define a representation, $\delta,$ of $SL$ on  $\cal C$ by
$$
\delta(f)=f\circ d^\prime_{A^{-1}}.
$$

In  $\cal C$ we also define an hermitian product by
$$
\langle f,\, f^\prime \rangle=  \int_{{
SL}/(G_\alpha)_{SL}}  \overline{f}  f^\prime   \
{\mbox{$\omega$}} ,
$$
 where $\omega$\ is an invariant volume element on $SL/(G_\alpha)_{SL}.$

With this inner product, $\cal C$\ becomes a prehilbert space.

The invariance of  {\mbox{$\omega$}} enable us to write 
$$
\langle \delta(A)(f),\, \delta(A)(f^\prime) \rangle=\langle f,\, f^\prime \rangle
$$
so that the representation $\delta$ is unitary.

In case  $(C_\alpha)_{SL}\neq{\bf 1},$\ , we assume that  we have a trivialization, $(\rho,\ {L},\ z_0),$ and we define ${\cal B}$ and ${\cal C}$ as in the preceeding section.

Of course, also in case $(C_\alpha)_{SL}={\bf 1},$\ we can have a trvialization and thus apply all that follows.

 In ${\cal C}$ we  define a hermitian product:
$$
\langle f,\, f^\prime \rangle=  \int_{{
SL}/(G_\alpha)_{SL}}  \overline{f}  f^\prime   \
{\mbox{$\omega$}} , $$
where by $ \overline{f}  f^\prime  $\ we means the function defined on $SL/(G_\alpha)_{SL},$\  by $$
\overline{f}  f^\prime  (m)=\overline{f}(z)  f^\prime (z)
$$
for all $m\in  SL/(G_\alpha)_{SL},$\ where $z$ is arbitrary in $r^{-1}(m),$\ and $\omega$\ is an invariant volume element on $SL/(G_\alpha)_{SL}.$

Also in this case, $\cal C$\ becomes a prehilbert space.

A representation, $\delta,$\  of $SL$ on $\cal C,$\ is defined by
$$
\delta(A): f\in {\cal C} \longrightarrow f\circ \rho(A^{-1})\in {\cal C}.
$$
for all $A\in SL.$

If $A\in SL$\ we have
$$
\langle \delta(A)(f),\, \delta(A)(f^\prime) \rangle= \int_{{ SL}/(G_\alpha)_{SL}} \left((\overline{f}  f^\prime)\circ  d'_{A^{-1}} \right)
\ {\mbox{$\omega$}}
$$
so that the invariance of  {\mbox{$\omega$}} enable us to write 
$$
\langle \delta(A)(f),\, \delta(A)(f^\prime) \rangle=\langle f,\, f^\prime \rangle
$$
We see that the representation $\delta$ is, also in this case, unitary.

Also we have a representation, $\delta^\prime,$\  of  $SL \oplus H(2)$ \ on $ {\cal C}$\ defined by:
$$
(\delta^\prime(A,H)\cdot f)(z)= f( \rho(A^{-1})\cdot z) Exp\left(-i\pi Tr
\left(\,P( r(z)) \varepsilon \overline{H} \varepsilon \right)\right),
$$
where $(A,H)\in SL \oplus H(2),\ f\in {\cal C},$\ and $z\in {\cal B}.$

To prove that $\delta^\prime$\ is a representation , notice that 
\begin{enumerate}
\item $r$\ is equivariant {\it i.e.} 
\begin{equation}\label{trans ro}
r(\rho(A)\cdot z)=d'_A(r(z))
\end{equation}
for all $A\in SL.$\ 

\item Formula \eqref{linear momentum}, leads to
$$
P(d'_B(A(G_\alpha)_{SL}))=P(BA(G_\alpha)_{SL})=-BAkA^*B^*=BP(A(G_\alpha)_{SL})B^*
$$
so that
\begin{equation}\label{trans lin mom}
P(r(\rho(A)\cdot z))=P(d'_A(r(z)))=AP(r(z))A^*.
\end{equation}

\item For all $A\in SL,$ 
\begin{equation}\label{Aeps}
A\, \varepsilon= \varepsilon\,{}^t\negmedspace A^{-1}
\end{equation}

\end{enumerate}

Now, we can prove that $\delta^\prime$ is a representation as follows.

 If $(B,K),(A,H)\in SL \oplus H(2),\ z\in {\cal B},$\ then
$$
\begin{gathered}
\delta^\prime(B,K)\cdot\left(\left(\delta^\prime(A,H)\cdot f\right)\right)(z)=\left(\delta^\prime(A,H)\cdot f\right)(\rho(B^{-1})\cdot z)  \\
Exp\left(-i\pi Tr[P(r( z))\varepsilon \overline K  \varepsilon ]\right)=f(\rho(A^{-1})\cdot (\rho(B^{-1})\cdot z))\\
Exp\left(-i\pi Tr[P(r(\rho(B^{-1})\cdot z))\varepsilon \overline H  \varepsilon+ P(r( z))\varepsilon \overline K  \varepsilon ] \right)=\\
=f(\rho((BA)^{-1})\cdot z) Exp\left(-i\pi Tr[P(r(z))\varepsilon \overline{(B HB^* + K) }\varepsilon]\right)=\\
=\left(\delta^\prime\left(\left(B,K\right)\left(A,H\right)\right)\cdot f\right)\left( z\right)
\end{gathered}
$$

The infinitesimal generator of $\delta$ associated to $a\in sl(2,\bf C)$\  is the linear map from $\cal C$ into itself, $d\,\delta(a)$ given by
		$$
	(d\,\delta(a) \cdot f)	(z)=\left(\frac{d}{dt}\right)_{t=0} \left(\delta\left(e^{t\,a}\right)\cdot f \right) \left(z\right). 
	$$
	
Then
$$
	(d\,\delta(a) \cdot f)	(z)=\left(\frac{d}{dt}\right)_{t=0} \left(f  \left(\rho\left(e^{-t\,a}\right) \cdot z\right) \right),
	$$
	and thus we see that $d\,\delta(a)$ acts on $f$ as the vector field, $X_a,$\ infinitesimal generator  of the  action on $\cal B,$ defined by
	$$A*z=\rho(A) \cdot z,$$ 
	associated to $a$.

We can give a more explicit form of $X_a$\ as follows.

 Let us fix a basis, $\beta=\{ e_1, \dots,e_q\},$\ of $L.$\ By means of $\beta$\ we identify $L$ with ${\bf C}^q$. Let us denote by ${}^t\negthinspace (z^1,\dots,z^q)$ the matrix of $z\in L$\ in the basis $\beta$\ and $(d\rho(a))_i^j,$\ the element in the file $j$ column $i$ of the matrix of $(d\rho(a))$\ in the basis $\beta.$\ We denote by $\beta^*=\{ w^1, \dots,w^q\}$\ the system of (complex) coordinates associated to $\beta$\ ({\it i.e.} $\beta^*$\ is the dual basis of $\beta.$)

If $w^j_x$\ (resp $w^j_y$) is the real (resp. imaginary) part of $w^j,$\  it is usually writen
$$
\frac{\partial}{\partial w^j}=\frac{1}{2}\left(\frac{\partial}{\partial w^j_x}-i\ \frac{\partial}{\partial w^j_y}\right),
$$
$$
\frac{\partial}{\partial \overline w^j}=\frac{1}{2}\left(\frac{\partial}{\partial w^j_x}+i\ \frac{\partial}{\partial w^j_y}\right).
$$

Thus, when $f$ is a complex valued differentiable function on an open subset, M, of ${\bf C}^q$, and $v(t)$ a differentiable map from an open neigbourhood of $0$ in ${\bf R}$\ into M, we have, using summation convention
$$
\begin{gathered}
\left( \frac{d}{dt}\right)_0 (f\circ v) =\left(\frac{\partial \, f}{\partial w^j_x}\right)_{v(0)}\, (w_x^j(v(t)))'(0)
+\left(\frac{\partial \, f}{\partial w^j_y}\right)_{v(0)}\, (w_y^j(v(t)))'(0)=\\
=\left(\frac{\partial \, f}{\partial w^j}\right)_{v(0)}\, (w^j(v(t)))'(0)+\left(\frac{\partial \, f}{\partial \overline {w}^j}\right)_{v(0)}\, (\overline {w}^j(v(t)))'(0).
\end{gathered}
$$

Since 
$$
\left(\frac{d}{dt}\right)_{t=0} \left(f  \left(\rho\left(e^{-t\,a}\right) \cdot z\right) \right)
		=\left(\frac{d}{dt}\right)_{t=0} \left(f  \left(\left(e^{-t\,d\rho\left(a\right)}\right) \cdot z\right) \right) 
$$
it follows that
$$
\begin{gathered}
\left(X_a)_z(f) \right)=\left(\frac{\partial \, f}{\partial w^j}\right)_{z}\, (-(d\rho(a))^j_i\,z^i)+\left(\frac{\partial \, f}{\partial \overline {w}^j}\right)_{z}\, ({-(\overline {d\rho(a)})} ^j_i\,\overline {z}^i)
\end{gathered}
$$
so that an extension of $X_a$\ from $\cal B$\ to $\mathbb{C}^q$\ is

\begin{equation}\label{equissuba}
X_a=-\left(w^i \,(d\rho(a))^j_i\,\frac{\partial}{\partial w^j}+\overline{w}^i \,(\overline{d\rho(a)})^j_i\,\frac{\partial}{\partial \overline{w}^j} \right).
\end{equation}

If $(C_\alpha)_{SL}={\bf 1},$\  , and do not use a trivialization, we have similar results, but $X_a$\ is the infinitesimal generator
of the canonical action of $SL$\ on $SL/(G_\alpha)_{SL},$\  associated to $a.$\ 

Thus, in all cases 
$$
d\,\delta(a) \cdot f=X_{a}(f),
$$
or simply, as linear maps on $\cal C$, 
$$
d\,\delta(a) =X_{a}.
$$

On the other hand, the infinitesimal generator of $\delta '$ associated to $$(a,h)\ \ \in sl(2,\bf C)\oplus H(2)$$ is the endomorphism, $d\,\delta'(a,h)$ of $\cal C$\ given by 
$$
	(d\,\delta'(a,h) \cdot f)	(z)=\left(\frac{d}{dt}\right)_{t=0} \left(\delta'\left(Exp({t\,(a,h))}\right)\cdot f \right) \left(z\right),
	$$
but the exponential map in $SL\oplus H(2)$\ is given by \eqref{exp}, so that
$$
(d\,\delta'(a,h) \cdot f)	(z)=\left(\frac{d}{dt}\right)_{t=0} \left(\delta'\left(\left(e^{ta},\int_0^t e^{sa}\,h\, e^{sa^*}\,ds \right)\right)\cdot f \right) \left(z\right),
	$$
		and a short computation thus leads to
	$$
	\begin{gathered}
	d\,\delta'(a,h) \cdot f= X_a(f)-i\pi Tr[P(r(\cdot))\varepsilon \overline{h}\varepsilon]\ f=\\
	=X_a(f)+2\pi  i \langle  P(r(\cdot)),h  \rangle \ f,
\end{gathered}
$$
where I have used the same symbol for any hermitian matrix and the corresponding element in ${\bf R}^4,$\ and $\langle \  ,\  \rangle$ is Minkowski product.

Also, we can write
\begin{equation}\label{diferencialcompleta}
d\,\delta'(a,h)=X_a+2\pi  i \langle  P(r(\cdot)),h  \rangle.
\end{equation}
where $X_a$ is the endomorphism given by the vectorfield \eqref{equissuba}	, and  $$2\pi  i\langle  P(r(\cdot)),h  \rangle $$
is the endomorphism given by ordinary multiplication by this function.

Instead of the infinitesimal generators $d\,\delta'(a,h),$\ we can use
\begin{equation}\label{operadhomog1}
(a,h)^{\theta}\stackrel{\mathrm def}{	=}\frac{1}{2 \pi i}\ d\,\delta'(a,h) ,
\end{equation}
so that

\begin{equation}\label{operadhomog2}
(a,h)^{\theta}	=\frac{1}{2 \pi i}\,X_a+ \langle  P(r(\cdot)),h  \rangle .
\end{equation}

The endomorphism $(a,h)^{\theta},$\ is hermitian for our hermitian product and is the \bf Quantum Operator associated to the Dynamical Variable (a,h),\ \rm when Quantum States are represented by $S$-homogeneous functions. 

When Quantum States are represented by Wave Functions, the Quantum Operators acquires its usual form in Quantum Mechanics (\it c.f. \rm   \eqref{operclasic}).

In the particular case of the  Linear Momentum (\it c.f. \rm  \eqref{limoang}), 
one sees that
\begin{equation}\label{linear.mom.general}
(P^k)^{\theta}\cdot f=(P^k \circ r)\,f,\ \ k=1,\,2,\,3,\,4.
\end{equation}
 where the  $P^k$\ in the left hand side are the components of Linear momentum as a dynamical variable, and the ones in the right hand side, the components of Linear momentum considered as a  function on 
  $SL/(G_\alpha)_{SL},$ given by \eqref{linear momentum}.

In what concerns to the Quantum Operators corresponding to Angular Momentum we have
\begin{equation}\label{ang.mom.uno}
(l^j)^{\theta}= \frac{1}{4\pi i}  \,X_{i\,\sigma_j}, \ \                                   j=1,\,2,\,3.
\end{equation}
\begin{equation}\label{ang.mom.dos}
(g^j)^{\theta}=\frac{1}{4\pi i}  \,X_{\sigma_j}, \ \                                   j=1,\,2,\,3.
\end{equation}

Of course, the representations on ${\cal C}$\ lead to equivalent representations on the isomorphic vector spaces $ {\cal V},\ \widetilde{\cal V},$\  and ${\cal PW}$.

We do that in detail in  ${\cal PW}$\ and ${\cal V}$  as follows:

\subsubsection{Representation on Prewave Functions.}\label{repprewave}

In  ${\cal PW}$\   we define an hermitian product by
$$
\langle \psi_f ,\psi_{f'} \rangle \stackrel{def}{=}\langle f ,{f'} \rangle, 
$$
and thus becomes a prehilbert space, isomorphic as inner product  spaces to ${\cal C.}$

The hermitian product can be given in terms of the
prewave functions
themselves, as follows.

Let $\beta$ be a sesquilinear form
on $L,$ \ 
nonvanishing on ${\cal B}$. We define
$$ \psi_f\, \beta\,
\psi_{f^\prime} \,: m \in SL/(G_\alpha)_{SL} \mapsto
\frac{\beta( \psi_f ( H,\,m),
\psi_{f^\prime} ( H,\,m ))}
{\beta(z,\ z )} $$
 where $z$ is arbitrary in ${\bf r}^{-1}(m)$ and
 $H$ is arbitrary in $H(2)$. Thus

$$\langle  \psi_f,\
\psi_{f^\prime} \rangle=
\int_{{
SL}/(G_\alpha)_{SL}}
\psi_f\, \beta\,
\psi_{f^\prime}\ \omega.
$$

The representation of $G$\ on ${\cal PW}$\  equivalent to $\delta^\prime$\ under $\psi$\ is given by
$$
\delta^{pw}(A,H)\cdot \psi_f=\psi_{\delta^\prime(A,H)  \cdot  f}.
$$

The  infinitesimal generator associated to $(a,h)\ \ \in sl(2,\bf C)\oplus H(2)$\ is
$$
d\delta^{pw}(a,h)\cdot \psi_f=\psi_{d\delta'(a,h)\cdot f}.
$$

The Quantum Operators for Prewave Functions must be defined as 
\begin{equation}
(a,h)^{\upsilon}\stackrel{\mathrm def}{=}\ \frac{1}{2\pi i}\,d\delta^{pw}(a,h),
\end{equation}
and we have
\begin{equation}
(a,h)^{\upsilon}\cdot \psi_f{=}\ \psi_{(a,h)^{\theta}\cdot f}.
\end{equation}

On the other hand, we have a natural action on $H(2)\times SL/(G_\alpha)_{SL}$\ defined by
$$
(B,K)* (H,A\,(G_\alpha)_{SL})=(BHB^*+K,BA (G_\alpha)_{SL}).
$$

Thus we define an, a priori different, representation, $\delta_{pw},$ on ${\cal PW}$\  by means of 
$$
\delta_{pw}(A,H)  \cdot  \psi_f =\rho (A) \circ \psi_f \circ \left((A,H)*\right)^{-1},
$$
\emph{i.e.}, for all $(K,m)\in H(2)\times SL/(G_\alpha)_{SL}$
\begin{equation}\label{rep.pw}
\left(\delta_{pw}(A,H)  \cdot  \psi_f \right)(K,m) =\rho (A) \cdot \psi_f  \left((A,H)^{-1}*(K,m)\right).
\end{equation}

Using \eqref{trans lin mom} and \eqref{Aeps} one can see that
\begin{equation}\label{deltapw}
\delta_{pw}=\delta^{pw}
\end{equation}
as follows

$$
\begin{gathered}
\left(\delta^{pw}(A,H)  \cdot  \psi_f \right)(K,m)=
 \psi_{\delta^\prime\left( A,H\right)\cdot f}(K,m)= f(\rho(A^{-1})\cdot z) \\
 Exp\left(- i\pi Tr [ P( m))\varepsilon \overline{H} \varepsilon ]\right) 
 Exp\left( i\pi Tr [ P( m)\varepsilon \overline{K} \varepsilon ]\right)  z=(*)
\end{gathered}
$$
 where  $z\in r^{-1}(m).$\ If we denote  $ z'=\rho(A^{-1})\cdot z ,$\ we have
 $$
\begin{gathered}
(*)= f( z')  Exp\left( i\pi Tr [ P( r(\rho(A)\cdot z'))\varepsilon \overline{(K-H)} \varepsilon ]\right)( \rho(A)\cdot z')  =\\
=f( z')  Exp\left( i\pi Tr [ P( r( z'))\varepsilon \overline{(A^{-1}(K-H)(A^{-1})^*)} \varepsilon ]\right) (\rho(A)\cdot z')=\\
=\rho(A) \psi_f((A,H)^{-1}*(K,m)  =
\left(\delta_{pw}(A,H)  \cdot \psi_f \right)(K,m).
\end{gathered}
$$

\subsubsection{Representation on Wave Functions.}

Now, let us consider ${\cal WF},$ the complex vector space composed by the Wave Functions $\widetilde{ \psi_f}$\ such that $f\in {\cal C}.$

A "translation" of the preceeding representations  of $SL\oplus H(2)$\ to ${\cal WF},$ is given by
$$
\delta_{w}(A,H)  \cdot  \widetilde{\psi_f} = \{\delta_{pw}(A,H) \cdot \psi_{ f}\}{\widetilde{}}= \{\psi_{\delta ^{\prime}(A,H)\cdot f}\}{\widetilde{}},
$$
where $\{\psi\}{\widetilde{}}$ means $\widetilde{\psi}.$

Then 
\begin{eqnarray*}
\delta_{w}(A,H)  \cdot  \widetilde{\psi_f}(K)&=& \{\delta_{pw}(A,H) \cdot \psi_{ f}\}{\widetilde{}}(K)\\
&=& \int_{SL/(G_\alpha)_{SL}}\rho(A) \cdot  \psi_f((A,H)^{-1}*(K,m))\omega_m=\\
&=& \rho(A) \cdot\int_{SL/(G_\alpha)_{SL}}  \psi_f((A,H)^{-1}*K,d'_{A^{-1}}m))\omega_m=\\
&=& \rho(A) \cdot \int_{SL/(G_\alpha)_{SL}}  \psi_f((A,H)^{-1}*K,m))\omega_m
\end{eqnarray*}
because of the invariance of $\omega.$ Thus
\begin{equation}
\left(\delta_{w}(A,H)  \cdot  \widetilde{\psi_f}\right)(K)=\rho (A) \cdot \widetilde{\psi_f } \left((A,H)^{-1}*K)\right).
\end{equation}
Compare to \eqref{rep.pw}.

Let us denote by $d\delta_{w}\cdot (a,h)$ \ the infinitesimal generator 
of the representation $\delta_w$\ associated to  $ (a,h) \in sl(2, C)\oplus
{  H(2)},$\  and
$$
\widehat{ (a,h)}  \stackrel{def}{=}\frac{1}{ 2 \pi i } (d\delta_{w}\cdot (a,h) ).
$$

The endomorphism $\widehat{ (a,h)}$\ of ${\cal WF}$\ is the Quantum Operator corresponding to the Dynamical Variable $(a,h)$ when the Quantum States are represented by Wave Functions.

A straightforward computation leads to the following expresions
for the operators corresponding to Linear and Angular Momentum

\begin{eqnarray}\label{operclasic}
\widehat{ P^k} \cdot \widetilde{ \psi _f }&=&\frac 1{2\pi i}
\frac{\partial}{\partial x^k} \ \widetilde{ \psi _f }    \nonumber   \\
\widehat{ P^4} \cdot \widetilde{ \psi _f } &=&\frac i{2\pi }
\frac{\partial}{\partial x^4} \ \widetilde{ \psi _f }     \nonumber \\
\widehat{ l^k} \cdot \widetilde{ \psi _f }&=&\frac 1{2\pi i}
\left( d\rho\left( {\frac{i\sigma_k}{2}}\right)+\sum_{j,r=1}^3
\varepsilon_{kjr} x^j
\frac{\partial}{\partial x^r} \right) \ \widetilde{ \psi _f }    \\
\widehat{ g^k} \cdot \widetilde{ \psi _f } &=&\frac 1{2\pi i}
 \left( d\rho\left(\frac{\sigma_k}{2}\right)- \left( x^4
\frac{\partial}{\partial x^k}+ x^k
\frac{\partial}{\partial x^4}   
\right)\right)\ \widetilde{ \psi _f }  \nonumber              
\end{eqnarray}
where $\varepsilon_{ijk}$ \ are the components of an
antisymmetric tensor such that $\varepsilon_{123}=1$.

\subsubsection{Representation on Pseudotensorial  Functions in the contact manifold.}\label{representcontacto}

Recall the isomorphism \eqref{pseud-contacto}
\begin{equation} \nonumber
\Phi:f\in {\cal C} \rightarrow \Phi_f \in {\cal V}
\end{equation}

If $f\in {\cal C},$\ we have
$$
\Phi_f((A,H)KerC_\alpha)=\,f(A\,Ker(C_\alpha)_{SL}) \,e^{i\pi \,Tr(P (A\,(G_\alpha)_{SL})   \,\varepsilon\,     \overline    H\,
\varepsilon)}.
$$

When $(C_{\alpha})_{SL}={\bf 1},$ we have $Ker(C_\alpha)_{SL}=(G_\alpha)_{SL},$ so that
\begin{equation*}
\Phi_f((A,H)KerC_\alpha)=\,f(A\,(G_\alpha)_{SL}) \,e^{i\pi \,Tr(P (A\,(G_\alpha)_{SL})   \,\varepsilon\,     \overline    H\, \varepsilon)}.
\end{equation*}

In the case $(C_{\alpha})_{SL}\neq{\bf 1},$ we assume the existence of a trivialization,  
  $(L,\,\rho,\,z_0),$\   and we identify  $SL/Ker\,(C_\alpha)_{SL}$\ with ${\cal B}.$\ Then  the pseudotensorial function $\Phi_f$ becomes
\begin{equation}
\Phi_f\left(\left(A,H\right)\,Ker C_\alpha \right)=
\,f(\rho(A)\cdot z_0) \,e^{i\pi \,Tr( P(r (\rho(A)\cdot z_0)   )\,\varepsilon\,     \overline    H\,
\varepsilon)}.
 \end{equation}

A natural representation of $G$ on $V$ is the $\delta^c$ given by
\begin{equation}\label{prirepcont}
\delta^c(A,H) \cdot \Phi_f=\Phi_f \circ ((A,H)^{-1}*)
\end{equation}
where $*$ is the canonical action on $G/Ker\, C_\alpha.$

Let us prove that
\begin{equation}\label{reprcontacto}
\delta^c(A,H) \cdot \Phi_f=\Phi_{\delta'(A,H) \cdot f} 
\end{equation}
$i.e.$ that $\Phi$ is equivariant for the representations $\delta'$ in ${\cal C}$ and $\delta^c$ in ${\cal V}$.

In fact, we have
$$
\begin{gathered}
\left(  \delta^c(A,H) \cdot \Phi_f\right) \left((B,K)Ker\, C_\alpha \right)=\Phi_f\left( A^{-1} B, A^{-1}(K-H)(A^{-1})^*   \right)=\\
=f\left( \rho(A^{-1})\cdot (\rho(B)\cdot z_0)\right)\,e^{i\pi \,Tr(P(r (\rho(A^{-1})\cdot (\rho(B)\cdot z_0))   )\,\varepsilon\,     \overline{ (   A^{-1}(K-H)(A^{-1})^*)}\,\varepsilon)}=\\
=f\left( \rho(A^{-1})\cdot (\rho(B)\cdot z_0)\right)\,e^{i\pi \,Tr(A^{-1}\,P(\rho(B)\cdot z_0)(A^{-1})^* A^*\varepsilon\,     \overline { (K-H)}\,\varepsilon          \,A)}=\\
= f\left( \rho(A^{-1})\cdot (\rho(B)\cdot z_0)\right)\,e^{i\pi \,Tr(P(\rho(B)\cdot z_0)\,\varepsilon\,     \overline { (K-H)}\,\varepsilon )  }=\\  
=\left( \delta'(A,H)\cdot f\right)\left(\rho(B) \cdot z_0 \right)\,e^{i\pi \,Tr(P(\rho(B)\cdot z_0)\,\varepsilon\,     \overline { K}\,\varepsilon   }=\\
=\Phi_{\delta'(A,H)\cdot f}\left((B,K)Ker\, C_\alpha \right).
\end{gathered}
$$

As a particular consecuence, for all $(a,h)\in \underline G,$\ we have for the infinitesimal generators of $\delta^c$
$$
d\delta^c(a,h) \cdot \Phi_f=\Phi_{d\delta^\prime(a,h)\cdot f}.
$$

Equation \eqref{prirepcont}, tell us that \it the infinitesimal generator of the representation $\delta^c$ associated to $(a,h)\in \underline G,$\ acts on $\Phi_f$ as the (vector field)infinitesimal generator of the action on $G/Ker\,C_\alpha,$\ \rm what will be denoted by
$
X_{(a,h)}^c.
$

If we denote by $(a,h)^c,$\ the Quantum Operator
$$
(a,h)^c \stackrel{def}{=}\frac{1}{2\pi i}\,d\delta^c(a,h)
$$
we have
$$
(a,h)^c \cdot \Phi_f=\Phi_{(a,h)^\theta \cdot f}.
$$
Thus, our results in subsection  \ref{rephomog}, on Quantum Operators in that case, gives us results on Quantum Operators in our present case.

\part{Explicit Construction}

\parindent=1cm
\parskip=5mm

\section{Quantizable forms in $ SL(2,\mathbb{C}) \oplus H(2)$.}\label{concreteqforms}

In \cite{adm95} I give a classification  of the coadjoint orbits of the group under consideration. The orbits are divided into 9 Types, and a canonical representative of each orbit is given, according with its type. 

Let  $\alpha =\{a,k\} $ \  be  a nonzero element of the dual of the Lie algebra of $ SL \oplus H(2),$\ and denote by $W$ and $P$
the Pauli- Lubanski and Linear momentum respectively at $\alpha$ \ (cf. section \ref{sec-grupo} ).

Figure 3 gives what type of coadjoint orbit is the one of $\alpha,$\ according with the values of $W$ and $P.$\

\begin{figure}
  \centering
\begin{tabular}{|c|c|c|c|c|c|}\hline
Type & $\vert$ P$\vert$  &$\vert$ W $\vert$  & P &W & Det a  \\\hline
1&0&0&0&0&0 \\\hline
2&0&0&0&0& $\neq$ 0  \\\hline
3&0&$<0$&$\neq$ 0&$\neq$ 0& \\\hline
4&0&0&$\neq$ 0&   &  \\\hline
5&$>0$&$\leq 0$&$\neq 0$& & \\\hline
6&$<0$&0&$\neq 0$&0&   \\\hline
7&$<0$&$>0$&$\neq 0$&$\neq 0$& \\\hline
8&$<0$&$<0$&$\neq 0$&$\neq 0$& \\\hline
9&$<0$&0&$\neq 0$&$\neq 0$&     \\\hline
\end{tabular}

\caption{Types of coadjoint orbits}

\label{tipos}

\end{figure}

\parindent=1cm
\parskip=5mm

When one knows the type of $\alpha,$\ Figure 4 gives us the canonical representative of the orbit of $\alpha.$\

\begin{figure}  
  \centering 
		
\begin{tabular}{|l|p{8.4cm}|p{3.83cm}|}\hline
     & \textbf{Representative}    &\textbf{Conditions}  \\\hline
1&$\left\{\left( \begin{array}{cc} 0 &0  \\ 1 &0 \end{array}  \right),0  \right\}$ &      \\\hline
2&$  \left\{ \sqrt{-Det\,a}\left(\begin{array}{cc} 1 &0  \\ 0 &-1 \end{array}  \right),0  \right\}$   & \begin{tabular}[h]{p{3.4cm}         } $     Im\sqrt{-Det\,a} \in {\bf R ^+}$\\    \text{or}\\ $  \sqrt{-Det\,a} \in{\bf R ^+} $\end{tabular}  \\\hline
3 &  $ \left\{ \sqrt{-\vert W \vert} \left(\begin{array}{cc} 0 &0 \\ 1 &0 \end{array} \right),\ -sig(Tr(P))\left(\begin{array}{cc} 1 &0 \\ 0 &0 \end{array} \right) \right\} $ &$ \sqrt{-\vert W \vert}  \in {\bf R ^+} $\\\hline
4&  $  \left\{ {\frac{is}{2}}\ \left(  \begin{array}{cc}1 &0 \\0 &-1\end{array} \right), -sign (Tr\left( P \right)  )\left(  \begin{array}{cc}1 &0 \\0 &0\end{array}  \right)\right\} $  & 
$ W=sP$\\\hline
5& \begin{tabular}[h]{c} $ \left\{   {\frac{i}{2}} \sqrt {\frac{-\vert W \vert}{\vert P \vert}}\left(  \begin{array}{cc}1 &0 \\0 &-1 \end{array}                  \right)\right.$,\\   $ \left.    -sign (Tr\left( P \right)  )\sqrt{|P|}\ I \right\} $\end{tabular} &      \begin{tabular}[h]{p{3.34cm}      }$   \sqrt {\frac{-\vert W \vert}{\vert P \vert}} \in {\bf R ^+}\cup \{ 0\}$\\ $\sqrt{|P|}\in{\bf R ^+} $ \end{tabular} \\\hline
6& $\left\{ 0, \sqrt{-|P|}\left( \begin{array}{cc}1 &0 \\ 0 &-1 \end{array}\right)\right\} $ &$ \sqrt{-\vert P \vert} \in {\bf R ^+} $ \\\hline
7& \begin{tabular}[h]{c} $\left\{\frac {-i}{2} sign (Tr\left( W \right))\sqrt {\frac{-\vert W \vert}{\vert P \vert}} \left(  \begin{array}{cc}1 &0 \\0 &-1\end{array}  \right) \right. ,$\\ $ \left. \sqrt{-|P|}\left(  \begin{array}{cc}1 &0 \\ 0 &-1 \end{array}  \right)\right\} $ \end{tabular} & \begin{tabular}[h]{c}$\sqrt{\frac{-\vert W \vert}{\vert P \vert}} \in {\bf R ^+}$ \\  $\sqrt{-\vert P \vert} \in {\bf R ^+}$ \end{tabular} \\\hline
8&$\left\{\sqrt {\frac{\vert W \vert}{\vert P \vert}}\left( \begin{array}{cc}0 &0 \\ 1 &0 \end{array}\right),\sqrt{-|P|}\left( \begin{array}{cc}1 &0 \\ 0 &-1 \end{array}\right) \right\}$ & \begin{tabular}[h]{c} $ \sqrt{\frac{\vert W \vert}{\vert P \vert}} \in {\bf R ^+}$\\$\sqrt{-\vert P \vert} \in {\bf R ^+}$ \end{tabular}     \\\hline
9& $ \left\{\left( \begin{array}{cc}i\eta &0 \\ 0 &-i\eta \end{array}\right),
\sqrt{-\vert P \vert}\left( \begin{array}{cc}1 &0 \\ 0 &-1 \end{array}\right) \right\}$&\begin{tabular}[h]{c}$\sqrt{-\vert P \vert}\in {\bf R ^+} $\\$ \eta=\pm1 $ \end{tabular} \\\hline
\end{tabular}

\caption{Canonical Representatives.} 
\label{segunda}

\end{figure}

\parindent=1cm
\parskip=5mm
The     $\mathbb R $-quantizable  orbits
are  all  of  the  types  3,  6,  8,  9 and these of type\  5
corresponding to the case $|W|=0$

                   Quantizable   but   not  {\mbox{$\bf   R
\rm$}}-quantizable   are   the   orbits   whose   canonical
representatives  are:
 $$
 \left\{\frac{iT}{8 \pi} \left(
\begin{array}{cc}  1  &0  \\  0  &-1 \end{array} \right) ,0
\right\} \ \ (type\  2),$$

$$\left\{ \frac{i\chi T}{8 \pi}\
\left(  \begin{array}{cc}
1 &0 \\
0 &-1
\end{array}                 \right)
,
 -sign (Tr\left( P \right)  )
\left(  \begin{array}{cc}
1 &0 \\
0 &0
\end{array}                 \right)
\right\}    \ \  (type\  4),  $$

$$\left\{  \frac{iT}{8 \pi}
\left(  \begin{array}{cc}
1 &0 \\
0 &-1
\end{array}                 \right)
,
 -sign (Tr\left( P \right)  )
\sqrt{|P|}\ I \right\}\ \ (type\  5), $$

$$ \left\{ \frac{i\chi T}{8 \pi}
\left(  \begin{array}{cc}
1 &0 \\
0 &-1
\end{array}  \right)
,
 \sqrt{-|P|}\
\left(  \begin{array}{cc}
1 &0 \\
0 &-1
\end{array}                 \right)
\right\}\ \   (type\  7).$$
where, in all cases, $T \in \bf Z^+\rm, \chi=1,-1.$\

The concrete   particles we  study in this paper are those corresponding to Type 5 (massive particles) and Type 4 (massless particles such that the Pauly-Lubanski fourvector is proportional to Impulsion-Energy). Results concerning  other types of particles  will be published elsewhere.

\section{Guide to the explicit construction of Wave Functions.}\label{guide}

Let $\alpha$\ be a quantizable element of $\underline G^*.$

To determine the explicit form of the Wave Functions of the corresponding free particles, one can proceed as follows

1. Evaluate the isotropy subgroup at $\alpha,$\  of the coadjoint representation, $G_{\alpha}$. 

The coadjoint orbit of $\alpha,$ that is the main symplectic manifold associated to  $\alpha,$ is identified to the homogeneous space $G/G_{\alpha}.$

2. Find a surjective homomorphism, $C_{\alpha},$ from $G_{\alpha}$\ onto the unit circle {\bf S}$^1,$\ whose differential is $\alpha.$

3. Evaluate $(G_{\alpha})_{SL},$ that is composed by the $g\in SL$ such that $(g,h)\in G_{\alpha}$ \ for some $h\in H(2).$

4. Determine  $(C_{\alpha})_{SL},$  defined in \eqref{calfasl}, and $S\equiv (C_{\alpha})_{SL}\left((G_{\alpha})_{SL}\right).$

5. Find an action of $SL(2,{\mathbb C})$ on a manifold, such that the isotropy subgroup at some point, $p$, is $(G_{\alpha})_{SL}.$ 

Identify  $SL/(G_{\alpha})_{SL}$ with the orbit of $p,$ ${\cal K}$.

6. Determine a volume element on ${\cal K},$ $\omega,$ invariant under the action of $SL$.

7. If $S=\{1\},$ the Prewave Functions are the 
$$
\psi_f(H,K)= f(K)\,e^{i\pi  Tr(P(K  )\varepsilon\,     \overline    H\, \varepsilon) }
$$
where $(H,K)\in H(2)\times {\cal K},$ $f$ is a function on ${\cal K},$ and $P$ is the dynamical variable \emph{linear momentum}, on ${\cal K},$ given by \eqref{linear momentum}.

8. If $S\neq\{1\},$  find a Trivialization of $ C_\alpha $ ($c.f.$ \eqref{trivialization}). 

Let ${\cal B}$ the orbit of $z_0$, identified to $SL/Ker(C_{\alpha})_{SL},$
$$
 r:{\cal B} \rightarrow {\cal K}
$$
the canonical map from $SL/Ker(C_{\alpha})_{SL})$ onto $SL/(G_{\alpha})_{SL}$.

The Prewave Functions are given by  ($c.f.$ \eqref{prefunc})
$$
\psi_f(H,K)= f(z)\,e^{i\pi Tr(P(K  )\varepsilon\,     \overline    H\, \varepsilon) }z
$$
where $(H,K)\in H(2)\times {\cal K},$ $z\in r^{-1}(K),$ and $f$ is a function on ${\cal B}$ homogeneous of degree -1 under product by modulus one complex numbers.

9. The Prewave Functions corresponding to continuous with compact support $f$ define Wave Functions by means of ($c.f.$ \eqref{WF})
$$
\widetilde     {\psi_f}(H)=\int_{\cal K} {\mbox{$\psi$}}_f(H,\cdot)\ {\mbox{$\omega$}} .
$$

\section{Massive particles.}\label{massive}

\subsection{Massive particles with $T=0.$ }

\subsubsection{Wave Functions. Klein-Gordon equation.}\label{WFKG}

Let us consider a particle whose movement space is the
coadjoint orbit of
$$
{\alpha_o}=\left
\{
0,\  \eta \,m\,I\right \},\ \
\ m \in {\mathbb{R}}^+, \eta=\pm 1.
$$

This orbit is a
$\mathbb{R} \rm$-quantizable orbit of the type 5, in the
notation of section \ref{concreteqforms}.

 The number $(-\eta)$ is the sign  of
energy \it i.e. \rm the sign
of the value of the dynamical variable $P^4$ ($c.f.$ section \ref{sec-grupo}) at any point of the orbit. According with the usual interpretation of the determinant of momentum-energy, $\vert P \vert,$\  as mass square, the number $m$\ must be interpreted as being the mass of the particle.

By direct computation, one sees that
$$
{\mbox{$G_{\alpha_o}$}}     = \{\ (A
, \ hI ): \   A\in SU(2),\ h\in \mathbb{R} \}.$$

The unique homomorphism onto $
{\mathbb{R}}$ whose differential is
${\alpha_o}$ is given by
$$ C^\prime_{\alpha_o}   (
A ,\ hI)=
-\eta m h .$$

The unique homomorphism onto ${\bf S}^1$ whose differential
is
$\alpha_o$ is given by
$$C_{\alpha_o}  (
A ,\ hI)=e^{-2\pi i\eta m h}.$$

 Then we have
 $$
\begin{gathered}
Ker\,C^\prime_{\alpha_o}=\{(A,0):A\in SU(2)\}=SU(2)\oplus \{0\},\\
Ker\,C_{\alpha_o}=\{(A,\,\frac{\eta   N}{m}\, I):A\in SU(2),\ N \in \mathbb{Z}\},\\
{\mbox{$\widetilde C_{\alpha_o}$}}    (
A ,\ H)=e^{-\pi i\eta m TrH}          ,\\
{\mbox{$(G_{\alpha_o})_{SL}$}}=SU(2) ,\\
{\mbox{$(C_{\alpha_o})_{SL}$}}={\bf 1 } ,\\
 SL_1 =SL_2
 = {\mbox{$Ker\,(C_{\alpha_o})_{SL}$}}
={\mbox{$(G_{\alpha_o}  )   _{SL}$}},\end{gathered} $$

Thus, in the commutative diagram of figure \ref{diagr2}, section \ref{wave}, 
 the four spaces on the left are the same. All of them represent State Space,
$$
H(2)\times  \frac{SL}{(G_{\alpha_o})_{SL}\cap SL_1}
$$
what in our present case becomes
$$
H(2) \times \frac{SL}{SU(2)},
$$
and can be obviously identified to
$$
\frac{G}{SU(2) \oplus \{0\}},
$$
by means of the diffeomorphism
$$
(H,A\,SU(2))\rightarrow (A,H)(SU(2) \oplus \{0\}).
$$

 Then, in this case,   besides the canonical map from State Space onto Movement Space (\emph{c.f.}  figure \ref{diagr2}),
\begin{equation}\label{nudos}
\nu_2: (H,ASU(2))\in H(2) \times \frac{SL}{SU(2)} \rightarrow (A,H)G_\alpha \in \frac{G}{G_{\alpha_o}}
\end{equation}
we have a natural map from State Space onto $G/Ker\,C_{\alpha_o}$
\begin{equation}\label{nuuno}
\nu_1:(A,H)(SU(2) \oplus \{0\})\in \frac{G}{SU(2) \oplus \{0\}} \rightarrow (A,H){Ker\,
C_{\alpha_o}} \in \frac{G}{Ker\,C_{\alpha_o}}.
\end{equation}

\begin{figure}
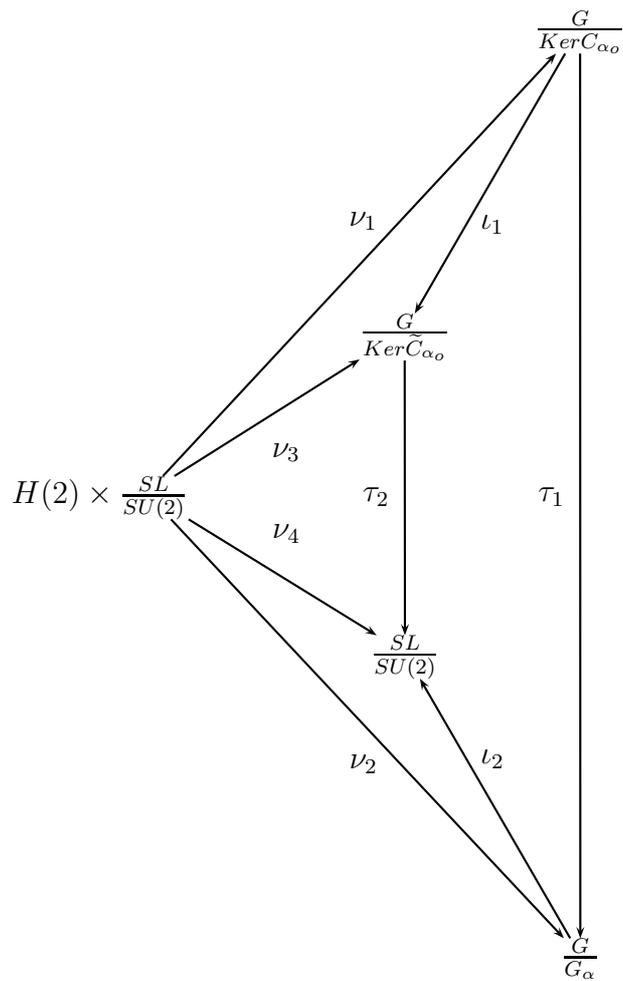

  \centering
$
\begin{psmatrix}[colsep=1cm,rowsep=1.5cm]
 \ \ \ \ \ \ \ \ \ \  & \ \ \ \ \ \ \ \ \ \  & \frac{G}{KerC_{\alpha_o}}   \\
 \ \ \ \ \ \ \ \ \ \  &  \ \ \ \ \ \ \ \ \ \ & \ \ \ \ \ \ \ \ \ \   \\
	 \ \ \ \ \ \ \ \ \ \ &   \frac{G}{Ker \widetilde C_{\alpha_o}}&   \ \ \ \ \ \ \ \ \ \   \\
H(2) \times \frac{SL}{SU(2)} \ \ \ \ \ \ \ \ \ \ &  \ \ \ \ \ \ \ \ \ \ &   \ \ \ \ \ \ \ \ \ \ \\
     \ \ \ \ \ \ \ \ \ \ & \frac{SL}{SU(2)}  & \ \ \ \ \ \ \ \ \ \      \\
		 \ \ \ \ \ \ \ \ \ \  &  \ \ \ \ \ \ \ \ \ \ & \ \ \ \ \ \ \ \ \ \   \\
	 \ \ \ \ \ \ \ \ \ \   &  \ \ \ \ \ \ \ \ \ \    &  \frac{G}{G_\alpha} 
\ncline{->}{4,1}{1,3}^{\nu_1}
\ncline{->}{4,1}{7,3}_{\nu_2}
\ncline{->}{4,1}{3,2}_{\ \ \nu_3}
\ncline{->}{4,1}{5,2}^{\ \ \nu_4}
\ncline{->}{3,2}{5,2}<{\tau_2}
\ncline{->}{1,3}{7,3}<{\tau_1}
\ncline{->}{1,3}{3,2}_{\iota_1}
\ncline{->}{7,3}{5,2}^{\iota_2}
\end{psmatrix}
$
\caption{ Diagram for Klein-Gordon particles.}\label{figuraKG}
\end{figure}

The
diagram of figure \ref{diagr2},  becomes that of figure  \ref{figuraKG}.

Let ${\cal H}^m$ be the positive mass hyperboloid
$$
{\cal H}^m=\{ H \in H(2):\ det\, H\,=\,m^2,\ Tr\,
H\,>\,0\,\}.
$$

In ${\cal H}^m$ we consider the action of  $SL(2,C)$ given
by
\begin{equation} \label{ac}
A* H=A\,H\,A^*,
\end{equation} 
for all $H\in {\cal H}^m$ and $A \in  SL(2,C).$ In these conditions we have $Tr(A\,H\,A^*)>0$ as a consecuence of the fact that $SL(2,C)$ is connected.

Since the group of restricted homogeneous Lorenz transformations is transitive on ${\cal H}^m,$\ so does $SL(2,\mathbb C).$\ 
 The
isotropy subgroup at
$mI$ is
$SU(2).$\  Thus $SL/(G_{\alpha_o})_{SL}$ is identified to
${\cal H}^m,$\ by means of the diffeomorphism
$$
A\, SU(2)\in \frac{SL}{SU(2)} \longrightarrow m\,A\,A^* \in {\cal H}^m.
$$

 If  $K=  m\,A\,A^* \in {\cal H}^m,$\  then $K$\ is identified to $A\, SU(2)\in \frac{SL}{SU(2)},$\ and the function P thus becomes 
$$
P(K)=P(ASU(2))=-A(\eta \,m\,I)A^*=-\eta K.
$$

The manifold ${\cal H}^m$\ has the global parametrization
$$
\varphi:(p^1,\ p^2,\ p^3)\in {{\mathbb{R}} ^3}\ \ \mapsto\ \
h(p^1,\ p^2,\ p^3,\
(m^2+  p^2)^{\frac 12} )\ \in\ {\cal H}^m
$$
where
$$
p^2= \sum_{i=1}^3  (p^i)^2.
$$

In particular it is orientable.

On the other hand, the restriction of the pseudoriemannian metric defined on ${\mathbb R}^4$ by Minkowski metric, to ${\cal H}^m,$\  is negative definite so that its opposite is a riemannian metric, that    provides us with a canonical volume element, $\nu$. A computation of the matrix of the riemannian metric in the chart $\varphi$\ leads to
$$
\nu =\frac {dp^1 \wedge dp^2 \wedge
dp^3}{\sqrt
{m^2+ p^2}}
$$
where the  $p^i$ are the coordinates in the chart $\varphi.$

The action on $H(2)$\ defined as in \eqref{ac} preserves Minkowski metric, so that the action on ${\cal H}^m$\ preserves $\nu$. 

As a consequence,  $\nu$\ is an invariant volume element under the action \eqref{ac}.

Since $(C_{\alpha_o})_{SL}={\bf 1}$, $\cal C$\ is composed by functions on ${\cal H}^m.$\  

 The prewave functions have the
form
$$
\psi _f:(H,\ K) \in   H(2) \times
{\cal H}^m
\
\mapsto \ \
f\left(K
 \right) e^{-i
\pi \eta Tr(K \varepsilon \overline H \varepsilon)  }.
$$
where f is in $\cal C.$

The corresponding wave function is
$$
\tilde \psi _f(X)=\int_{{\cal H}^m}
\psi_f(h(X),\,\cdot\,)\ \nu
$$
By direct computation one sees that the prewave and wave functions
satisfy \bf Klein-Gordon \rm equation.

If $f^\prime$ is another continuous  with compact
support function on ${\cal H}^m$, the hermitian product of
the quantum states corresponding to $\psi _f$ and $\psi
_{f^\prime},$\ defined in section \ref{repprewave}, can be writen
$$
\langle \psi_f,\ \psi_{f^\prime} \rangle
=\int_{{\cal H}^m }
\psi_f^*\,\psi_{f^\prime}\,\nu.
$$

\subsubsection{The Homogeneous Contact and Symplectic Manifolds for Klein-Gordon particles}\label{contKG}



Let us consider the action of $G$ on ${\cal H}^m  \times \mathbb{R}^3  \times \mathbb{S}^1$ \ given by
\begin{equation}
(A,H)*(K,\overrightarrow y,z)=(AKA^*,\overrightarrow x(A,H,\overrightarrow y,K), e^{2\pi i \eta m \ell(A,H,\overrightarrow y,K)}z)
\end{equation}
where $\overrightarrow x(A,H,\overrightarrow y,K)$\ and $\ell(A,H,\overrightarrow y,K)$\ are given by
\begin{equation}
	h(\overrightarrow x(A,H,\overrightarrow y,K),0)=A\,h(\overrightarrow y,0)\,A^*+H+\ell(A,H,\overrightarrow y,K) A\,\frac{K}{m}\,A^*.
\end{equation}
Obviously we have

\begin{equation}
	\ell(A,H,\overrightarrow y,K)=-m\frac{Tr(A\,h(\overrightarrow y,0)\,A^*+H)}{Tr(AKA^*)}.
\end{equation}

This is a transitive  action, as can be proved by direct computation.

The isotropy subgroup at $(mI,\,\overrightarrow 0,1),$\ is $Ker\,C_{\alpha_0},$\ so that the homogeneous space $G/Ker\,C_{\alpha_0},$\ can be identified to ${\cal H}^m  \times \mathbb{R}^3  \times \mathbb{S}^1$ \ by means of the map
\begin{equation}
	(A,H) Ker\,C_{\alpha_0} \longrightarrow  (A,H)*(mI,\overrightarrow 0,1).
\end{equation}

We know  (\emph{c.f.} section \ref{quantizable}) that the left-invariant 1-form on $G$\ whose value at $(I,0)$\ is $\alpha_o$,\ $\tilde \alpha_o,$\ projects in a 1-form , $\Omega,$\ in $G/Ker\,C_{\alpha_o}$\ which is a  contact form, and is homogeneous in the sense that it is preserved by the diffeomorphisms corresponding to the canonical action of $G$ on  $G/Ker\,C_{\alpha_o}.$\ 

One way to give explicitly $\Omega,$\  is to use coordinate domains in $G/Ker\,C_{\alpha_0}$\ that are also domains of sections of the canonical map from the group $G$ onto $G/Ker\,C_{\alpha_0}.$\ The pull back of $\tilde \alpha_0$\ by that section, is the restriction of    $\Omega$\ to the domain, and we can give its local expresion in the coordinate system.

 With our identification , the just cited canonical map becomes
\begin{equation}\label{proy.can5}
\mu:(A,H)\in G \longrightarrow (A,H) * (mI,\overrightarrow 0,1)\in {\cal H}^m  \times \mathbb{R}^3  \times \mathbb{S}^1.
\end{equation}

In ${\cal H}^m  \times \mathbb{R}^3  \times \mathbb{S}^1$ \ be define a coordinate system for each $\tau \in \mathbb{R}$ as the inverse of the parametrization
$$
\begin{gathered}
	\phi_\tau: (k_1,k_2,k_3,x^1,x^2,x^3,t)\in \mathbb{R}^6 \times  \left(-\frac{1}{2},\frac{1}{2}\right) \rightarrow  \\
	\rightarrow   \left(  m\ h(k_1,k_2,k_3,k_4),x^1,x^2,x^3,e^{2\pi i(t+\tau)}\right)
\in   \\  
	\in  {\cal H}^m  \times \mathbb{R}^3  \times \mathbb{S}^1
\end{gathered}
$$
where $k_4=\sqrt{1+k_1^2+k_2^2+k_3^2}.$


Now, let us consider the map
$$
\begin{gathered}
\sigma_\tau:\phi_\tau (k_1,k_2,k_3,x^1,x^2,x^3,t)  \rightarrow\\  \rightarrow \left( \left( \begin{array}{cc}
	\sqrt{\frac{k_4+k_3}{1+k_1^2+k_2^2}}&\sqrt{\frac{k_4+k_3}{1+k_1^2+k_2^2}}(k_1-ik_2)\\
	0&\sqrt{\frac{1+k_1^2+k_2^2 }{  k_4+k_3}}
	\end{array}\right), 
	h(\overrightarrow x,0)-\frac{\eta}{m}(t+\tau)) h(\overrightarrow k,k_4)\right)  \\   \in  G.\hspace{12cm}
\end{gathered}
$$

We have 
$$
\mu \circ \sigma_\tau(\phi_\tau (R))=\sigma_\tau(\phi_\tau (R))*(mI,\overrightarrow 0,1)=\phi_\tau (R),
$$
for all $R\in \mathbb{R}^6 \times  (-1/2,\ 1/2),$\ so that $\sigma_\tau$ is a section of the canonical map $\mu$.

Then, on the image of   $\phi_\tau$\ we have
$$
\Omega=\sigma_\tau^* \tilde \alpha_0.
$$

In order to obtain, for exemple, the value
$$
(\Omega)_{\left(  \phi_\tau\left(   R   \right)  \right)}\cdot \left(   \frac{\partial}{\partial k_1}  \right)_{\left(  \phi_\tau\left( R    \right)  \right)},
$$
we must find  the value of $\alpha_0$\ on the tangent vector to the curve
\begin{equation}
\gamma(s)= \left( \sigma_\tau ( \phi_\tau (R)  )\right)^{-1}
           \left( \sigma_\tau ( \phi_\tau (R+(s,0,0,0,0,0,0))\right) 
\end{equation}
at $0.$\ 

Since $$\alpha_0=\left
\{
0,\  \eta \,m\,I\right \},
$$
the first component of the tangent vector at 0 of $\gamma$\ has no incidence in the value of $\alpha_0$\ on it.

The preceding computation gives us 
$$
\langle (o,\ \eta m I),\stackrel{.}{\gamma}_0   \rangle=0.
$$

A similar computation for each of the other coordinates, or a common reasoning with a curve $\gamma(s)=\phi_\tau(c(s)),$\ leads 
to
\begin{equation}
	\Omega=d\,t+\eta \,m\,k_i\,dx^i.
\end{equation}

Obviously
$$
d\Omega=\eta \,m \,dk_i\wedge\,dx^i,
$$
so that
$$
\Omega\wedge\left(d\Omega\right)^3=\frac{\eta m^3}{16}\ dt\wedge dk_1\wedge dk_2\wedge dk_3\wedge dx^1\wedge dx^2\wedge dx^3
$$
what confirms the fact that $\Omega$ is a contact form.

The characteristic vectorfield $Z(\Omega)$\ defined by
$$
i(Z(\Omega))\Omega =1,\ \  i(Z(\Omega))\,d\Omega =0,
$$
has  flow, ${\phi_t},$\  given by
$$
\phi_t(K,\overrightarrow x,z)=(K,\overrightarrow x, e^{2\pi i t}z).
$$
As a consecuence, the period of all its integral curves is $T(\Omega)=1,$\ so that $\Omega$\ is a connection form, and $d\Omega$\ the corresponding curvature form.

Now, let us consider the action of $G$ on ${\cal H}^m  \times \mathbb{R}^3  $ \ given by
\begin{equation}\label{accion simpl}
(A,H)*(K,\overrightarrow y)=(AKA^*,\overrightarrow x(A,H,\overrightarrow y,K))
\end{equation}
where
\begin{equation}
	h(\overrightarrow x(A,H,\overrightarrow y,K),0)=A\,h(\overrightarrow y,0)\,A^*+H+\ell(A,H,\overrightarrow y,K) A\,\frac{K}{m}\,A^*.
\end{equation}

This is also a transitive  action.The isotropy subgroup at $(mI,\,\overrightarrow 0),$\ is $G_{\alpha_0},$\ so that  $G/G_{\alpha_0},$\ can be identified to ${\cal H}^m  \times \mathbb{R}^3  $ \ by means of the map
\begin{equation}
	(A,H) G_{\alpha_0} \longrightarrow  (A,H)*(mI,\overrightarrow 0).
\end{equation}

The left-invariant 2-form on $G$\ whose value at $(I,0)$\ is $d\alpha_0$,\ $d\tilde \alpha_0,$\ projects in a 2-form , $\omega,$\ in $G/G_{\alpha_0}$\ which is a  simplectic form, and is homogeneous in the sense that it is preserved by the diffeomorphisms corresponding to the canonical action of $G$ on  $G/G_{\alpha_0}.$\ 
With our identifications, the canonical map
$$
(A,H) Ker\,C_\alpha \rightarrow (A,H) G_\alpha
$$
becomes
$$
(K,\overrightarrow x,z )\in {\cal H}^m  \times \mathbb{R}^3 \times \mathbb{S}^1 \rightarrow (K,\overrightarrow x)\in {\cal H}^m  \times {\mathbb R}^3 
$$
and, since $T(\Omega)=1,$\  $\omega$\  must be a projection of $d\Omega,$\ so that
$$
\omega= {\eta \,m} \,d{k}_i \wedge \,d{x}^i,
$$
where now $\{ k_1,\dots,x^3\}$\ are the coordinates corresponding to the  following global parametrization of ${\cal H}^m  \times {\mathbb R}^3 $
$$
\begin{gathered}
	\underline{\phi}: ({k}_1,{k}_2,{k}_3,{x}^1,{x}^2,{x}^3)\in {\mathbb R}^6  \rightarrow  \\
	\rightarrow   \left(  m\ h({k}_1,{k}_2,{k}_3,{k}_4),{x}_1,{x}_2,{x}_3\right)
   \\  
	\in  {\cal H}^m  \times {\mathbb R}^3  
\end{gathered}
$$
where ${k}_4=\sqrt{1+{k}_1^2+{k}_2^2+{k}_3^2}.$

In this case we have a global section of the canonical map
$$
\begin{gathered}
	\underline{\sigma}(\underline\phi({k}_1,{k}_2,{k}_3,{x}^1,{x}^2,{x}^3))=\\
	=\left( \left( \begin{array}{cc}
	\sqrt{\frac{k_4+k_3}{1+k_1^2+k_2^2}}&\sqrt{\frac{k_4+k_3}{1+k_1^2+k_2^2}}(k_1-ik_2)\\
	0&\sqrt{\frac{1+k_1^2+k_2^2 }{  k_4+k_3}}
	\end{array}\right), 
	h(\overrightarrow x,0) \right).
\end{gathered}
$$
The map $\underline{\sigma}$ is a section since
$$
\underline{\sigma}(\phi(R))*(mI,\overrightarrow 0)=\phi(R).
$$

The coadjoint orbit of $\alpha_0,$\ becomes identified to $ {\cal H}^m  \times {\mathbb R}^3  $\ by means of
 $$
Ad^*_{(A,H)}\cdot \alpha_0  \rightarrow (A,H)*(mI,\overrightarrow 0),
$$
whose inverse is given by
$$
(K,\overrightarrow x) \rightarrow Ad^*_{\underline{\sigma}(K,\overrightarrow x)}  \cdot \alpha_0.
$$

Thus, the $(e,g)\in \underline{G},$\  what are dynamical variables on the coadjoint orbit, becomes functions on $ {\cal H}^m  \times \mathbb{R}^3 , $\ denoted  by   $D_{(e,g)}$\  in section \ref{quantizable}, defined as  follows
$$
D_(e,g)(\underline{\phi}(k_1,\dots,x_3))=(Ad^*_{\underline{\sigma}(\underline{\phi}(k_1,\dots,x_3))}\alpha_0)( e,g).
$$

In particular, the Linear and Angular Momentum, given in \eqref{limoang}, are, as functions on  $ {\cal H}^m  \times \mathbb{R}^3, $\   
\begin{eqnarray}\nonumber
P(m\,k,\overrightarrow x)&=&- \eta\,m\ k, \\  \label{KGvardin}
\overrightarrow l(m\,k,\overrightarrow x)&=&\eta\,m\,\overrightarrow k \times \overrightarrow x=\overrightarrow x \times \left(\overrightarrow  P(m\,k,\overrightarrow x) \right) \\   \nonumber
\overrightarrow g(m\,k,\overrightarrow x)&=&-\eta\,m\,k^4 \overrightarrow x=\left( P^4(m\,k,\overrightarrow x) \right) \ \overrightarrow x ,
\end{eqnarray}
where we have denoted $D_{P^k},\ D_{l^i},$\ {and} $D_{g^j}$\ by $P^k, l^i$ {and} $g^j$ respectively.

Also, each $a\in \underline{G}, $\  define an infinitesimal generator, $X^s_{a},$\  of the action \eqref{accion simpl}.

These infinitesimal generators are defined by means of its flow, and its local expresions can be obtained  directly from the definition,
 but it is also possible to use formula  \eqref{hamilt} to obtain the following local expresions in the chart $\underline \Phi ^{-1}$
\begin{eqnarray}
X^s_{P^j}&=&\frac{\partial}{\partial x^j}\\
X^s_{P^4}&=&\sum_{j=1}^3\frac{k_j}{k_4}\ \frac{\partial}{\partial x^j}\\
X_{l^k}^s &=&\sum_{j,r=1}^3 \varepsilon_{kjr} k_j \frac{\partial}{\partial k_r} +\sum_{j,r=1}^3 \varepsilon_{kjr} x^j
\frac{\partial}{\partial x^r} \\
X^s_{g^j}&=&x^j\, X_{P^4}^s-k_4\frac{\partial}{\partial k_j}
\end{eqnarray}
where $ \varepsilon_{kjr}$ is as in \eqref{operclasic}.

Equation \eqref{hamilt} proves that these vector fields are globally hamiltonian, and the corresponding hamiltonian is, with our actual notation, the function appearing in the subindex in each case.

The infinitesimal generator corresponding to $a\in \underline G$\ for the action in the contact manifold is denoted by $X_a^c.$\ We have in the charts $\Phi_\tau^{-1}$
\begin{eqnarray}\label{inf gen cont}
X^c_{P^j}&=&\frac{\partial}{\partial x^j}\nonumber\\
X^c_{P^4}&=&\frac{1}{k_4}\left(\sum_{j=1}^3\,k_j\, \frac{\partial}{\partial x^j}+\eta m  \frac{\partial}{\partial t} \right)\nonumber \\
X_{l^k}^c &=&\sum_{j,r=1}^3 \varepsilon_{kjr} k_j \frac{\partial}{\partial k_r} +\sum_{j,r=1}^3 \varepsilon_{kjr} x^j
\frac{\partial}{\partial x^r} \\
X^c_{g^j}&=&x^j\, X_{P^4}^c-k_4\frac{\partial}{\partial k_j}.\nonumber
\end{eqnarray}

The Quantum Operators representing Linear and Angular Momentum for Quantum States on the Contact Manifold are $\frac{1}{2 \pi i}$\ times the vectorfiels in \eqref{inf gen cont}.

If $f$\ is a $C^\infty$\ function on ${\cal H}^m$\ the corresponding \bf Quantum State \rm in the contact manifold is
$$
\Phi_f((A,H)\,Ker\,C_\alpha)=f(A\,(G_\alpha)_{SL})Exp[i \pi Tr(P(A\,(G_\alpha)_{SL})\varepsilon \overline{H} \varepsilon)]
$$
or, with the identifications we have made
$$
\Phi_f((A,H)* (mI,\overrightarrow 0,1))=f(mAA^*) Exp[i \pi Tr((-\eta mAA^*)\varepsilon \overline{H} \varepsilon)]
$$
but
$$
\Phi_f((A,H)* (mI,\overrightarrow 0,1))=\Phi_f(mAA^*,h^{-1}( H+\ell AA^*), Exp[2 \pi i \eta m \ell]),
$$
where $ \ell $\ is such that $Tr(H+\ell AA^*)=0.$
Thus
$$
\Phi_f(K,\overrightarrow x,z)=f(K)  Exp[i \pi Tr((-\eta K)\varepsilon \overline{(h(\overrightarrow x,0)-\ell \frac{K}{m})} \varepsilon)]
$$
where $\ell$\ is such that $z= Exp[2 \pi i \eta m \ell].$\ Then
$$
\Phi_f(K,\overrightarrow x,z)=f(K) Exp[-i \eta \pi Tr( K \varepsilon \overline{h(\overrightarrow x,0)}\varepsilon]
Exp[i \ell \eta \pi Tr( K \varepsilon \overline{\frac{K}{m}}\varepsilon]
$$
and one sees that, if $K=m h(\overrightarrow k,k_4),$
$$
\Phi_f(K,\overrightarrow x,z)=f(K)\, Exp[-2 \pi i \eta  m \langle \overrightarrow k ,\overrightarrow x \rangle ] \,\overline{z} .
$$

In the coordinate system associated to $\phi_\tau,$\ we have 
$$
\Phi_f \circ \phi_\tau (\overrightarrow k ,\overrightarrow x ,t)=f(mh(\overrightarrow k,k_4))\, \, Exp[-2 \pi i ( \eta  m \langle \overrightarrow k ,\overrightarrow x \rangle +t+\tau)].
$$

\parindent=1cm
\parskip=5mm

\subsection{Massive particles with $T\geq 1$.}

\subsubsection{Wave Functions for $T=1.$  Dirac equation.}\label{WFDIRAC}

Let us consider a particle whose movement space is the
coadjoint orbit of
\begin{equation}
\alpha_1=\left
\{ \ \frac  {i}{8\pi}\left( \begin{array}{cc} 1&0  \\ 0&-1
\end{array} \right) \ ,\  \eta \,m\,I\right \},
 \label{eq:rep5.1}
\end{equation}
where $
 m \in {\bf R}^+, \eta=\pm 1.
$
This orbit is a quantizable, not
$\bf R \rm$-quantizable orbit, of the type 5, in the
notation of section \ref{concreteqforms}.

In this case we have
$$
{\mbox{$G_{\alpha_1}$}}     = \{\ (\ \left( \begin{array}{cc}
z&0 \\
0&\overline z
\end{array} \right)
, \ hI ): \   z\in S^1,\ h\in {\mbox{$\bf R \rm$}} \}.
$$

The
unique homomorphism from $G_{\alpha_1}$ onto ${\bf S}^1$ whose
differential is ${\alpha_1}$ is given by
$${\mbox{$C_{\alpha_1}$}}    (\ \left( \begin{array}{cc}
e^{2\pi i\phi}&0 \\
0&e^{-2\pi i\phi}
\end{array} \right) ,\ hI)=e^{2\pi i(\phi-\eta m h)}.
$$

Then

$
\begin{array}{l}
Ker\, C_{\alpha_1}=\{(\ \left( \begin{array}{cc}
e^{2\pi i\eta m h}&0 \\
0&e^{-2\pi i\eta m h}
\end{array} \right) ,\ hI):h\in \mathbb R \},\\
\vspace{.1in}
(G_{\alpha_1}  )   _{SL}=\{\ \left( \begin{array}{cc}
z&0 \\
0&\overline z
\end{array} \right)
:\ z\in S^1\},     \\
\vspace{.1in}
{{\mbox{$(C_{\alpha_1})_{SL}$}}(\
\left( \begin{array}{cc}
z&0 \\
0&\overline z
\end{array} \right)
)=z}, \\
\vspace{.1in}
 SL_1 \cap    SL_2 ={\mbox{$(G_{\alpha_1}  )   _{SL}$}}, \\
\vspace{.1in}
 SL_1 \cap {\mbox{$Ker\,(C_{\alpha_1})_{SL}$}}
={\mbox{$Ker\,(C_{\alpha_1})_{SL}$}}=\{ I \}.    \\
\end{array}
$

Let us denote 
$(G_{\alpha_1})_{SL}$ by $[S^1],$  $G_{\alpha_1}$ by $[S^1]\oplus \mathbb R,$ and $Ker\, C_{\alpha_1}$ by $[R].$

\begin{figure}[h]
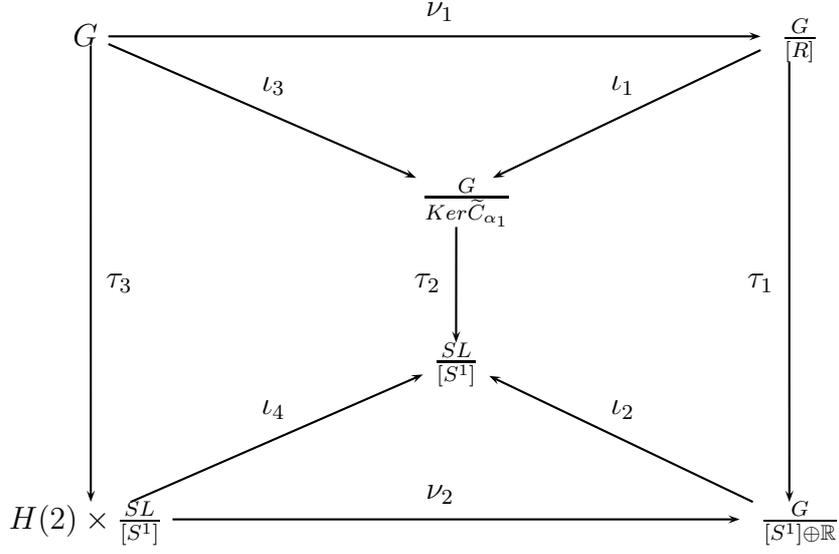

  \centering
$
\begin{psmatrix}[colsep=3cm,rowsep=1.5cm]
G\ \    &                                          &\ \   \frac{G}{[R]}   \\
  \ \   &\ \    \frac{G}{Ker \widetilde C_{\alpha_1}}  &  \\
  \ \   & \ \frac{SL}{[S^1]}\ \          &  \\
H(2) \times \frac{SL}{ [S^1]}\ \   &  &\ \  		\frac{G}{[S^1]\oplus \mathbb R}
\ncline{->}{1,1}{1,3}^{\nu_1}
\ncline{->}{4,1}{4,3}^{\nu_2}
\ncline{->}{1,1}{4,1}>{\tau_3}
\ncline{->}{2,2}{3,2}<{\tau_2}
\ncline{->}{1,3}{4,3}<{\tau_1}
\ncline{->}{1,1}{2,2}^{\iota_3}
\ncline{->}{1,3}{2,2}^{\iota_1}
\ncline{->}{4,1}{3,2}^{\iota_4}
\ncline{->}{4,3}{3,2}^{\iota_2}
\end{psmatrix}
$
\caption{Fibre Bundles for Dirac Particles}
\label{diagrDirac2}
\end{figure}

Then, in the conmutative diagram of Figure \ref{diagr2}, $\iota_3$ and $\iota_4$ become identical maps, $\tau_3=\tau_4$ and the diagram becomes, with obvious conventions, that of Figure    \ref{diagrDirac2}.

The homogeneous space $SL/[S^1]$\  can be characterised
as follows.

Let  ${\bf P}_1(\mathbb{ C})$  be the
complex projective space corresponding to $ \mathbb{ C}^2$
( $i.e.$\ the space $ {\bf P}_1(\mathbb{ C}) $\ consists of 
the one dimensional complex subspaces of $ {\mathbb{ C}}^2$).

In ${\cal H}^m\times  {\bf P}_1(\mathbb{ C})$\
 we can consider the action of   $ SL(2, C)$
given by
$$
A\,*\,(H,[z])=(AHA^*,\,[Az])
$$
where $[z]\in {\bf P}_1(\mathbb{ C})$
 is the vector subspace generated by $z \in\  {\mathbb{ C}}^2-\{(0,  0) \}$.

We know that the partial action on ${\cal H}^m$\ is transitive. Now let us see that the complete  action on  ${\cal H}^m\times  {\bf P}_1(\mathbb{ C})$ is also transitive.

 Let $(K,[w])\in  {\cal H}^m\times  {\bf P}_1(\mathbb  C).$\ Then, there exist $A\in SL$\ such that $A*(K,[w])=
(mI,[w']),$ for some $w'\in {\mathbb C}^2-\{ (0,0)\},$\ but there exist obviously an element $B$ of $SU(2)$\ such that
$$
\left[B  \left(\begin{array}{c}1\\0 \end{array}\right) \right]=[w'],
$$
and we see that
 $$(B^{-1}A)*(K,[w])=\left( mI,\left[ \left(\begin{array}{c}1\\0 \end{array}\right) \right] \right).
$$
Transitivity follows.

 On the other hand, the isotropy subgroup at 
$$
(mI,\ [ \left( \begin{array}{c} 1 \\ 0 \end{array}\right) ])$$
is $[S^1] $,\ so that, since the action is
 transitive,
one can identify $SL/[S^1]$\ to
${\cal H}^m\times  {\bf P}_1(\mathbb{ C})$.

It is a well known fact that ${\bf P}_1(\mathbb{C})$\ is diffeomorphic to the sphere $S^2.$\ A diffeomorphism can be described as follows.

Let $C^+$\ be the future lightcone, composed by the hermitian matrices having 0 determinant and positive trace. The subset, $S_2,$\  of  $C^+$\ composed by the elements whose trace is 2 is
$$
S_2=\{ h(x,y,z,1):1-x^2+y^2+z^2=0 \}
$$
that can be identified to the sphere $S^2,$\ by means of
$$
\delta:(x,y,z)\in S^2\subset\mathbb{R}^3    \longleftrightarrow  h(x,y,z,1)\in S_2\subset C^+.
$$

The map
$$
\beta_0:[z]\in {\bf  P}_1(\mathbb{ C}) \longrightarrow  \frac{2}{z^*\,z}\,z\,z^* \in S_2,
$$
is bijective. Its inverse is
$$
\beta_0^{-1}: C \in S_2\subset C^+  \longrightarrow [w]\in  {\bf P}_1(\mathbb{ C}),
$$
where $w$ \ is an eigenvector of $C$ corresponding to the eigenvalue 2. This is a consecuence of the fact that
\begin{equation}
\begin{split}	
&\left(\frac{2}{z^*\,z}\,z\,z^*\right)z=2z\\
&\left(\frac{2}{z^*\,z}\,z\,z^*\right)\varepsilon \overline{z}=0
\end{split}	
\end{equation}

Thus
\begin{equation}\label{defbeta}
\beta \stackrel {def}=	\delta^{-1}\circ \beta_0:  {\bf P}_1(\mathbb{ C}) \rightarrow S^2,
\end{equation}
can be described by 
\begin{equation}\label{otradefbeta}
\beta	([z])=\vec u \       \Longleftrightarrow         \  \frac{2}{z^*\,z}\,z\,z^*=h(\vec u,1).
\end{equation}
that is practical in order to use $\beta.$

 More practical in order to use $\beta^{-1}$ is
\begin{equation}\label{otramasdefbeta}
\beta	([z])=\vec u \       \Longleftrightarrow         \  \ h(\vec u,-1)z=0.
\end{equation}

If $p_s$\ and $p_n$ denote the stereographical projections from poles $s=(0,0,-1)$\ and $n=(0,0,1)$\ respectively, we have
\begin{eqnarray}
p_s\left(\beta\left[ \left( \begin{array}{c} z^1\\z^2 \end{array}\right)\right]\right)= \frac{z^2}{z^1}\ \ \  \mathrm{if}\  z^1\neq 0,\\
p_n\left(\beta\left[\left( \begin{array}{c} z^1\\z^2 \end{array}\right)\right]\right)= \frac{\overline z^1}{\overline z^2}\ \ \   \mathrm{if}\  z^2\neq 0.  \nonumber	
\end{eqnarray}
This can be used to prove easily the differentiability of $\beta $\ and $\beta^{-1}. $

Thus, it is possible  to change in all that follows  ${\bf  P}_1(\mathbb{C})$\ by $S^2,$\  but, at this moment, I  prefer  to use ${\bf  P}_1( \mathbb{C}).$

In order to  describe Prewave Functions in this case ,  one can
consider the trivialisation $(\rho,\  {\bf
C}^4,z_0)$,\ where $z_0={}^t\negmedspace (1, 0, 1, 0)$ and $\rho$ is given by
\begin{equation}\label{roendir}
\rho (A)=\left(
\begin{array}{cc}
A&0 \\
0&(A^*)^{-1}
\end{array}
\right)
\end{equation}

The orbit of $z_0$ is
\begin{equation}
 {\cal B}=\left \{
\left( \!
\begin{array}{c}
 w \\
z
\end{array}
 \! \right)
 \ :\ w,\ z \in  \mathbb{ C }^2,\
z^*\,w
=1 \right \}.   \label{eq:B}
\end{equation}

This can be seen by consideration of  the map
\begin{equation}\label{fi0}
\phi_0:
\left( 
\begin{array}{c}
 w \\
z
\end{array}
  \right) \in {\cal B} \longrightarrow \left(w \vert -\varepsilon \overline z   \right)\in SL(2,\bf C) 
\end{equation}
 (it is a diffeomorphism), and the fact that
\begin{equation}\label{ro}
\rho(w|-\varepsilon \overline z )\left( 
\begin{array}{c}
 1 \\0\\1\\0
\end{array}
  \right)=\left( 
\begin{array}{c}
 w \\
z
\end{array}
  \right).
	\end{equation}

When one identifies $SL / Ker(C_{\alpha_1})_{SL}$ \ to $\cal B$,
and
$SL / (G_{\alpha_1} ) _{SL}$\ to ${\cal H}^m \times {\bf  P}_1( \mathbb{C})$,\  by means of the preceeding actions, the
canonical map between
these homogeneous spaces becomes a map, ${\mbox{\boldmath $ r$}}$,
from $\cal B$ onto
${\cal H}^m \times {\bf P}_1(\mathbb{C})$.\

As a consecuence of \eqref{ro}, we have
$$
(w \ |\,-\varepsilon \overline z)* (mI,\left[\left(  \begin{array}{c}
 1 \\0
\end{array}
  \right)\right])={\mathbf r}\left( 
\begin{array}{c}
 w \\
z
\end{array}
  \right)
	$$
and this leads easily to
\begin{equation}\label{erreDirac}
{{\mbox{\boldmath $ r$}}} \left(
 \!
\begin{array}{c} w\\  z
\end{array}
 \! \right)
=
(m(ww^*\,-\,
\varepsilon \overline {zz^*} \varepsilon ),\, [w]).
\end{equation}

On the other hand, if we define
\begin{equation}\label{sigmakw}
	\sigma(K,w)=\frac{1}{\sqrt{mw^*K^{-1}w}}\left(w\ \vline \ -\frac{1}{m}K \varepsilon \overline w\right), 
\end{equation}
where the vertical bar separates the two columns of the $2\times 2$ matrix, we have
\begin{equation}\label{accionsigma}
\sigma(K,w)*(mI,\left[\left( \begin{array}{c}
 1 \\0
\end{array}  
  \right)\right])=(K,[w]).
\end{equation}

Notice that if $z\ \mathrm{and}\ z'$\ are representatives of the same element of $ P^1(\mathbb{C}),$\ $\sigma\left(K, z \right)$\ and $\sigma\left(K, z' \right)$\ are in general different.

Thus, since
$$
{\mathbf r}^{-1}(mI,\left[\left( \begin{array}{c}
 1 \\0
\end{array}
  \right)\right])=\{ s {\left( 
\begin{array}{c}
 1 \\0\\1\\0
\end{array}
  \right)}:s\in S^1 \}
	$$
	we have
	
	\begin{equation}
{\mathbf r}^{-1}(K,[w])=\rho (\sigma(K,w))	\{ s \left( {
\begin{array}{c}
 1 \\0\\1\\0
\end{array}  }
  \right):s\in S^1 \}	
	\end{equation}
which leads to
\begin{equation}	
{{\mbox{\boldmath $ r$}}}^{-1}(K,\ [a])=\left \{ s \left(
 \!
\begin{array}{c} I\\
m\,K^{-1}
\end{array}
 \! \right)
 \frac {a}{\sqrt{ma^*K^{-1}a}}:\ s \in  {\bf S}^1 \right \}
\end{equation}

Also we have 
$$
P(K,\ [w])\,=P(\sigma(K,w)G_{(\alpha_1)_{SL}})=\,-\eta K.
$$

If $f$ is a continuous function  with
compact support in $\cal B,$ and is $S^1$-homogeneous
of degree -1 (\it i.e. \rm $f\in {\cal C}$\ in the notation of section \ref{rephomog})
, the corresponding prewave function is
$$
\psi _f:(H,\ K,\ [a]) \in   H(2) \times
{\cal H}^m \times   {{\mbox{\boldmath $ P_1$}}}( {\bf C})\
\
\mapsto \ \
f\left(
 \!
\begin{array}{c} w \\  z  \end{array}  \! \right) e^{-i
\pi \eta \ Tr(K\,\varepsilon \overline H \varepsilon)  }\
\left(
 \!
\begin{array}{c} w\\  z  \end{array}  \! \right)
$$
where $\left(
 \!
\begin{array}{c} w\\ z \end{array}  \! \right) $\ is
arbitrary in ${\mbox{\boldmath $r$}}^{-1}(K,\ [a])$.

When one considers the Dirac matrices in the representation
$$
\gamma ^4=\left( \begin{array}{cc}
0&I \\
I&0
\end{array} \right)
\ \ \ ,\gamma ^k=\left( \begin{array}{cc}
0&-\sigma_k  \\
\sigma_k&0
\end{array} \right)
\ \ \ ,k=1,\ 2,\ 3,
$$
a straightforward computation proves that that these prewave functions satisfy Dirac Equation
$$
(\gamma^\nu\ \partial_\nu\ -2 \pi i \eta m)\ \psi _f \ =\ 0.
$$

A sesquilinear form on ${\bf C}^4$ whose value on ${\cal
B}$ is 1, is defined by $$ \Phi\left( Z,\
Z^\prime\right)=\frac{1}{2} Z^* \gamma^4 Z.
$$
Thus, if $f,\ f' \in {\cal C},$\  the hermitian  product  of  $\psi_f$ and $\psi_{f^\prime}$  can be writen
$$
\langle \psi_f,\ \psi_{f^\prime} \rangle
=\frac 1 2 \int_{{\cal H}^m \times
P_1(C)} 
\psi_f^*\,\gamma^4\,\psi_{f^\prime}\,\mu.
$$

To describe Wave Functions we need an invariant volume element on ${\cal H}^m\times  {\bf  P}_1 (\mathbb  C).$\

Let us consider
the 5-form in ${\cal H}^m \times  (  {\mathbb C}^2-\{(0,0)\})$
\rm
,\ given by
$$
(\mu _0)_{(K,\ z)}=
\nu \ \wedge \frac
{(z^1\,dz^2-z^2\,dz^1)\,\wedge \,
 {(\overline{z^1}\,d\overline{z^2}-\overline{z^2}\,d\overline{z^1})}}{(z^* {\mbox{$\varepsilon$}}
 \overline K {\mbox{$\varepsilon$}} z)^2},
$$
where the $z^k$ are the two canonical projections of $
{\mathbb C \rm }^2$
onto ${\mathbb C}$.

This form is well defined since, for all $K\in {\cal H}^m, $ the hermitian product defined in ${\mathbb C}^2$ by
$$
\langle x,y \rangle=x^* \overline K y
$$
is positive definite.

This differential form
 projects to an invariant volume element, $\mu$,\ in
${\cal H}^m \times   {\mbox{\boldmath $ P_1$}}(\mathbb C)$.\rm To prove this fact, one must proceed in many steps. 

Let us denote by $\tau$ the canonical map 
$$
\tau:(K,z)\in {\cal H}^m \times  (  {\mathbb C}^2-\{(0,0)\}) \rightarrow (K,[z])\in {\cal H}^m \times   {\mbox{\boldmath $ P_1$}}(\mathbb C).
$$

The triple ${\cal H}^m \times   (  {\mathbb C}^2-\{(0,0)\})({\cal H}^m \times {\mbox{\boldmath $ P_1$}}(\mathbb C), {\mathbb C}-\{0\})$ \ is a principal fibre bundle with projection $\tau.$ To prove that there exist a form, $\mu,$\ such that 
$$
\tau^*\mu=\mu_0
$$
it is  enought to see that $\mu_0$\ is invariant under the bundle action, what is obvious, and that $\mu_0$\ vanishes on vertical vectors, what can be proved as follows.

The vertical vectors at $(K,z)$\ are tangent at $t=0$ to the curves 
$$
\gamma(t)=(K,\lambda(t)z)
$$
with $\lambda$\ a $C^\infty$\ map from $\mathbb R$\ into $\mathbb C$\ such that
 $\lambda(0)=1.$

Let $X_{(K,z)}$\ be the tangent vector to $\gamma$\ at $0.$\

We have
$$
X_{(K,z)}=\lambda^\prime (0)\left(z^1 \frac{\partial}{\partial z^1}+z^2 \frac{\partial}{\partial z^2}
\right)+\overline {\lambda^\prime (0)}\left(\overline {z^1} \frac{\partial}{\partial \overline {z^1}}+\overline {z^2} \frac{\partial}{\partial \overline {z^2}}
\right),
$$
and it is easily seen that 
$$
i_{X_{(K,z)}}{(\mu_0)_{(K,z)}}=0.
$$

This finishes the proof of the existence of $\mu.$

Let us denote
$$
\Vert p \Vert^2=\sum_{i=1}^3 (p^i)^2 
$$
and
\begin{eqnarray} 
q_1(p_1,p_2,p_3,z)&=&(h(p_1,p_2,p_3,(m^2+ \Vert p \Vert^2 )^{\frac 12} ),\left[   \left(\begin{array}{c}1\\z \end{array}\right) \right]) \label{paramDirac}     \\
q_2(p_1,p_2,p_3,z)&=&(h(p_1,p_2,p_3,(m^2+ \Vert p \Vert^2)^{\frac 12} ),\left[  \left(\begin{array}{c}z\\1 \end{array}\right) \right])  \nonumber
\end{eqnarray}
for all $p_1,p_2,p_3,z\in {\mathbb R}^3 \times \mathbb C.$\ 

The maps $q_1$ and $q_2$ are parametrizations, whose inverses are local charts that compose an atlas of  ${\cal H}^m \times   {\mbox{\boldmath $ P_1$}}(\mathbb C).$

The local expressions of $\mu$ in these charts  can be obtained as the reciprocal image of $\mu_0$ by the maps
$$
\sigma_1(p_1,p_2,p_3,z)=(h(p_1,p_2,p_3,(m^2+  {\mbox{$\sum_{i=1}^3$}}  \ (p^i)^2)^{\frac 12} ),\left(\begin{array}{c}1\\z \end{array}\right) ),
$$
and
$$
\sigma_(p_1,p_2,p_3,z)=(h(p_1,p_2,p_3,(m^2+  {\mbox{$\sum_{i=1}^3$}}  \ (p^i)^2)^{\frac 12} ),\left(\begin{array}{c}z\\1 \end{array}\right) ).
$$

These local expressions  are given by
$$
\sigma_1^*\mu_0=\frac{\nu \wedge dz \wedge d \overline z}{2\Re(z(p_1-ip_2))+(1-\vert z \vert ^2)p_3-(1+\vert z \vert ^2)(m^2+ \Vert p \Vert^2 )^{\frac 12}},
$$
$$
\sigma_2^*\mu_0=\frac{\nu \wedge dz \wedge d \overline z}{2\Re(z(p_1+ip_2))+(\vert z \vert ^2-1)p_3-(1+\vert z \vert ^2)(m^2+ \Vert p \Vert^2 )^{\frac 12}}.
$$

As a particular consequence, $\mu$ is a volume element.

Now, let us consider the action of $SL$\ on ${\cal H}^m \times  (  {\mathbb C}^2-\{(0,0)\})$\ given by
$$
A*(K,z)=(AKA^*,Az).
$$

We already know that $\nu$\ is invariant under the action on ${\cal H}^m ,$ and it is not a difficult matter to see that $z^1\,dz^2-z^2\,dz^1,\ {\overline{z^1}\,d\overline{z^2}-\overline{z^2}\,d\overline{z^1} } $\ and $z^* {\mbox{$\varepsilon$}}
 \overline K {\mbox{$\varepsilon$}} z$\ are invariant under this action.

 As a consecuence, $\mu_0$\ is invariant under the same action. It follows that $\mu$\ is an invariant volume element on  ${\cal H}^m \times   {\mbox{\boldmath $ P_1$}}(\mathbb C).$

The  wave functions have the form
$$
\tilde \psi _f(X)=\int_{{\cal H}^m \times
P_1(C)}
\psi_f(h(X),\cdot,\cdot)\ \mu
$$
and also satisfies Dirac Equation
$$
(\gamma^\nu\ \partial_\nu\ -2 \pi i \eta m)\ \tilde \psi _f \ =\ 0.
$$

\subsubsection{Wave functions for $T>1$.}\label{generconmasa}

Now, let us consider a particle whose movement space is the
coadjoint orbit of
\begin{equation}
\alpha_T=\left
\{ \ \frac  {iT}{8\pi}\left( \begin{array}{cc} 1&0  \\ 0&-1
\end{array} \right) \ ,\  \eta \,m\,I\right \},
 \label{eq:rep5}
\end{equation}
where $
T\in
{\bf Z}^+,\ m \in {\bf R}^+, \eta=\pm 1.
$

The case of the preceeding section is the particular one given by $T=1.$

For all $T$ the orbit is a quantizable, not
$\bf R \rm$-quantizable orbit, of the type 5.

Now  we have
$$
{\mbox{$G_{\alpha_T}$}}     = \{\ (\ \left( \begin{array}{cc}
z&0 \\
0&\overline z
\end{array} \right)
, \ hI ): \   z\in S^1,\ h\in {\mbox{$\bf R \rm$}} \}.$$
The
unique homomorphism from $G_{\alpha_T}$ onto ${\bf S}^1$ whose
differential is $\alpha_T$ is given by
$${\mbox{$C_{\alpha_T}$}}    (\ \left( \begin{array}{cc}
e^{2\pi i\phi}&0 \\
0&e^{-2\pi i\phi}
\end{array} \right) ,\ hI)=e^{2\pi i(\phi T-\eta m h)}.$$
Then
$$
\begin{array}{l}
\vspace{.1in}
{{\mbox{$(G_{\alpha_T }  )   _{SL}$}}=\{\ \left( \begin{array}{cc}
z&0 \\
0&\overline z
\end{array} \right)
:\ z\in S^1\} },     \\
\vspace{.1in}
{{\mbox{$(C_{\alpha_T})_{SL}$}}(\
\left( \begin{array}{cc}
z&0 \\
0&\overline z
\end{array} \right)
)=z^T}, \\
\vspace{.1in}
 SL_1 \cap    SL_2 ={\mbox{$(G_{\alpha_T}  )   _{SL}$}}, \\
\vspace{.1in}
 SL_1 \cap Ker\,(C_{\alpha_T})_{SL}
=Ker\,(C_{\alpha_T})_{SL}=\{\ \left( \begin{array}{cc}
z&0 \\
0&\overline z
\end{array} \right)
:\ z\in \sqrt[T]{\mathbf 1}\}.    \\
\end{array}
$$
where $\sqrt[T]{\mathbf 1}$\ is the subgroup of ${\mathbb C}^*$ composed by the roots of order $T$ of 1.

The homogeneous space
SL/${\mbox{$(G_{\alpha_T}  )   _{SL}$}}$, is the same as in the preceeding section so that it can be identified to ${\cal H}^m\times  {\bf P}_1(\mathbb  C)$, and consider on it the invariant volume element $\mu.$

 In order to give the wave functions in the case of arbitrary ${T}$, the following results are useful.

                   Let    ${\beta},\ \beta^\prime$    be
quantizable  elements   of  $\underline   G^*$  with 
${\beta^\prime}$  not {\bf R}-quantizable, 
 $(G_{\beta})_{SL}
=
(G_{{\beta^\prime}})_{SL}$,\     $(C_{\beta})_{SL} =\left(
(C_{{\beta^\prime}})_{SL}\right)^T,$\ where $T  \in {\bf Z}^+$.  

Notice that, as a consecuence of the fact that $\beta'$ is not ${\mathbb R }$ -quantizable, we must have $(C_{\beta^\prime})_{SL}((G_{{\beta^\prime}})_{SL})=S^1.$ \

 If $(\rho,  L,  z_0)$  is  a  trivialization  of   $C_{\beta^\prime}$,
 we consider the triple
$$(\rho^{\otimes T},\ L^ {\otimes T},\ z_0^ {\otimes T}),$$
 where $L^ {\otimes
T}=L   \otimes   \stackrel{(T}{\cdots}  \nolinebreak
\otimes L,\ z_0^
{\otimes T}=z_0 \otimes \stackrel{(T}{\cdots} \otimes z_0$,
and  $\rho^{\otimes  T}$  is  the  representation such that
$\rho^{\otimes    T}(A)(z_1    \otimes    \cdots    \otimes
z_T)=\rho(A)(z_1)  \otimes  \cdots  \otimes  \rho(A)(z_T).$

In lemma 6.1 of \cite{adm96} I have proved that, under these circunstances,  \emph{if $(G_{\beta})_{SL}$ is connected,
$(\rho^{\otimes T},\ L^ {\otimes T},\ z_0^ {\otimes T})$
 is a trivialization of $C_\beta$.}

           Let us assume that
$(\rho^{\otimes T},\ L^ {\otimes T},\ z_0^ {\otimes T})$
 is a trivialization of $C_\beta$ and  let
${\cal B}_T$ be the orbit of
$ z_0^
{\otimes T}$.  The pullback by $z \in {\cal B} \mapsto
z^
{\otimes T} \in  {\cal B}_T$, establishes  a one to  one
map from the set  of the $S^1$-homogeneous  functions
of  degree  -1 on $ {\cal B}_T$,\   onto  the  set  of the ${S^1}$-homogeneous
functions of degree -T on  ${\cal B}$. If $f$ is one of these functions,
the corresponding prewave function of
particles  corresponding to  $\beta$ has the form
$$ 
{\mbox{$\psi$}}_f(H,\ m)\,=\,f(z)\,e^{i\pi Tr(P(m)\,\varepsilon \overline H \varepsilon)}z^{\otimes T}
$$
where
$z \in {  \bf r}^{-1}(m).$

Now we  apply these results to the particles of type 5 with $T> 1.$ 

Let $\beta'=\alpha_1$\ and $\beta=\alpha_T.$\ The remark just made leads  to the following Prewave Function

$$
\psi _f:(H,\ K,\ [a]) \in   H(2) \times
{\cal H}^m \times   {{\mbox{\boldmath $ P_1$}}}( {\bf C})\
\
\mapsto \ \
f\left(
 \!
 \!
\begin{array}{c} w \\  z  \end{array} \! \! \right) e^{-i
\pi \eta \ Tr(K\,\varepsilon \overline H \varepsilon)  }\
\left(
 \!
 \!
\begin{array}{c} w\\  z  \end{array}  \!\! \right)^{\otimes
T}
$$
where $\left(
 \!
 \!
\begin{array}{c} w\\ z \end{array}  \!\! \right) $\ is
arbitrary in ${\mbox{\boldmath $r$}}^{-1}(K,\ [a])$, and f
is a function on ${\cal B}$, $C^\infty,$ with compact
support and homogeneous of degree -T under multiplication
by complex numbers of modulus one.

The Wave Functions are obtained by integration as usual.

\subsubsection{Movement Space for massive particles with $T\geq 1$.}\label{movT>1}

We consider the action of $G$ on ${\cal H}^m \times {\bf P}_1(\mathbb{C})\times \mathbb{R}^3$\ given by
\begin{equation}
(A,H)*(K,[z],\overrightarrow y)=(AKA^*,[Az],	\overrightarrow x(A,H,\overrightarrow y,K))
\end{equation}
where $\overrightarrow x(A,H,\overrightarrow y,K)$\ is given by
\begin{eqnarray}
	h(\overrightarrow x(A,H,\overrightarrow y,K),0)=Ah(\overrightarrow y,0)A^*+H+\ell(A,H,\overrightarrow y,K) A\frac{K}{m}A^* \\
	\ell(A,H,\overrightarrow y,K)=\frac{-mTr(Ah(\overrightarrow y,0)A^*+H)}{Tr( AKA^*)}.
\end{eqnarray}

This is a transitive action. The isotropy subgroup at $(mI,[{}^t\negthinspace(1,0 
 )], \overrightarrow 0),$\ is found to be $G_{\alpha_T}.$ Thus, the map
	$$
	\lambda:(A,H) G_{\alpha_T}\in G/G_{\alpha_T} \leftrightarrow (A,H)*\left(mI,\left[\left( 
\begin{array}{c}
 1 \\0
\end{array}
  \right)\right], \overrightarrow 0\right)\in {\cal H}^m \times {\bf P}_1(\mathbb{C})\times \mathbb{R}^3
	$$
	enables us to identify the coadjoint orbit with ${\cal H}^m \times {\bf P}_1(\mathbb{C})\times \mathbb{R}^3.$
	
A parametrization whose image is a neighborhood of $(mI,[{}^t\negmedspace(1,0)], \overrightarrow 0),$\ is
\begin{eqnarray}
\Gamma_s:(\overrightarrow k,\overrightarrow x,z)\in \mathbb{R}^3 \times 	 \mathbb{R}^3\times \mathbb{C}\rightarrow && 
\left(mh(\overrightarrow k,k_4),\left[\left( 
\begin{array}{c}
 1 \\z
\end{array}
  \right)\right], \overrightarrow x\right)\\
	&&\in {\cal H}^m \times {\bf P}_1(\mathbb{C})\times \mathbb{R}^3.\nonumber
\end{eqnarray}
where $k_4=\sqrt{1+k_1^2+k_2^2+k_3^2}.$	

The corresponding local chart is given by
$$
\Gamma_s^{-1}:\left(K,\left[\left( 
\begin{array}{c}
 z^1 \\z^2
\end{array}
  \right)\right],\overrightarrow x\right)\in \{z^1\neq 0 \} \rightarrow \left(\frac{1}{m}h^{-1}(K-Tr(K)I),\overrightarrow x,\frac{z^2}{z^1}\right)
$$

Another parametrization is
\begin{eqnarray}
\Gamma_n:(\overrightarrow k,\overrightarrow x,z)\in \mathbb{R}^3 \times 	 \mathbb{R}^3\times \mathbb{C}\rightarrow && 
\left(mh(\overrightarrow k,k_4),\left[\left( 
\begin{array}{c}
 z \\1
\end{array}
  \right)\right], \overrightarrow x\right)\\
	&&\in {\cal H}^m \times {\bf P}_1(\mathbb{C})\times \mathbb{R}^3.\nonumber
\end{eqnarray}
with $k_4=\sqrt{1+k_1^2+k_2^2+k_3^2},$\ whose inverse is given by
\begin{equation}
\Gamma_n^{-1}:\left(K,\left[\left( 
\begin{array}{c}
 z^1 \\z^2
\end{array}
  \right)\right],\overrightarrow x\right)\in \{z^2\neq 0 \} \rightarrow \left(\frac{1}{m}h^{-1}(K-Tr(K)I),\overrightarrow x,\frac{z^1}{z^2}\right)	\nonumber
\end{equation}

Obviously, these two charts compose an atlas.

The identification of ${\cal H}^m \times {\bf P}_1(\mathbb{C})\times \mathbb{R}^3$\ with the coadjoint orbit is given by
$$
(A,H)*((mI,[{}^t\negthinspace(1,0)], \overrightarrow 0)\longleftrightarrow  Ad^*_{(A,H)}\alpha_T.
$$

Now, let  $$\left(K,\left[ z\right],\overrightarrow x\right)\in {\cal H}^m \times {\bf P}_1(\mathbb{C})\times \mathbb{R}^3.
	$$
If $\sigma\left(K, w \right)$\ is given by \eqref{sigmakw}, then
\[
\left(\sigma\left(K, z \right),h(\overrightarrow x,0)\right)*(mI,[{}^t\negthinspace(1,0)], \overrightarrow 0)=\left(K,\left[ z\right],\overrightarrow x\right),
\]
so that $\left(K,\left[ z\right],\overrightarrow x\right)$\ must be identified to $Ad^*_{\left(\sigma\left(K, z \right),h(\overrightarrow x,0)\right)}\cdot \alpha_T.$\ 

Now let $F$\ be a dynamical variable on the coadjoint orbit. $F$\ becomes a function on ${\cal H}^m \times {\bf P}_1(\mathbb{C})\times \mathbb{R}^3$. 

To give explicit expresions of these functions, is now preferable to use the sphere $S^2$\ instead of ${\ bf P}_1(\mathbb{C}) $\  thus writing
$$
F(K,[z],\vec x) =F( Ad^*_{\left(\sigma\left(K, z \right),h(\overrightarrow x,0)\right)}\cdot \alpha_T)=  F(K,\vec u,\vec x),        
$$
where $\vec u\in S^2$ \ is related to $[z]\in \mathbf{P_1(\mathbb{C})}$ by (\emph{c.f.} \eqref{otradefbeta})
$$
\frac{2}{z^*z}zz^*=h(\vec u,1).
$$

Let us  evaluate
$$
Ad^*_{\left(\sigma\left(K, z \right),h(\vec x,0)\right)}\cdot \alpha_T.
$$

To do that computation we need some of the equations \eqref{conmherm}  to \eqref{casicom}, and
\begin{equation}\label{sigma-1}
	(\sigma(K,z))^{-1}=\frac{1}{\sqrt{mz^*K^{-1}z}}   \left( \frac{ mz^*K^{-1}}{ z \varepsilon}\right),
\end{equation}
where the horizontal line separates the two files of the $2\times 2$\ matrix. This leads to
\begin{eqnarray*}
Ad^*_{\left(\sigma\left(K, z \right),h(\vec x,0)\right)}\cdot \alpha_T&=&\left\{ \left[ -\frac{T}{8\pi}\frac{1}{\langle k_4\vec{ u}-\vec{ k},\vec u \rangle }h(\vec{k}\times\vec{u},0)+\frac{\eta m}{2}k_4h(\vec x,0) \right]+ \right. \\        \nonumber
&+&i \left[\frac{T}{8 \pi} \frac{1}{\langle k_4\vec{ u}-\vec{ k},\vec u \rangle }h( k_4\vec{ u}-\vec{ k},0)+  \frac{\eta m}{2} h(\vec{k}\times\vec{x},0)\right], \\  \nonumber  && \left. \eta m h(\vec{k},k_4) \right\},
\end{eqnarray*}
where
$$
h(\vec{k},k_4)=\frac{1}{m}K.
$$

Thus, using \eqref{paralgP}, we obtain 
\begin{equation}\begin{split}\label{momentosdiracabajo}
P(mh(\vec{k},k_4),\vec{u},\vec{x})&=-\eta m h(\vec{k},k_4),\\
\vec{ l}(mh(\vec{k},k_4),\vec{u},\vec{x})&=\frac{T}{4\pi}\frac{k_4\vec{u}-\vec{k}}{k_4-\langle \vec{ k},\overrightarrow u \rangle}+\eta m \vec{k}\times\vec{x},\\
\vec{g}(mh(\vec{k},k_4),\vec{u},\vec{x})&=\frac{T}{4\pi}\frac{\vec{k}\times\vec{u}}{k_4-\langle \vec{ k},\overrightarrow u \rangle}-\eta m k_4 \vec{x}.
\end{split}
\end{equation}

Notice that
$$
k_4-\langle \vec{ k},\overrightarrow u \rangle=\langle  k_4 \overrightarrow u - \vec{ k},\overrightarrow u \rangle.
$$

If we denote $P(K,\vec{u},\vec{x}),\vec{l}(K,\vec{u},\vec{x})$\ and $\vec{g}(K,\vec{u},\vec{x})$\ simply by $P,\ \vec{l}$\ and $\vec{g}$\ respectively when no danger of confusion exist, and $\vec{P}=(P^1,P^2,P^3),$\ we have the following relations between these dynamical variables
\begin{eqnarray}\label{ldiracabajo}
\vec{ l}&=&\frac{T}{4\pi}\frac{P^4\vec{u}-\vec{P}}{P^4-\langle\vec{ P},\vec u \rangle }+ \vec{x}\times\vec{P}\label{l2dirac},\\   \label{gdiracabajo}
\vec{g}&=&\frac{T}{4\pi}\frac{\vec{P}\times\vec{u}}{ P^4-\langle\vec{ P},\vec u \rangle }+P^4\vec{x}\label{g2dirac}
\end{eqnarray}

The value of the Pauli-Lubanski fourvector at $(mh(\vec{k},k_4),\vec u,\vec x)$\  can be calculated using \eqref{P-L3} , \eqref{P-L4},\eqref{momentosdiracabajo},...,\eqref{g2dirac} and is found to be
\begin{equation}   \label{plvecdiracabajo}   \begin{split}
\overrightarrow{W}&=P^4\frac{T}{4\pi}\frac{P^4\vec{u}-\vec{P}}{\langle P^4\vec{u}-\vec{P},\vec u \rangle}=-\eta m k_4\frac{T}{4\pi}     \frac{k_4\vec{u}-\vec{k}}{\langle k_4\vec{u}- \vec{ k},\vec u \rangle }   \\
W^4&=\frac{T}{4\pi}\frac{\langle P^4\vec{u}-\vec{P},\vec P \rangle}{\langle P^4\vec{u}-\vec{P},\vec u \rangle}= -\eta m
\frac{T}{4\pi}     \frac{\langle k_4\vec{u}-\vec{k},\vec k \rangle}{\langle k_4\vec{u}- \vec{ k},\vec u \rangle  },
\end{split}
\end{equation}
so that we can also write
\begin{equation}
\vec{ l}=\frac{1}{P^4}\overrightarrow{W}+ \vec{x}\times\vec{P}	.
\end{equation}

\subsubsection{Contact manifold for Dirac particles}\label{contactdirac}

Now we pay attention to  the contact manifold in the case $T=1$.

We consider the action of $G$ on ${\cal B}\times {\mathbb R}^3$\ given by
$$
(A,H)*(w,z,\overrightarrow y)=\left( e^{2\pi i\eta m \ell}Aw,e^{2\pi i\eta m \ell}(A^*)^{-1}z,\overrightarrow x(A,H,w,z,\overrightarrow y)\right)
$$
where
$$
h(\overrightarrow x(A,H,w,z,\overrightarrow y),0)= Ah(\overrightarrow y,0)A^*+H+\ell   A(ww^*-\varepsilon \overline{zz^*} \varepsilon)A^*.
$$

Obviously, we have
$$
\ell=-\frac{Tr(Ah(\vec x,0)A^*+H)}{Tr( A(ww^*-\varepsilon \overline{zz^*} \varepsilon)A^*)}.
$$

This is a transitive action and the isotropy subgroup at 
$$
\left(\left(\begin{array}{c}1\\0 \end{array}\right),\left(\begin{array}{c}1\\0 \end{array}\right),  \vec 0\right)
 $$
 is $Ker\,C_{\alpha_1},$\ so that we can identify $G/Ker\,C_{\alpha_1}$ \ with ${\cal B}\times {\mathbb R}^3$\ by means of
$$
(A,H)Ker\,C_{\alpha_1} \longleftrightarrow (A,H)* \left(\left(\begin{array}{c}1\\0 \end{array}\right),\left(\begin{array}{c}1\\0 \end{array}\right),  \vec 0\right).
$$

When $w\in {\mathbb C}^2-\{ 0 \},$\ the $z$\ such that $(w,z)\in {\cal B}$\ compose the set
$$
\left\{ \frac{1}{w^*w}(w+y\varepsilon \overline{w}): y\in {\mathbb C}  \right\}.
$$

Thus, the map
\begin{eqnarray}
\Delta_0:(w,y)\in ({\mathbb C}^2-\{ 0 \})\times {\mathbb C} \longrightarrow \left( w , z(w,y)\right)\in {\cal B},
\end{eqnarray}
where
\begin{equation}\label{zwy}
	z(w,y)=\frac{1}{w^*w}(w+y\varepsilon \overline{w}),
\end{equation}
 is a bijection, and in fact a diffeomorphism. Its inverse is given by
$$
(\Delta_0)^{-1}:(w,z)\in  {\cal B} \longrightarrow (w, {}^tw \varepsilon z)\in ({\mathbb C}^2-\{ 0 \})\times {\mathbb C}.
$$

Then, $(\Delta_0)^{-1}$\ is a global complex coordinate system of ${\cal B},$\ and
\begin{equation}
	\Delta:(w,y,\vec x)\in ({\mathbb C}^2-\{ 0 \})\times {\mathbb C}\times{\mathbb R}^3 \longrightarrow \left( w , z(w,y),\vec x\right)\in {\cal B}\times {\mathbb R}^3,
\end{equation}
is a parametrization of the Contact Manifold defined in an  an open  subset of ${\mathbb C}^3  \times {\mathbb R}^3.$

Notice that, since we know a  diffeomorphism, $\phi_0,$\ from ${\cal B}$ onto $SL(2,\mathbb C)$\ (\emph{c.f.}  \eqref{fi0}), we obtain also a complex parametrization of this group by means of
\begin{equation}\label{paramsl}
\phi_0 \circ \Delta_0:(w,y)\in ({\mathbb C}^2-\{ 0 \})\times {\mathbb C} \longrightarrow   \left( w\ \vline\ \frac{1}{w^*w}({\overline y}w-\varepsilon  \overline{w}\right)\in SL(2,\mathbb C).
\end{equation}

The inverse of $\phi_0 \circ \Delta_0$ is the global complex chart of $SL(2,\mathbb C)$ given by
$$
\phi:(w|v)\in SL(2,\mathbb C) \longrightarrow (w, {}^tw \overline v)\in ({\mathbb C}^2-\{ 0 \})\times {\mathbb C}.
$$

The canonical map of $G$ onto $G/Ker\,C_{\alpha_1}$\ becomes
\begin{equation}\label{canondirac}
(A,H)\in G \longleftrightarrow (A,H)* \left(\left(\begin{array}{c}1\\0 \end{array}\right),\left(\begin{array}{c}1\\0 \end{array}\right),  \vec 0\right),
\end{equation}
and the map
\begin{equation}\label{sigmadirac}
	\Sigma:(w,z,\vec x)\in  {\cal B}\times {\mathbb R}^3     \longrightarrow \left( \left( w\ \vline \ -\varepsilon \overline{z}\right),h(\vec x,0)\right)\in G,
\end{equation}
is a section of this canonical map  , as a consecuence of the fact that
$$
\Sigma(w,z,\vec x)*\left(\left(\begin{array}{c}1\\0 \end{array}\right),\left(\begin{array}{c}1\\0 \end{array}\right),  \vec 0 \right)=(w,z,\vec x).
$$

The canonical map from $G/Ker\,C_\alpha$\ onto $G/G_\alpha$\ becomes
\begin{eqnarray*}
\tilde{\mathbf r}&:&(w,z,\vec x)\in  {\cal B}\times {\mathbb R}^3   \longrightarrow  \Sigma(w,z,\vec x)* (mI,[\left(\begin{array}{c}1\\0 \end{array} \right)],\vec 0)=\\&=&(m(ww^*-\varepsilon \overline{zz^*}\varepsilon),[w],\vec x)\in {\cal H}^m\times {\mathbf P_1(\mathbb C)}\times  {\mathbb R}^3,
\end{eqnarray*}
\emph{i.e.} coincides with ${\bf r}\times Id_{{\mathbb R}^3}.$

The Linear Momentum, Angular Momentum and Pauli-Lubanski four\-vector, whose descriptions in the  symplectic manifold are given by \eqref{momentosdiracabajo} and \eqref{plvecdiracabajo}, when composed with $\tilde{\mathbf r},$\ give  functions $P_c,\ \vec l_c,\vec g_c,W_c,$\  on the Contact Manifold, that are the translation  of these dynamical variables to this homogeneous space. 

In order to give explicit expressions for these functions ($c.f.$ \eqref{vardindiraccont}), let us denote
 $$
\tilde {\mathbf r}(w,z,\vec x)=(m\,h(\vec k,k_4),\vec u, \vec x)\in {\cal H}^m \times S^2 \times {\mathbb R}^3
$$
we have
$$
h(\vec u,1)= \frac{2}{w^*w}w\,w^*,
$$
and
$$
h(\vec k,k_4)=ww^*-\varepsilon \overline{zz^*}\varepsilon,
$$
but
\begin{equation}
	-h(\vec u,1)\varepsilon \overline{h(\vec k,k_4)}\varepsilon=h(k_4\vec u-\vec k,k_4-\langle \vec k,\vec u \rangle) + i h(\vec k \times \vec u,0)
\end{equation}
and
\begin{equation}
-h(\vec u,1)\varepsilon \overline{h(\vec k,k_4)}\varepsilon=-	\frac{2}{w^*w}ww^*\varepsilon \overline{(ww^*-\varepsilon \overline{zz^*}\varepsilon)}\varepsilon=\frac{2}{w^*w}w\,z^*,
\end{equation}
 leads to
$$
h(k_4\vec u-\vec k,k_4-\langle \vec k,\vec u \rangle) + i h(\vec k \times \vec u,0)=\frac{2}{w^*w}w\,z^*,
$$
so that
\begin{equation}
\begin{split}
&h(k_4\vec u-\vec k,k_4-\langle \vec k,\vec u \rangle) =\frac{1}{w^*w}(w\,z^*+zw^*),\\
&h(\vec k \times \vec u,0)=i\frac{1}{w^*w}(zw^*-w\,z^*)
\end{split}
\end{equation}
and
\begin{equation}
\begin{split}
&\langle k_4\vec u-\vec k,\vec u \rangle=k_4-\langle \vec k,\vec u \rangle=\frac{1}{w^*w}\\
&h\left(\frac{k_4\vec u-\vec k}{k_4-\langle \vec k,\vec u \rangle},1\right)=w\,z^*+zw^*\\
&h\left(\frac{\vec k \times \vec u}{k_4-\langle \vec k,\vec u \rangle},0\right)=i(zw^*-w\,z^*)
\end{split}
\end{equation}

Thus
\begin{equation}\label{vardindiraccont}
\begin{split}
&P_c(w,z,\vec x)=-\eta m\,(ww^*-\varepsilon \overline{zz^*}\varepsilon)\\
&h(\vec l_c(w,z,\vec x),0)=\frac{T}{4\pi}(wz^*+zw^*-I)+h(\vec x \times \vec P_c,0)\\
&h(\vec g_c(w,z,\vec x),0)=\frac{iT}{4\pi}	(zw^*-wz^*)+P^4 h(\vec x,0)\\
&h(\vec W_c(w,z,\vec x),0)=P^4\frac{T}{4\pi}(wz^*+zw^*-I)\\
&W^4_c(w,z,\vec x)=\frac{-\eta m T}{8\pi}(w^*w-z^*z)
\end{split}
\end{equation}
The formula for $W_c^4$\ can be obtained as follows
\begin{equation}
\begin{split}
&\langle \vec l,\vec P \rangle \circ  \mathbf{\tilde{r}}=\left\langle \frac{T}{4\pi}\frac{k_4\vec u-\vec k}{\langle k_4\vec u-\vec k,\vec u \rangle},\vec P \right\rangle   \circ  \mathbf{\tilde{r}}= \\&=
\frac{1}{2}Tr\left(h\left(\frac{T}{4\pi}\frac{k_4\vec u-\vec k}{\langle k_4\vec u-\vec k,\vec u \rangle},0\right)P  \right)   \circ  \mathbf{\tilde{r}}=\\ &=\frac{1}{2}Tr\left(\left(\frac{T}{4\pi}(wz^*+zw^*-I)\right)(-\eta m)(ww^*-\varepsilon \overline{zz^*}\varepsilon)\right)=\\ &=\frac{-\eta mT}{8\pi}(w^*w-z^*z).
\end{split}
\end{equation}

The local expresions of these functions in the chart $\Delta^{-1}$ are obtained simply by changing in the above expresions $z$ by the $z(w,y)$\ given in \eqref{zwy}.

Let $\Omega_1$\ be the contact form and $\tilde \alpha_1$\ the left invariant differential form on $G$ whose  value at $(I,0)$\ is $\alpha_1.$\ We have
$$
\Omega_1=\Sigma^* \tilde \alpha_1,
$$
and this formula enables us, by similar procedures to those of section \ref{contKG}, to see that
\begin{equation}\label{Omegadirac}
	\Omega_1=\frac{iT}{4\pi}\left(z^1 d\overline {w^1}-\overline{ z^1} d w^1+z^2 d\overline{ w^2}-\overline{ z^2} d w^2\right)-P^1dx^1-P^2 dx^2-P^3 d x^3,
\end{equation}
where (\emph{c.f.} \eqref{zwy})
$$
\left(\begin{array}{c}
z^1\\
z^2
\end{array}\right)=z(w,y).
$$

This equation must be  interpreted as follows: the right hand side of \eqref{Omegadirac} is a differential 1-form in $\mathbb{C}^4 \times \mathbb{R}^3$\ refered to coordinates $w^1,\dots,x^3,$ and $\Omega_1$\ is the restriction of this form to the submanifold ${\cal B}\times \mathbb{R}^3.$\ 

Obviously, under the same conditions
\begin{equation}\label{dOmega}
\begin{split}
d\Omega_1=\frac{iT}{4\pi}&\left(dz^1\wedge d\overline {w^1}-d\overline{ z^1}\wedge d w^1+dz^2 \wedge d\overline{ w^2}-d\overline{ z^2} \wedge d w^2\right)\\&-dP^1\wedge  dx^1-dP^2 \wedge  dx^2-dP^3  \wedge d x^3.	
\end{split}
\end{equation}

Now, to obtain the symplectic form we only need  sections of the map $\tilde{\mathbf r}.$\ 

On the domain, $U_s,$\  of the local chart $\Gamma_s^{-1},$\ the map 
\begin{equation}
{\tilde \sigma}_s: \Gamma_s(\vec k,\vec x,t) \rightarrow \left( \sigma \left(mh(\vec k,k_4),\left( \begin{array}{c} 1\\ t\end{array}\right)\right),h(\vec x,0)\right)*\left(\left(\begin{array}{c}1\\0 \end{array}\right),\left(\begin{array}{c}1\\0 \end{array}\right), \vec 0 \right),	
\end{equation}
where $\sigma(K,w)$\ is given by \eqref{sigmakw},  is a section. Then

$$
{\tilde \sigma}_s \circ \Gamma_s(\vec k,\vec x,t) =  
\left(
\frac{1}{\sqrt{(1,\overline t)(h(\vec k,k_4))^{-1}\left(\begin{array}{c}1\\t \end{array}\right)}}  \left(\begin{array}{c}1\\t \end{array}\right),
\right.
$$
$$
\left.  \frac{1}{\sqrt{(1,\overline t)(h(\vec k,k_4))^{-1}\left(\begin{array}{c}1\\t \end{array}\right)}}(h(\vec k,k_4))^{-1}\left(\begin{array}{c}1\\t \end{array}\right), \vec x \right).	
$$

If $\omega_1$\ is the symplectic form, we have on $U_s$\ 
\begin{equation}\label{forsym s}	
\omega_1={\tilde \sigma}_s ^* d\Omega_1.
\end{equation}

In the same way, on the domain, $U_n,$\ of the chart $\Gamma_n,$\ we define a section, $\tilde \sigma_n,$\  by
\begin{equation}
{\tilde \sigma}_n \circ \Gamma_n(\vec k,\vec x,t)
=\left( \sigma \left(mh(\vec k,k_4),\left( 
\begin{array}{c} t\\ 1\end{array}\right)\right),h(\vec x,0)\right)*\left(\left(\begin{array}{c}1\\0 \end{array}\right),\left(\begin{array}{c}1\\0 \end{array}\right), \vec 0 \right).
\end{equation}
 Then
$$
{\tilde \sigma}_n \circ \Gamma_n(\vec k,\vec x,t) =  
\left(
\frac{1}{\sqrt{(\overline t,1)(h(\vec k,k_4))^{-1}\left(\begin{array}{c}t\\1 \end{array}\right)}}            \left(\begin{array}{c}t\\1 \end{array}\right),        
\right.  
$$$$
 \left.  \frac{1}{\sqrt{(\overline t,1)(h(\vec k,k_4))^{-1}\left(\begin{array}{c}t\\1 \end{array}\right)}}(h(\vec k,k_4))^{-1}\left(\begin{array}{c}t\\1 \end{array}\right), \vec x \right).	
$$

 On $U_n$\ we have
\begin{equation}\label{forsym n}
\omega_1={\tilde \sigma}_n ^* d\Omega_1.
\end{equation}

Since $\{U_s,U_n\}$\ is an open covering of the symplectic manifold, \eqref{forsym s} and \eqref{forsym n}, determine $\omega $\ everywhere.

\subsubsection{Contact manifold for $T > 1.$}

Let us denote by $r_1,\dots, r_T$ the elements of $\sqrt[T]{\mathbf 1}.$\

We define a properly discontinuous free action, $\cdot,$ of $\sqrt[T]{\mathbf 1}$\ on 
${\cal B}\times  {\mathbb R}^3 $  by means of
$$
r_j\cdot (w,z,\overrightarrow x)=(r_j w,r_j z,\overrightarrow x).
$$

The quotient space is denoted by $({\cal B}\times {\mathbb R}^3)/\sqrt[T]{\mathbf 1},$  and the canonical map from ${\cal B}\times {\mathbb R}^3$ onto $({\cal B}\times {\mathbb R}^3)/\sqrt[T]{\mathbf 1},$ is denoted by $\pi_T.$ Thus $\pi_T(w,z,\overrightarrow x)$ is the orbit of $(w,z,\overrightarrow x)$, considered as an element of $({\cal B}\times {\mathbb R}^3)/\sqrt[T]{\mathbf 1}.$ 

The map $\pi_T$ is a T-fold covering map and, since ${\cal B}\times {\mathbb R}^3$ is simply connected, the Universal covering map of the quotient space.

The action of $G$ on ${\cal B}\times {\mathbb R}^3$ conmutes with that of $\sqrt[T]{\mathbf 1}.$ As a consequence, the action on  $({\cal B}\times {\mathbb R}^3)/\sqrt[T]{\mathbf 1}$ given by
$$
(A,H)*((w,z,\overrightarrow x)\sqrt[T]{\mathbf 1})=((A,H)*(w,z,\overrightarrow x))\sqrt[T]{\mathbf 1}
$$
is well defined, and obviously makes the covering map $\pi_T$ equivariant.

The isotropy subgroup at 
$$
\left(\left(\begin{array}{c}1\\0 \end{array}\right),\left(\begin{array}{c}1\\0 \end{array}\right),  \vec 0\right)\sqrt[T]{\mathbf 1}
 $$
 is $Ker\,C_{\alpha_T},$\ so that we can identify $G/Ker\,C_{\alpha_T}$ \ with $({\cal B}\times {\mathbb R}^3)/\sqrt[T]{\mathbf 1}$\ by means of
$$
(A,H)Ker\,C_{\alpha_T} \longleftrightarrow (A,H)* \left(\left(\left(\begin{array}{c}1\\0 \end{array}\right),\left(\begin{array}{c}1\\0 \end{array}\right),  \vec 0\right)\sqrt[T]{\mathbf 1}\right).
$$

The contact form on ${\cal B}\times {\mathbb R}^3,\ \Omega_1,$ is invariant under the action $``\cdot",$  so that there exist an unique contact form, $\Omega_T,$ on $({\cal B}\times {\mathbb R}^3)/\sqrt[T]{\mathbf 1},$ such that
$$
\pi_T^*\Omega_T=\Omega_1.
$$

The action of $G$ on $({\cal B}\times {\mathbb R}^3)/\sqrt[T]{\mathbf 1}$  is transitive and preserves $\Omega_T.$

The manifold $({\cal B}\times {\mathbb R}^3)/\sqrt[T]{\mathbf 1},$ when provided  with the contact form $\Omega_T$ and the cited action of $G,$ is the \bf homogeneous contact manifold that correspond to massive particles with \rm $T\geq 1.$

\section{Massless  Type 4 particles. }\label{waveNula}

\rmdefault

In this section we consider particles whose movement
space is the coadjoint orbit of
$$
\alpha =\left\{ \frac{i \chi T}{8 \pi}
\left (\begin{array}{cc} 1 & 0\\
0 & -1
\end{array}
\right )
,
 \eta\,
\left (\begin{array}{cc} 1 & 0\\
0 & 0
\end{array}
\right )
\right\}\ \ \ \ ,\ \chi,\ \eta \in \{\pm 1\},\ T \in
{\bf Z}^+ ,
$$
where $\eta=-sign(Tr(P)),$\ (see Table 2 of section \ref{concreteqforms} ). So, at least at the classical level, $\eta$ must be interpreted as the opposite of the sign of energy. We will see in the following subsections  that this interpretation
 is also exact at the quantum level.

Since $|P|$ is mass square and 
$$
Det\left( \eta\,
\left (\begin{array}{cc} 1 & 0\\
0 & 0
\end{array}
\right ) \right)=0
$$
these orbits correspond to particles with zero mass.

These are  quantizable, not $\bf
R\rm$-quantizable orbits of the type 4.

The group  $G_\alpha$ is connected, so that there exists at most
 one homomorphism from $G_\alpha$ onto ${\bf S}^1$ whose differential
is $\alpha$. In fact we have

$$
\begin{array}{l}
\vspace{.1in}
G_\alpha = \left\{ \left(
\left (\begin{array}{cc}
z&a  \\
0&{\overline z}
\end{array}
\right )
,\ \left (\begin{array}{cc}
b&i \chi \eta T a z/2 \pi   \\
\overline {i \chi \eta T a z / 2 \pi}&0
\end{array}
\right )
\right )
:\ z \in   {\bf S^1},\right.  \\
 \left. \ \ \ \ \ \ \ \ \ b\in  {\bf R},\ a \in
{\bf C}
\right\} \\
\      \\
\mathrm{and\ the\ homomorphism}          \\
\     \\
C_\alpha \left( \left( \left (\begin{array}{cc}
z&a \\
0&{\overline z}
\end{array}
\right )
,\ \left (\begin{array}{cc}
b&{i \chi \eta T a z/2 \pi} \\
{\overline {i \chi \eta T a z/2 \pi}}&0
\end{array}
\right )
\right )
\right )
=z^{\chi T}
\end{array}
$$
has differential $\alpha.$

Other  computations   give us
$$
\begin{array}{l}
({\mbox{$G_\alpha$}} )_{SL}=\left\{ \left(
\begin{array}{cc} z&a \\
0&{\overline z}
\end{array}
\right )
: z \in {\bf S^1},\  a \in {\bf C} \right\}    \\
\vspace{.1in}
(C_\alpha)_{SL}\left(\
\left (\begin{array}{cc}
z&a \\
0&{\overline z}
\end{array}
\right )
\right )=\ z^{\chi T}  \\
\vspace{.1in}
SL_1=
\left\{
\left (\begin{array}{cc}
a&0  \\
0&{1/a}
\end{array}
\right ):\ a \in {\bf C}
\right \}        \\
\vspace{.1in}
SL_2=(G_\alpha)_{SL}          \\
\vspace{.1in}
SL_1 \cap SL_2=
\left\{
\left (\begin{array}{cc}
z&0  \\
0&\overline z
\end{array}
\right ):\ z \in   {\bf S}^1
\right \}
\end{array}.
$$

The isotropy subgroup at \tiny $\left (\begin{array}{cc}
1&0  \\
0&0

\end{array}
\right )
$,\ \normalsize for the usual action of   {\mbox{$SL$}}(2,{\mbox{$
C$}}) \  on H(2) (that is $A\ast m=AmA^*$)\ is ${\mbox{$(G_\alpha)$}}_{SL}$.\
Thus SL/${\mbox{$(G_\alpha)$}}_{SL}$ will be identified
to the orbit of \tiny
$\left (\begin{array}{cc}
1&0  \\
0&0
\end{array}
\right ),
$\ \normalsize
which is  $${\mbox{$\bf C^+$}}=\{H \in   {H(2)}:\
Det\,H=0,\ Tr\,H>0\}.$$

 This set is composed by the hermitian matrices that represent space-time points in  the future lightcone, so that it will be called itself, future lightcone.

When this identification is made, the function $P$ on
SL/${\mbox{$(G_\alpha)$}}_{SL}$ becomes $P(H)=-\eta\,H$.

An invariant volume element in ${\mbox{$\bf C^+$}}$ is
\begin{equation}\label{omegaC+}
\omega=\frac {1}{\Vert \vec{ p}\Vert }   \ 
dp^1\wedge\,dp^2\wedge\,dp^3,
\end{equation}
where $(p^1,\ p^2,\ p^3)$
is the coordinate system corresponding to the parametrization
$$
\phi :(p^1,\ p^2,\ p^3) \in {\bf R^3}-\{\,0\,\} \mapsto
h(\vec{ p},\Vert{\vec{ p}\Vert}) \in
{\mbox{$\bf C^+$}}
$$
where $\vec{p}= (p^1,\ p^2\ p^3),$\ and 
$$
\Vert{\vec{ p}}\Vert=+ \sqrt {  \sum_{i=1}^3  (p^i)^2}.
$$

In this parametrization, the linear momentum is given by 
$$
P(\phi(\vec{ p}))=-\eta\ h(\vec{ p},\Vert{\vec{ p}\Vert}).
$$

Before to proceed to the study of the general case, we
shall consider two particular ones.

\subsection{Massless antineutrino}\label{antineutrino}

Let us consider the case  {$ {T=1,\ \chi=1,\
\eta=\,-\,1}$.}

A trivialization is given by  $$\left(\rho_+ ,\ {\bf C}^2,\ \left(\begin{array}{c}
1 \\
0
\end{array}
\right) \right),$$
\ where $\rho_+$(A) is multiplication by A.

The orbit of  $$\left(
\begin{array}{c} 1\\
0
\end{array}
\right)$$
 is ${\bf C}^2-\{0\}$. This space can thus
be identified to SL/${\mbox{$Ker\,(C_\alpha)$}}_{SL}$
so that the canonical map $$
  SL/{\mbox{$Ker\,(C_\alpha)$}}_{SL} \longrightarrow SL/(G_\alpha)  $$ becomes  a  map $$r_+ :
{\bf C}^2-\{0\} \mapsto  {\bf C}^+ .$$
This map must be equivariant, so that, for all $A\in SL$ \[
r_+\left(A \left(\begin{array}{c}
1 \\
0
\end{array} \right) \right) =A\left (\begin{array}{cc}
1&0  \\
0&0

\end{array}
\right ) A^*.
\]
Thus $r_+$ is explicitly given by\[
r_+(z)=z\,z^* \in
{\mbox{$\bf C^+$}} .
\]

 Since $z$ and $\varepsilon \overline z$ are
eigenvectors of $z\,z^*$
corresponding to the eigenvalues $\Vert z\Vert ^2$ and 0 respectively,
$r_+^{-1}(H)$\ is composed by the eigenvectors of $H$ corresponding to
the positive eigenvalue, whose norm is the square root of that eigenvalue.

The principal $  {\bf S}^1$-bundle whose projection is $r_+$ is
related to the
{\bf Hopf fibration} as follows. The image of the restriction of $r_+$ to the
sphere
$S^3(R)=\{ z \in { {\bf C}}^2\,:\,\Vert z\Vert ^2=R^2\}$ is composed
by the elements of ${\mbox{$\bf C^+$}}$ whose trace is $R^2$.\ Since the image of this
subset
by the preceeding chart is the sphere of radius $R^2$/2, we obtain
 maps from spheres $S^3$\ onto spheres $S^2$.\ Each one of these
mappings is, up to the radius and a reflection, the Hopf fibration.

For each $S$ -homogeneous of degree -1 function on ${\bf C}^2-{0},\ f, $\ we have the following prewave function 
\begin{equation}\label{pwa1}
{\psi} _f^+:\,(N,\ H) \in {\mbox{$\bf C^+$}}\times  {H(2)}
\mapsto f(z)\
e^{i\pi \,Tr(\,N \varepsilon
 \overline H \varepsilon) } z \in { {\bf C}}^2,
\end{equation}
where z is an arbitrary element of $r_+^{-1}(N)$. Here we use the fact that, since $\eta=-1$,\ the linear momentum is given in $C^+$ by $P(N)=N.$

If $X$ and $Q$ are the elements of ${\bf R}^4$ corresponding to $H$ and $N$,\ one can write
\begin{equation}\label{pwa2}
{\psi} _f^+(Q,X) = f(z)\
e^{-2\pi i\langle Q,X \rangle } z ,
\end{equation}
where $z$ is arbitrary in $r_+^{-1}(Q).$
The corresponding wave function , if $f$ is continuous with compact support, is
\[
{\widetilde {\psi}} _f^+(H)=               \int_{C^+}    {\psi} _f^+(\cdot,\ H) \omega .
\]

By direct computation one can see that
$$
\left (\ \sigma_1 \ \frac {\partial}{\partial x^1}+
\ \sigma_2 \ \frac {\partial}{\partial x^2}+
\ \sigma_3 \ \frac {\partial}{\partial x^3}+
\ \sigma_4 \ \frac {\partial}{\partial x^4} \right )\
{\psi}_f^+(N,\ h(x))\,=\,0
$$

The corresponding {wave function} thus satisfy the same equation,
which is the  Weyl equation (positive energy), usually admised as corresponding to the antineutrino.

If $f$ and $f^\prime$ are pseudotensorial functions, we
have
$$({\psi} _f^+(N,H))^*\ {\psi}
_{f^\prime}^+(N,H)=\overline{f(z)}\, f^\prime(z)\  Tr\,N.
$$ 
Thus, the hermitian product of the
corresponding quantum states ($cf.$ section \ref{wave}) can be
written as follows
$$
{\frac 12}\
\int_{\bf C ^+} \frac{({\psi}^+ _f)^*\ {\psi}^+_{f^\prime}}{
\sum_{i=1}^3 (p^i)^2}\ dp^1\,dp^2\,dp^3.
$$

To obtain prewave functions directly from functions on $C^+,$ \ we  use Remark \ref{bajar}, as follows.

Let $$ 
U= \phi({\bf R}^3-\{p^1=p^2=0, p^3<0\}),
$$$$
 V= \phi({\bf R}^3-\{p^1=p^2=0, p^3>0\}).
$$   
 Then the maps

\begin{equation}
\sigma_U : \phi(p^1,p^2,p^3)\in U \mapsto \left({\begin{array}{c} \sqrt{\Vert\vec{ p}\Vert+p^3}  \\ 
\frac{p^1+i p^2}{ \sqrt{\Vert\vec{ p}\Vert+p^3}}  \end{array}}\right) \in {\bf C}^2-{0} 
\end{equation}
\[
\sigma_V : \phi(p^1,p^2,p^3)\in V \mapsto  \left( {\begin{array}{c} \frac{p^1-i p^2}{ \sqrt{\Vert\vec{ p}\Vert-p^3}} \\  
 \sqrt{\Vert\vec{ p}\Vert-p^3}  \end{array}}\right) \in {\bf C}^2-{0}
\]
are sections of $r_+,$\ so that we obtain prewave functions having the form
\[
\psi(\phi(p^1,p^2,p^3), X)=F(p^1,p^2,p^3) \left({\begin{array}{c} \sqrt{\Vert\vec{ p}\Vert+p^3}  \\ 
\frac{p^1+i p^2}{ \sqrt{\Vert\vec{ p}\Vert+p^3}}  \end{array}}\right)
\]
\[   \displaystyle
e^{-2\pi i (\Vert\vec{ p}\Vert X^4-p^1X^1-p^2X^2-p^3X^3)}
\]
where $F$ is a complex valued function continuous on ${\bf R}^3-\{ 0\},$\  with compact support in ${\bf R}^3-\{p^1=p^2=0, p^3<0\},$\ and prewave functions having the form
\[
\theta(\phi(p^1,p^2,p^3), X)=J(p^1,p^2,p^3) \left( {\begin{array}{c} \frac{p^1-i p^2}{ \sqrt{\Vert\vec{ p}\Vert-p^3}} \\  
 \sqrt{\Vert\vec{ p}\Vert-p^3}  \end{array}}\right)
\]
\[  \displaystyle
e^{-2\pi i (\Vert\vec{ p}\Vert X^4-p^1X^1-p^2X^2-p^3X^3)}
\]
where $J$ is a complex valued function continuous on ${\bf R}^3-\{ 0\},$\  with compact support in ${\bf R}^3-\{p^1=p^2=0, p^3>0\}.$

If one is interested in the case {$T=1,\ \chi =1,\ \eta =+1$,} everything is as in the case $\eta=-1$, but the exponent of $e$ in the prewave functions changes its sign, thus giving wave functions for quantum states of negative energy.   Remenber that $\eta$ is the opposite of the sign of energy at the clasical level.

\subsection{Massless neutrino}

Now let us consider the case in which {
$T=1,\ \chi =-1,\ \eta =-1$.}

A  trivialisation in this case is  $$\left(\rho_- ,\ {\bf C}^2,\ \left(\begin{array}{c}
0 \\
1
\end{array}
\right) \right),$$
$ \rho_- $(A) being  multiplication by $(A^*)^{-1}.$

Thus we identify $SL/Ker\,C_{\alpha}$\ with the orbit of $$ \left(\begin{array}{c}
0 \\
1
\end{array}
\right),$$ which, also in this case, is ${\bf C}^2-\{0\}.$

Thus,the canonical map $$
  SL/{\mbox{$Ker\,(C_\alpha)$}}_{SL} \longrightarrow SL/(G_\alpha)  $$
	
	becomes  a  map
	
	$$r_- :
{\bf C}^2-\{0\} \mapsto  {\bf C}^+ .$$

Now, using equivariance  as in the case of antineutrino, we obtain $$r_-(z)\,=\,-\varepsilon
 \overline {z\,z^* }\varepsilon  .  $$

The prewave functions one obtains have exactly the same form that (\ref{pwa1}) and   (\ref{pwa2}), but now $z$ is in $(r_-)^{-1}(N)$
 or $(r_-)^{-1}(h(Q))$\ respectively.

The wave functions one obtains in this case satisfies the Weyl equation
that, according to Feynman, corresponds  to the {\bf neutrino}.

The map from the real vector space $\bf C \rm^2$ onto
itself defined by sending $z$ to $ \varepsilon \overline
z$, is a complex structure and its restriction to $\bf C
\rm^2-\{ 0 \}$, gives us an isomorphism of the principal
circle bundle corresponding to ${\bf r}_{-}$( resp.${\bf
r}_+$) onto
the principal
circle bundle corresponding to ${\bf r}_+$ ( resp. ${\bf
r}_-$). The isomorphism of the structural
group is defined by sending each element to its inverse.
Thus if $z\in r_+^{-1}(H)$\ then $\varepsilon \overline
z\in r_-^{-1}(H)$\ and conversely.

As a consequence, the sections of the map $r_+$\ defined in \ref{antineutrino} give rise to the following sections of $r_-$\ 
\[
\sigma^\prime _U(u)=\varepsilon \overline{\sigma_U(u)},\ \   \sigma^\prime _V(v)=\varepsilon \overline{\sigma_V(v)},
\]
for all $u\in U$\ and $v\in V,$\ and the prewave functions of antineutrino $\psi$\ and $\theta$\  give rise to prewave functions of neutrino, $\psi^\prime$\ and $\theta^\prime$,\  by  means of
\[
\psi^\prime(\phi(p^1,p^2,p^3), X)=F(p^1,p^2,p^3)\varepsilon \overline{\left({\begin{array}{c} \sqrt{\Vert\vec{ p}\Vert+p^3}  \\ 
\frac{p^1+i p^2}{ \sqrt{\Vert\vec{ p}\Vert+p^3}}  \end{array}}\right) }
\]
\[  
e^{-2\pi i (\Vert\vec{ p}\Vert X^4-p^1X^1-p^2X^2-p^3X^3)}
\]
\[
\theta^\prime(\phi(p^1,p^2,p^3), X)=J(p^1,p^2,p^3) \varepsilon \overline{\left( {\begin{array}{c} \frac{p^1-i p^2}{ \sqrt{\Vert\vec{ p}\Vert-p^3}} \\  
 \sqrt{\Vert\vec{ p}\Vert-p^3}  \end{array}}\right) }
\]
\[  
e^{-2\pi i (\Vert\vec{ p}\Vert X^4-p^1X^1-p^2X^2-p^3X^3)}
\]
that leads to
\begin{eqnarray*}
  \psi'(\phi(p^1,p^2,p^3), X)&=& F(p^1,p^2,p^3){\left({\begin{gathered}[c]  \frac{p^1-i p^2}{ \sqrt{\Vert\vec{ p}\Vert+p^3}}\\-\sqrt{\Vert\vec{ p}\Vert+p^3}    \end{gathered}}\right) }\\ &&
{e^{-2\pi i (\Vert\vec{ p}\Vert X^4-p^1X^1-p^2X^2-p^3X^3)}} \\  
\    \theta'(\phi(p^1,p^2,p^3), X)&=&J(p^1,p^2,p^3){\left( {\begin{gathered}  \sqrt{\Vert\vec{ p}\Vert-p^3}\\
{-\frac{p^1+i p^2}{ \sqrt{\Vert\vec{ p}\Vert-p^3}}}  \end{gathered}}\right)}\\ && {
e^{-2\pi i (\Vert\vec{ p}\Vert X^4-p^1X^1-p^2X^2-p^3X^3)}}.
\end{eqnarray*}

If one consider the case   {$T=1,\ \chi =-1,\ \eta =+1$} one obtain similar wave functions but corresponding to negative energy.

\subsection{General case}\label{casogeneral}

\parindent=1cm
\parskip=5mm

 We proceed as in section \ref{generconmasa}. 

In the case 
$\chi=+1$\  a trivialization is given by 
$$
\left((\rho_+)^{\otimes T},\ \left({\bf C}^2\right)^ {\otimes T},\ 
 \left(
\begin{array}{c} 1\\
0
\end{array}
\right)^{\otimes T}\right),
$$
and if $\chi=-1$\  a trivialization is given by 
$$
\left((\rho_-)^{\otimes T},\ \left({\bf C}^2\right)^{\otimes T},\  
 \left(
\begin{array}{c} 0\\
1
\end{array}
\right)^{\otimes T}\right),
$$

The   prewave   functions   are   given   by
functions
on  ${\bf C} \rm  ^2-
\{0\}$, which are $C^\infty$ with compact support and
 homogeneous of degree  -T under multiplication by modulus
one complex numbers.

 Let $f_T$ be one of these functions. 

If $\chi =1$, the corresponding
prewave function is given by
$$
\psi^+_{f_T}:\,(N,\ H) \in {\mbox{$\bf C^+$}}\times  {H(2)}
\mapsto f_T(z)\
e^{-i \pi \eta \,Tr(N \varepsilon
 \overline H \varepsilon) } z
^{\otimes T}
  \in ({\bf C}^2)^{\otimes T}
$$
where $z$ is an arbitrary element of $r_+^{-1}(N)$.

In the case $\chi =-1$,
 the
corresponding prewave function is
$$
\psi^-_{f_T}:\,(N,\ H) \in {\mbox{$\bf C^+$}}\times  {H(2)}
\mapsto f_T(z)\
e^{-i \pi \eta \,Tr(N \varepsilon
 \overline H \varepsilon )} z
^{\otimes T}
  \in ({\bf C}^2)^{\otimes T}
$$
but now, z is an arbitrary element of $r_-^{-1}(N)$.

The associated wave functions, 
\begin{eqnarray}           
{\widetilde\psi}^+_{f_T}(H)&=&\int_{C^+}  {\psi} ^+_{f_T}(\cdot,\ H) \omega ,\\
{\widetilde\psi}^-_{f_T}(H)&=&\int_{C^+}  {\psi} ^-_{f_T}(\cdot,\ H) \omega 
\end{eqnarray}
 satisfies \bf Penrose's wave equations,\rm  that we will describe here for the sake of
completeness.

Let us consider in $({\bf C}^2)^{\otimes T}$ the basis
$\{ e_A \otimes e_B \otimes \stackrel{(T}{\cdots}\ :\ A,\
B,\dots \in \{1,\ 2\}\}$, where $\{e_1,\ e_2\}$ is the
canonical basis of $\bf C \rm ^2$.

The prewave functions $\psi^\pm_{f_T}$ and the wave
functions $\tilde {\psi}^\pm_{f_T}$, have components in
this basis which will be denoted by $\{\psi_\pm^{A\,B
\dots} \}$ and
$\{\tilde{\psi}_\pm^{A\,B
\dots} \}$, respectively.

Let us consider the vector fields in $\bf R \rm ^4$ given
by
\begin{eqnarray*}
\nabla_{11}&=&\frac12 \left( \frac{\partial}{\partial x^3}+
\frac{\partial}{\partial x^4} \right)\\
\nabla_{12}&=&\frac12 \left( \frac{\partial}{\partial x^1}
-i\frac{\partial}{\partial x^2} \right)\\
\nabla_{21}&=&\frac12 \left( \frac{\partial}{\partial x^1}
+i\frac{\partial}{\partial x^2} \right)\\
\nabla_{22}&=&\frac12 \left( \frac{\partial}{\partial x^4}
-\frac{\partial}{\partial x^3} \right)
\end{eqnarray*}
and, for all $A, A^\prime \in \{1,\ 2\},\ \nabla^{A \
A^\prime}=\varepsilon ^{A\ B}\ \varepsilon ^{A^\prime\ B
^\prime
}\ \nabla_{B\ B^\prime}
$  (summation convention), where $\{ \varepsilon^{A\ B}\}$
are the elements of $-\varepsilon$.

We also define
\begin{eqnarray*}
\psi^\pm_{A\,B\dots}&=&
\varepsilon _{A\ A^\prime}\ \varepsilon _{B\ B^\prime}\dots
\psi_\pm^{A^\prime\,B^\prime\dots}  \\
\tilde\psi^\pm_{A\,B\dots}&=&
\varepsilon _{A\ A^\prime}\ \varepsilon _{B\ B^\prime}\dots
\tilde\psi_\pm^{A^\prime\,B^\prime\dots}
\end{eqnarray*}
where $\{ \varepsilon_{A\ B}\}$
are the elements of $\varepsilon$.

Thus we have for all $h(x) \in C^+$
\begin{eqnarray*}
\nabla^{A\ A^\prime}\ \psi^+_{A^\prime\,B\ C\dots}(h(x),\,
\cdot\, )&=&0\\
\nabla^{ A^\prime\ A}\ \psi^-_{A^\prime\,B\ C\dots}(h(x),\,
\cdot\,)&=&0
\end{eqnarray*}
so that, by derivation under the  integral sign, we
see that Penrose wave equations:
\begin{eqnarray*}
\nabla^{A\ A^\prime}\ \tilde{\psi}^+_{A^\prime\,B\
C\dots}&=&0\\
\nabla^{ A^\prime\ A}\
\tilde{
\psi}^-_{A^\prime\,B\ C\dots}&=&0
\end{eqnarray*}
are satisfied.

\subsubsection{Helicity}\label{helicidadgen}

In formula \eqref{operclasic} one sees that, if $(\rho,L,z_0)$\ is the  trivialization we use, the components of \bf spin operator \rm are given by
$$
\hat {s^k}:\widetilde{\psi}_f \longrightarrow   s^k \circ  \widetilde{\psi}_f,
$$
$s^k$\ being the following endomophism of $L$\ 
$$
 s^k=\frac 1{2\pi i}
  \frac{\widetilde{i\sigma_k}}{2}=\frac 1{4\pi i}\widetilde{i\sigma_k},
$$
where, 
$$
\widetilde{i\sigma_k}=d\rho({i\sigma_k}).
$$

Thus, in the case $\chi=1,$\ 
$$
\widetilde{i\sigma_k}=d\left((\rho_+)^{\otimes T}\right)({i\sigma_k}),
$$
and in the case $\chi=-1,$
$$
\widetilde{i\sigma_k}=d\left((\rho_-)^{\otimes T}\right)({i\sigma_k}).
$$

As a consecuence, in both cases, $\chi=+ 1$\ or $\chi= 1,$\  when acting on monomials we have
$$
\widetilde{i\sigma_k}\cdot z_1\otimes\dots\otimes z_T=\sum_{j=1}^T\ z_1\otimes \dots\otimes z_{j-1}\otimes i\, \sigma_k\,z_j\otimes z_{j+1}\otimes\dots\otimes z_T.
$$

We  consider the operators $\hat {s^k}$\ as also acting on prewave functions by the same formula that in the case of wave functions,  without the tilde.

On the other hand, Linear Momentum is given on $C^+$\ by
$$
P(K)=-\eta K,\ \text{\rm for all}\ K\in C^+.
$$
Then, with the notation $K=h(K^1,K^2,K^3,K^4)$\ and $P=h(P^1,P^2,P^3,P^4),$\ the $P^k$\ are functions on $C^+,$\ given by
$$
P^k(K)=-\eta K^k.
$$
We also denote
\begin{eqnarray*}
\overrightarrow{K}&=&(K^1,K^2,K^3)\\
\Vert \overrightarrow{K}\Vert &=&
 +\left(\sum_{k=1}^3({K^k})^2\right)^{1/2}\\
\overrightarrow{P}&=&(P^1,P^2,P^3)\\ \Vert \overrightarrow{P}\Vert &=& 
 +\left(\sum_{k=1}^3({P^k})^2\right)^{1/2}.
\end{eqnarray*}

The \bf Helicity Operator \rm is defined by
$$
\mathfrak{h}=\frac{1}{\Vert \overrightarrow{P}\Vert} {\sum_{k=1}^3}{P^k}\hat{s^k},
$$
which means
$$
\left(\mathfrak{h}\cdot \psi_{f_T}^{\pm }\right)(K,H)=\frac{1}{\Vert \overrightarrow {P}\Vert(K)} {\sum_{k=1}^3}{P^k(K)}{s^k}\left(\psi_{f_T}^{\pm }(K,H)\right).
$$

Then, if $z\in r^{-1}_{\pm}(K)$\ and $z_1=\dots=z_T=z,$\ we have
\begin{eqnarray*}
&&\left(\mathfrak{h}\cdot \psi_{f_T}^{\pm }\right)(K,H)=\frac{1}{K^4} \sum_{k=1}^3{P^k(K)}{s^k}\left(f_T(z)e^{-i\eta \pi TrK\epsilon \overline{H}\epsilon} z^{\otimes T}\right)=\\&&=\frac{1}{4 \pi K^4}\ f_T(z)e^{-i\eta \pi TrK\epsilon \overline{H}\epsilon}\\
&&\left(  \sum_{j=1}^T\ z_1\otimes \dots\otimes z_{j-1}\otimes (\sum_{k=1}^3{P^k(K)}\, \sigma_k\,z_j)\otimes z_{j+1}\otimes\dots\otimes z_T\right)=\\&&=
\frac{1}{4 \pi K^4} \ 
f_T(z)e^{-i\eta \pi TrK\epsilon \overline{H}\epsilon}       \\&&   \left(  \sum_{j=1}^T\ z_1\otimes \dots\otimes z_{j-1}\otimes (-\eta)((K-K^4\,I)z_j)\otimes z_{j+1}\otimes\dots\otimes z_T\right).
\end{eqnarray*}

In the case $\chi=+1,$\ we have $K= r_+(z)=zz^*,$\  so that $$ (K-K^4\,I)z=(zz^*-\frac{\Vert z \Vert}{2}\,I)z=K^4\,z,
$$
and then
\begin{eqnarray*}
\left(\mathfrak{h}\cdot \psi_{f_T}^{+}\right)(K,H)=\frac{-\eta T}{4\pi }\psi_{f_T}^{+}(K,H).
\end{eqnarray*}

On the other hand, in the case $\chi=-1,$\ we have $ K=-\epsilon \overline{zz^*}\epsilon.$\ Thus
$$
 (K-K^4\,I)z=(-\epsilon \overline{zz^*}\epsilon-K^4\,I)z=-K^4\,z,
$$
so that
\begin{eqnarray*}
\left(\mathfrak{h}\cdot \psi_{f_T}^{-}\right)(K,H)=\frac{\eta T}{4\pi }\psi_{f_T}^{-}(K,H).
\end{eqnarray*}

Thus \it $\psi_{f_T}^{\pm}$\ is an eigenvector  of the helicity operator, corresponding to the eigenvalue\rm $$-\eta \chi T/4 \pi.$$

In particular, \it the sign of helicity is $-\eta \chi.$\rm

																																															\subsection{The Homogeneous Contact and Symplectic Manifolds for  Massless particles of type 4.  Twistors.}\label{contactnula}

Let us denote by $\sqrt[T]{\mathbf 1}=\{ r_1, \dots ,r_T\},$\ the group composed by the roots of order $T$ of 1, and 
$$
\nu=\frac{- \eta \chi T}{4\pi} .
$$

In section \ref{helicidadgen} we have seen that all wave functions obtained in section \ref{casogeneral} are eigenvectors of helicity
 with eigenvalue $\nu.$\

The group $Ker\,C_\alpha,$\ has $T$ connected components
\begin{equation*}
	(Ker\,C_\alpha)_j   = 
 \left\{ \left(
\left (\begin{array}{cc}
r_j&a  \\
0&{\overline r_j}
\end{array}
\right )
,\ \left (\begin{array}{cc}
b&-2i\nu \,a\, r_j   \\
\overline {-2i\nu \,a\, r_j } &0
\end{array}
\right )
\right ):
 b\in  {\bf R},\ a \in
{\bf C}
\right\}    
\end{equation*}

The component of the identity is
\begin{equation*}
	(Ker\,C_\alpha)_o   = 
 \left\{ \left(
\left (\begin{array}{cc}
1&a  \\
0&{1}
\end{array}
\right )
,\ \left (\begin{array}{cc}
b&-2i\nu \,a   \\
\overline {\-2i\nu \,a}&0
\end{array}
\right )
\right ):
 b\in  {\bf R},\ a \in
{\bf C}
\right\}    
\end{equation*}
and  $(Ker\,C_\alpha)_j $\ is the left translation by 
$$
\tilde{\bf r}_j \stackrel{def}{=}\left( \left(  \begin{array}{cc}
r_j&0  \\
0&{\overline r_j}
\end{array}
\right ),\ 0 \right)
$$
of $(Ker\,C_\alpha)_o .$

As a consequence, the group $Ker\,C_\alpha/(Ker\,C_\alpha)_o$\ is isomorphic to $\sqrt[T]{\mathbf 1}$.

The canonical map $$
 \begin{gathered}
\tilde{\bf c} :    (A,H)\,(Ker\,C_\alpha)_o\in \frac{G}{(Ker\,C_\alpha)_o}   \longrightarrow \\
 \longrightarrow (A,H)\,Ker\,C_\alpha\in \frac{G}{Ker\,C_\alpha}.
\end{gathered}
$$
is a $T$-fold covering map, whose structural group is $Ker\,C_\alpha/(Ker\,C_\alpha)_o$. 

Thus, $G/Ker\,C_\alpha$\ is the quotient of $G/(Ker\,C_\alpha)_o$\ by a properly discontinuous with no fixed point action of 
 $Ker\,C_\alpha/(Ker\,C_\alpha)_o$. But, as we have seen, this group can be changed  to  $\sqrt[T]{\mathbf 1}$.

The action of  $\sqrt[T]{\mathbf 1}$\  on $G/(Ker\,C_\alpha)_o$\  corresponding to this construction is:
\begin{equation}\label{accionraicesT}
((A,H)\,(Ker\,C_\alpha)_o)* r_j= ((A,H)\,\tilde{\bf r}_j)\,(Ker\,C_\alpha)_o.
\end{equation}

In what follows, we identify $G/Ker\,C_\alpha$\  with the quotient space of $G/(Ker\,C_\alpha)_o$\ by this action.

We have the following conmutative diagram

\[
\begin{CD}
\frac{G}{(Ker\,C_\alpha)_o  }    @>  \tilde{\bf c}   >> \frac{G}{Ker\,C_\alpha  } \\
        @VVV                                                     @VVV             \\
\frac{G}{G_\alpha}                     @=                  \frac{G}{G_\alpha}
\end{CD}
\]
where the vertical arrow on the right is the bundle map of the contact manifold onto the symplectic manifold, for general $T.$\ If $T=1$ both vertical arrows are the same.

 The homomorphism
$$
\mu_1:(A, H) \in SL\oplus H(2)  \longrightarrow \left( \begin{array}{cc}
A&{-iH{A^*}^{-1}} \\
0&{A^*}^{-1}
\end{array} \right)
\in GL(4,\ \text{\bf C}).
$$
is a representation in ${\bf C}^4.$\

The isotropy subgroup at 
$$
q \stackrel{def}{=}\left( \begin{array}{c}0\\ 2\nu\\0\\ 1\end{array}\right)
$$
 is $(Ker\,C_\alpha)_o .$\

Let us denote by $\mathrm O_q$\ the orbit of $q$\ by this representation. 

In order to describe $\mathrm O_q$\  in another way, we
 consider in $\text{\bf C}^4$ the hermitian product
$$
<\binom {\hat w}{\hat z},\ \binom wz>=\frac 12 (\hat w^*z+\hat
z^*w)\ \ \ \ \ \forall \hat w,\,\hat z,\,w,\,z \in \text{\bf C}^2
$$
whose signature and quadratic form are respectively (+,\,+,\,-,\,-) and
$$\Phi\binom wz =Re\,z^*\,w.$$ 

The complex vector space $\text{\bf C}^4$\  provided with this
hermitian product is \bf Penrose's {Twistor Space}. \rm

The representation $\mu_1$\ preserves $\Phi$, so that any orbit must be contained in a subset of the form $\Phi=\mathnormal{constant}.$\ It follows that the orbit of $q$ is contained in the 
seven dimensional submanifold, $\mathcal O_\nu$,\  given by 
$$
\Phi=2\nu.
$$

Let $$\binom wz\in \text{\bf C}^4$$ be such that $$sign\left(\Phi\binom wz \right)=sign(\nu),$$ and define

\begin{eqnarray}\label{pretransit}
A(w,z)&=&\left(\frac{2\nu}{ \Phi\binom wz}\right)^{1/2}\left(\begin{array}{c} {}\\\epsilon \overline{z}\\{}  \end{array}\  \Bigg|   \  \frac{1}{2\nu}\left(i\frac{Im(z^*w)}{\Vert z \Vert^2}z-w\right)\right)\\ \label{pretransit22} 
 H(w,z)  &=& -\frac{Im(z^*w)}{\Vert z \Vert^2}\ I.
\end{eqnarray}

Then we have
\begin{eqnarray}\label{transit}
\mu_1(A(w,z),H(w,z)) q=\left(\frac{2\nu}{ \Phi\binom wz}\right)^{1/2}\ \binom wz.
\end{eqnarray}

This enables us to prove that each element of $\mathcal O_\nu$\  is in the orbit, so that  $\mathrm O_q=\mathcal O_\nu$.

Then, the map from $G/(Ker\,C_\alpha)_o$\ onto $ \mathcal O_\nu,$\ given by 
\begin{equation}\label{difeoorbit}
\tau_T:(A,H)\,(Ker\,C_\alpha)_o\in \frac{G}{(Ker\,C_\alpha)_o} \longrightarrow \mu_1(A,H)\,q\in \mathcal O_\nu,
\end{equation} 
is a diffeomorphism.

If we identify these spaces by $\tau_T,$\ 
the action of $\sqrt[T]{\mathbf 1}$\  on $G/(Ker\,C_\alpha)_o$\ given by \eqref{accionraicesT}, traslates to an action on $\mathcal O_\nu.$\ 

Since
\begin{eqnarray*}\label{accionraicesTCAL}
\tau_T((A,H)\,(Ker\,C_\alpha)_o) *r_j)&=&\mu_1 ((A,H)\,\tilde{\bf r}_j)q=\mu_1 (A,H)\mu_1(\tilde{\bf r}_j)q\\=\mu_1 (A,H)(\overline{r_j}q)&=&\overline{r_j}\ \tau_T((A,H)\,(Ker\,C_\alpha)_o) ,
\end{eqnarray*}
 the action of $\sqrt[T]{\mathbf 1}$ on $\mathcal O_\nu$\ is given by ordinary product by the conjugated:
\begin{equation}\label{acraiTO}
V*r_j=\overline{r_j}\ V
\end{equation} 

Let us denote by $\mathcal O_\nu/\sqrt[T]{\mathbf 1}$\ the quotient space of $\mathcal O_\nu$\ by this action. 

It follows from the preceeding construction that the \bf contact manifold, \rm $G/Ker\,C_\alpha,$ is diffeomorphic, and will be identified, to $\mathcal O_\nu/\sqrt[T]{\mathbf 1}$. The contact form will be determinated later on.

Let ${}^t\negthinspace(w^1,w^2,z^1,z^2)\in {\mathbb C}^4-\{0\},$\ and let us denote by $[{}^t\negthinspace(w^1,w^2,z^1,z^2)]$\ the complex vector subspace of ${\mathbb C}^4$\ generated by ${}^t\negthinspace(w^1,w^2,z^1,z^2).$\ The set composed by these subspaces  is the complex projective space of ${\mathbb C}^4,$\ ${\bf P}_3(\mathbb C),$\ and we consider it as provided with its canonical differentiable structure.

Penrose also considers the
subsets of twistor space, ${\bf T}^+,\  {\bf T}^-,\ {\bf T}^0,$\ given by $
\Phi\,>\,0,\ \Phi\,<\,0,\ \Phi\,=\,0$,\ respectively, and the subsets of
projective
space $ {\bf P}_3^+=\pi({\bf T}^+),\ {\bf P}_3^-=\pi({\bf T}^-),\ {\bf P}_3^0=\pi({\bf T}^0)$,\ where $$
\pi:{\mathbb C}^4 -\{0\} \longrightarrow {\bf P}_3(
{\mathbb  C}) $$
 is the canonical map.

Since $\Phi$\ is preserved by the representation, the subsets ${\bf T}^+,\  {\bf T}^-,\ {\bf T}^0,$\ are stable under $\mu_1$.

The representation $\mu_1$\ also defines an action of $G$ on ${\bf P}_3(\text
{\bf C}) $\ such that $\pi$\ is equivariant. Explicitly
\begin{equation}\label{accionproy}
(A,H)*\left[ \left( \begin{array}{c}w^1\\w^2\\z^1\\ z^2\end{array}\right)\right]=\left[\left( \begin{array}{cc}
A&{-iH{A^*}^{-1}} \\
0&{A^*}^{-1}
\end{array} \right)\left( \begin{array}{c}w^1\\w^2\\z^1\\ z^2\end{array}\right)\right].
\end{equation}

As a consecuence of formula \eqref{transit}, the open subsets of ${\bf P}_3(\text
{\bf C}) $\ denoted by ${\bf P}_3^+$ \ and ${\bf P}_3^-$\ are orbits of this action.

The point $[q]$\ of ${\bf P}_3(\text {\bf C}) $\ is obviously in ${\bf P}_3^{sign(\nu)}$\ so that its orbit is this open subset. 

The isotropy subgroup at $[q]$\ is $G_\alpha.$\ 

As a consecuence we have a diffeomorphism from ${\bf P}_3^{sign(\nu)}$\ onto  the coadjoint orbit of $\alpha,$\ given by
$$
\Pi: [\mu_1(A,H)\,q]\in {\bf P}_3^{sign(\nu)} \longrightarrow Ad^*_{(A,H)}\alpha \in {\mathcal MS} . 
$$
were I have denoted the coadjoint orbit by ${\mathcal MS},$\ because of its interpretation as Movement Space.

We have
\begin{equation}\label{Pi}
\Pi\left(\left[ \binom { w}{ z} \right]\right)= Ad^*_{(A(w,z),H(w,z))}\alpha
\end{equation}
so that, the formula for coadjoint representation in section \ref{sec-grupo},\eqref{coadjunta}, thus leads, after some computation, to
\begin{equation}
\Pi\left( \left[ \binom { w}{ z} \right] \right)=\frac{2\nu\eta }{ \Phi\binom { w}{ z}} \biggl \{\frac {i}{4}
(wz^*+\epsilon \overline{zw^*}\epsilon),\ - \epsilon \overline {zz^*}\epsilon \biggr
\}.
\end{equation}

We identify $P_3^{sign(\nu)}$\  to ${\mathcal MS},$\  by means of $\Pi.$\ 

Section \ref{sec-grupo} provides us with well defined expresions for linear and angular momentum in $\mathcal MS,$\  $c.f.$\ \eqref{vardin}, which, when composed with $\Pi,$\ give   expressions  for \it linear and angular momentum  \rm in  $P^{sign(\nu)}_3.$\ In its hermitian form, these expressions are
\begin{eqnarray}\label{Pproy}
h(P\left( \left[ \binom { w}{ z} \right] \right))&=& \frac{2\eta \nu}{\Phi\binom { w}{ z}} \epsilon \overline {zz^*}\epsilon \label{vardinO1}\\   \label{lproy}
h(\vec l \left( \left[ \binom { w}{ z} \right] \right),0)&=&\frac{\eta \nu}{2\Phi\binom { w}{ z}} (zw^*+wz^*+\epsilon \overline{(zw^*+wz^*)}\epsilon)\label{vardinO2}\\ \label{gproy}
h(\vec g \left( \left[ \binom { w}{ z} \right] \right),0)&=&\frac{i\,\eta \nu}{2\Phi\binom { w}{ z}}(zw^*-wz^*-\epsilon \overline{(zw^*-wz^*)}\epsilon)\label{vardinO3}
\end{eqnarray}

The Pauli-Lubanski four vector, when evaluated according with \eqref{P-L1}, is found to be
\begin{equation}\label{paul}
h(W\left( \left[ \binom { w}{ z} \right] \right)) = 
-\eta\nu\,
h(P\left( \left[ \binom { w}{ z}\right] \right)).     	
\end{equation}

In case T=1, the bundle map of the contact manifold onto the coadjoint orbit becomes the canonical map

$$
\pi_1:\binom { w}{ z} \in \mathcal O_1\ \longrightarrow \left[ \binom { w}{ z} \right]\in P^{sign(\nu)}_3,
$$
and in the general case, the bundle map is
$$
\pi_\nu:\binom { w}{z} \sqrt[T]{\mathbf 1}\in \mathcal O_\nu/\sqrt[T]{\mathbf 1} \longrightarrow \left[ \binom { w}{ z} \right]\in P^{sign(\nu)}_3.
$$

Then, the composition with $\pi_\nu$\ of the canonical dynamical variables, are also given by the right hand sides of \eqref{vardinO1}, \eqref{vardinO2} and \eqref{vardinO3}, but taking $\Phi=2\nu.$

We  obtain the following formulae for the components of these dynamical variables
\begin{eqnarray*}
P^1\left(\binom { w}{ z} \sqrt[T]{\mathbf 1}\right) &=\frac {\eta}{2}(z^1\overline z^2\,+\,z^2\overline z^1) \\
P^2\left(\binom { w}{ z} \sqrt[T]{\mathbf 1}\right)&=\frac {i\eta}{2}(z^1\overline z^2\,-\,z^2\overline z^1) 
\\
P^3\left(\binom { w}{ z} \sqrt[T]{\mathbf 1}\right)&=\frac {\eta}2(\vert z^1\vert ^2\,-\,\vert z^2\vert ^2)  \\
P^4\left(\binom { w}{ z} \sqrt[T]{\mathbf 1}\right)&=-\frac {\eta}2(\vert z^1\vert ^2\,+\,\vert z^2\vert ^2) 
\end{eqnarray*}
\begin{eqnarray*}
l^1\left(\binom { w}{ z} \sqrt[T]{\mathbf 1}\right)&=\frac {\eta}4 (z^1\overline w^2+w^1\overline z^2+
z^2\overline w^1+w^2\overline z^1)  \\
l^2\left(\binom { w}{ z} \sqrt[T]{\mathbf 1}\right)&=\frac {i\eta}4 (z^1\overline w^2+w^1\overline z^2-
z^2\overline w^1-w^2\overline z^1)  \\
l^3\left(\binom { w}{ z} \sqrt[T]{\mathbf 1}\right)&=\frac {\eta}4 (z^1\overline w^1+w^1\overline z^1-
z^2\overline w^2-w^2\overline z^2)  
\end{eqnarray*}
\begin{eqnarray*}
g^1\left(\binom { w}{ z} \sqrt[T]{\mathbf 1}\right)&=\frac {i\eta}4 (z^1\overline w^2-w^1\overline z^2+
z^2\overline w^1-w^2\overline z^1)  \\
g^2\left(\binom { w}{ z} \sqrt[T]{\mathbf 1}\right)&=-\frac {\eta}4 (z^1\overline w^2-w^1\overline z^2-
z^2\overline w^1+w^2\overline z^1)  \\
g^3\left(\binom { w}{ z} \sqrt[T]{\mathbf 1}\right)&=\frac {i\eta}4 (z^1\overline w^1-w^1\overline z^1-
z^2\overline w^2+w^2\overline z^2).  
\end{eqnarray*}

We can also consider the functions on ${\bf T}^{sgn(\nu)}$\ obtained by composing the dynamical variables in $\mathcal MS$\ with the canonical projection, $\pi,$\ from ${\bf T}^{sgn(\nu)}$\ onto ${\bf P}_3^{sgn(\nu)},$\ thus obtaining functions whose expresion is as the right hand sides of the preceeding formulae multiplied by
$$
\frac{2\nu}{ \Phi\binom { w}{ z}}
$$

These  expressions coincide, up to  notational
conventions, with the expressions that R.Penrose gives for its energy -
momentum and angular momentum in twistor space. We denote these functions by $\widetilde P^k,\ \widetilde l^k,\ \widetilde g^k.$\ 

In general, for all $(a,h)$\ in the Lie algebra of $G,$\ the function it defines on the coadjoint orbit, identified to ${\bf P}_3^{sgn(\nu)},$\  is denoted by the same symbol, $(a,h),$\ and its composition with $\pi,$\ $\widetilde {(a,h)}.$\ In a similar way, the  infinitesimal generator of the action on ${\bf T}^{sgn(\nu)}$\ defined by the representation $\mu_1,$\ associated to $(a,h),\ i.e.$\ the vector field whose flow is given by $\mu_1(Exp(-t(a,h))),$\ is denoted by
$$
\widetilde{X_{(a,h)} }.
$$

Let us consider
  the following one form on ${\bf T}^{sgn(\nu)}$\ 
$$
\omega_0=\frac {i\eta\nu}{2\Phi} (z^1d\overline w^1+w^1d\overline z^1+
z^2d\overline w^2+ w^2d\overline z^2 -
$$$$                                  
- \overline  z^1 d w^1- \overline w^1d z^1-
\overline z^2d w^2-\overline  w^2dz^2)
$$

A  computation lead us to
\begin{equation}\label{omegaX}
\omega_0\left( \widetilde{X_{(a,h)} }\right)= -\widetilde {(a,h)}.
\end{equation}

Let us denote by $\omega$\ the restriction of $\omega_0$\ to $\mathcal O_\nu$.

The one form $\omega$\ is invariant by the action of $\sqrt[T]{\mathbf 1},$\ so that it projects to a well defined one form on $\mathcal O_\nu/\sqrt[T]{\mathbf 1},$\ that 
 we denote by  $\omega_\nu.$\ 

We know that $\mathcal O_\nu/\sqrt[T]{\mathbf 1},$\ represent the \bf homogeneous contact manifold \rm corresponding to the kind of particle under consideration. As a consecuence of \eqref{omegaX} and \eqref{funcabaj} is not difficult to prove that  $\omega_\nu$\ is the \bf contact form.\rm

As a consecuence of the fact that the map of $\mathcal O_\nu$\ on $\mathcal O_\nu/\sqrt[T]{\mathbf 1}$\ is a covering map, $\omega$\ is also a contact form on $\mathcal O_\nu,$\ that becomes itself an homogeneous contact manifold.

The two form $d(\omega_\nu)$\  thus projects under $\pi_\nu$\ on the symplectic form on $P_3^{sign(\nu)},\ \Omega_\nu,$\ that corresponds to Kirillov form in  the identification of  $P_3^{sign(\nu)}$\  with the coadjoint orbit. 

Then, $d\omega$ \ also projects on $\Omega_\nu,$\ under the restriction of the canonical map $\pi$\ to $\mathcal O_\nu.$\ 

But $d\omega$\ is the restriction to  $\mathcal O_\nu$\ of  $d\omega_0$\ 
and we have
$$
d\omega_0=\frac {i\eta\nu}{ \Phi} (dz^1\wedge d\overline
w^1+dw^1\wedge d\overline z^1+ dz^2
\wedge d\overline w^2+dw^2 \wedge d\overline z^2 )+(d\Phi)\wedge \delta,
$$
where $\delta$\  is a one form.

Since $d\Phi$\ vanishes on $\mathcal O_\nu,$\  it follows that $d\omega$\ is also the restriction to $\mathcal O_\nu$\ of
$$
\Omega=\frac {i\eta\nu}{ \Phi} (dz^1\wedge d\overline
w^1+dw^1\wedge d\overline z^1+ dz^2
\wedge d\overline w^2+dw^2 \wedge d\overline z^2).
$$

Thus, we can obtain explicit expressions of $\Omega_\nu$\ as follows:
 for each differentiable section of $\pi$\ with values in $\mathcal O_\nu$\ 
$$
\sigma: U \rightarrow {\cal O}_\nu \subset {\bf T}^{sgn(\nu)},
$$
 we have on the open set $U$\ 
$$\Omega_\nu=\sigma^*\Omega_0,$$

Also we have
$$
\Omega_\nu=d\,\left(\sigma^*\omega_0 \right).$$

\subsubsection{Local expression of the symplectic form.}\label{coordenadasproy}

In ${{\mathbb C}}^{3}$\ we define
$$
{\cal D}=\{(t,u,v):\,sign(\nu)\,\Re(\phi(t,u,v)) > 0\},
$$
where $\phi(t,u,v)=u+\overline t v$\ and $\Re{}$\ stands for real part.

In ${\bf P }^{sgn(\nu)}_3$\ we define
$$
{\cal D}_1=\left\{\left[\begin{array}{c}w^1\\w^2\\z^1\\z^2 \end{array} \right]:w^1\neq 0,\ sign(\nu)\,\Re(\Phi(\left(\begin{array}{c}w^1\\w^2\\z^1\\z^2 \end{array}\right))) > 0\right\} 
$$$$
{\cal D}_2=\left\{\left[\begin{array}{c}w^1\\w^2\\z^1\\z^2 \end{array} \right]:w^2\neq 0,\ sign(\nu)\,\Re(\Phi(\left(\begin{array}{c}w^1\\w^2\\z^1\\z^2 \end{array}\right))) > 0\right\} 
$$$$
{\cal D}_3=\left\{\left[\begin{array}{c}w^1\\w^2\\z^1\\z^2 \end{array} \right]:z^1\neq 0,\ sign(\nu)\,\Re(\Phi(\left(\begin{array}{c}w^1\\w^2\\z^1\\z^2 \end{array}\right))) > 0\right\} 
$$$$
{\cal D}_4=\left\{\left[\begin{array}{c}w^1\\w^2\\z^1\\z^2 \end{array} \right]:z^2\neq 0,\ sign(\nu)\,\Re(\Phi(\left(\begin{array}{c}w^1\\w^2\\z^1\\z^2 \end{array}\right))) > 0\right\} .
$$

The ${\cal D}_k$\ compose an open cover of ${\bf P }^{sgn(\nu)}_3.$

The maps
$$
\psi_1: (t,u,v)\in {\cal D}  \rightarrow  \left[\begin{array}{c}1\\t\\u\\v \end{array} \right] \in {\cal D}_1 
$$$$
\psi_2: (t,u,v)\in {\cal D}  \rightarrow  \left[\begin{array}{c}v\\1\\t\\ u \end{array} \right] \in {\cal D}_2 
$$
$$
\psi_3: (t,u,v)\in {\cal D}  \rightarrow  \left[\begin{array}{c}u\\v\\1\\t \end{array} \right] \in {\cal D}_3 
$$$$
\psi_4: (t,u,v)\in {\cal D}  \rightarrow  \left[\begin{array}{c}t\\ u \\v\\1 \end{array} \right] \in {\cal D}_4
$$
are such that the $({\cal D}_k,(\psi_k)^{-1})$\ compose an atlas of ${\bf P }^{sgn(\nu)}_3.$ 

For all of these charts the coordinates will be denoted by $(t,u,v).$\

We also define  sections of $\pi_\nu,$$$
 \sigma_k:{\cal D}_k \rightarrow {\cal O}_\nu,\ \ \ \ k=1,\dots,4,
$$
by means of 
$$
\sigma_1\circ\psi_1: (t,u,v)\in {\cal D}  \rightarrow  F(t,u,v)\left(\begin{array}{c}1\\t\\u\\v \end{array} \right) \in {\cal D}_1 
$$$$
\sigma_2\circ \psi_2: (t,u,v)\in {\cal D}  \rightarrow F(t,u,v)\left(\begin{array}{c}v\\1\\t\\ u \end{array} \right) \in {\cal D}_2 
$$$$
\sigma_3\circ\psi_3: (t,u,v)\in {\cal D}  \rightarrow  F(t,u,v)\left(\begin{array}{c}u\\v\\1\\t \end{array} \right) \in {\cal D}_3
$$$$   
\sigma_4\circ \psi_4: (t,u,v)\in {\cal D}  \rightarrow  F(t,u,v)\left(\begin{array}{c}t\\  u \\v\\1 \end{array} \right) \in {\cal D}_4
$$
where
$$
F(t,u,v)=\sqrt{\frac{2\nu}{\Re(\phi(t,u,v))}}.
$$

For all $k$\ we have
\begin{equation}\label{omeg}
(\sigma_k\circ \psi_k)^*\omega = \frac{\eta \nu}{\Re(\phi)}(d(\Im(\phi))+i(v\,d\overline t-\overline v\,dt)),
\end{equation}
Where $\Im(\phi)$\ is the imaginary part of $\phi.$\ 

Then, the local expression of $\Omega_\nu$\ in all of these coordinate systems can be evaluated by means of
\begin{equation}\label{Omeg}
\Omega_\nu    \stackrel{loc}{=}d\,((\sigma_k\circ \psi_k)^*\omega).
\end{equation}
for all $k.$

Notice that $\omega $\ is not projectable on ${\bf P }^{sgn(\nu)}_3.$\ Thus \eqref{omeg} need not be local expressions of a well defined 1-form on all of ${\bf P }^{sgn(\nu)}_3.$

\subsection{Alternative form of Wave Functions for the Photon.}\label{alternphoton}

\subsubsection{Symmetric Wave Functions of the Photon.}

There are many proposals in the literature for Wave Functions of the Photon. Including someones that asserts that such a thing does not exist.

In this paper I have described the Wave Functions of the Massless Type 4 Particles, where the case $T=2$ corresponds to the Photon. These Wave Functions satisfies the  Penrose Wave Equations.

In this section I describe other forms of the Wave Functions of Photon. Equivalent representations.These forms enables us to relate directly the Wave Functions with the Electromagnetic Potential  and the Electromagnetic Field.
 
   \par         Photon is the name of four kinds of particles: the massless Type 4 particles with $T=2,$
\ $i.e.$ those  corresponding to
\begin{equation}\label{foton}
\gamma =\left\{ \frac{i \chi }{4 \pi}
\left (\begin{array}{cc} 1 & 0\\
0 & -1
\end{array}
\right )
,
 \eta\,
\left (\begin{array}{cc} 1 & 0\\
0 & 0
\end{array}
\right )
\right\}\ \ \ \ ,\ \chi,\ \eta \in \{\pm 1\},
\end{equation}

We already know that the value of $-\eta$\ is  the sign of energy and that $-\eta \chi$\  is the sign of helicity ($c.f.$\  section \ref{helicidadgen}). We denote $\ell=-\eta \chi$.

From section \ref{waveNula} we obtain

$$
\begin{array}{l}
\vspace{.1in}
G_\gamma = \left\{ \left(
\left (\begin{array}{cc}
z&a  \\
0&{\overline z}
\end{array}
\right )
,\ \left (\begin{array}{cc}
b&i \chi \eta  a z/ \pi   \\
\overline {i \chi \eta  a z /  \pi}&0
\end{array}
\right )
\right )
:\ z \in   {\bf S^1},\right.  \\
 \left. \ \ \ \ \ \ \ \ \ b\in  {\bf R},\ a \in
{\bf C}
\right\} \\
\      \\
\mathrm{and\ the\ homomorphism}          \\
\     \\
C_\gamma \left( \left( \left (\begin{array}{cc}
z&a \\
0&{\overline z}
\end{array}
\right )
,\ \left (\begin{array}{cc}
b&{i \chi \eta  a z/ \pi} \\
{\overline {i \chi \eta  a z/ \pi}}&0
\end{array}
\right )
\right )
\right )
=z^{2\chi }
\end{array}
$$

has differential $\gamma.$

$$
\begin{array}{l}
({\mbox{$G_\gamma$}} )_{SL}=\left\{ \left(
\begin{array}{cc} z&a \\
0&{\overline z}
\end{array}
\right )
: z \in {\bf S^1},\  a \in {\mathbb C} \right\}    \\
\vspace{.1in}
(C_\gamma)_{SL}\left(\
\left (\begin{array}{cc}
z&a \\
0&{\overline z}
\end{array}
\right )
\right )=\ z^{2\chi}  \\
\vspace{.1in}
SL_1=
\left\{
\left (\begin{array}{cc}
a&0  \\
0&{1/a}
\end{array}
\right ):\ a \in {\mathbb C}
\right \}        \\
\vspace{.1in}
SL_2=(G_\gamma)_{SL}          \\
\vspace{.1in}
SL_1 \cap SL_2=
\left\{
\left (\begin{array}{cc}
z&0  \\
0&\overline z
\end{array}
\right ):\ z \in   {\bf S}^1
\right \}
\end{array}.
$$

In section \ref{waveNula} we have identified the space $SL/(G_\gamma)_{SL}$  
to the future lightcone, ${\mbox{${\bf C}^+$}}=\{H \in   {H(2)}:\
Det\,H=0,\ Tr\,H>0\},$\ by means of the diffeomorphism
$$
A\, (G_\gamma)_{SL} \in SL/(G_\gamma)_{SL} \mapsto A \left (\begin{array}{cc} 1 & 0\\
0 & 0
\end{array}
\right ) A^* \in {\bf C}^+.
$$

In this section I give another description of the wave functions of a photon, 
by means of different trivialisations than the ones I have used in 
 section \ref{waveNula}. Of course these trivialisations are ``isomorphic'', in an obvious  
 sense. 

Let us consider first the cases $\chi=+1$\ $i.e.$ $\ell=-\eta.$

In these cases, 
a                                                                
trivialisation ($c.\,f.$\ section \ref{wave}) is
$$
(\mu_{+}, { \cal{S}},s_0^+),
$$
where $\cal S$\ is the vector subspace of $gl(2,\bf C),$\ composed by the 
symmetric matrices,
$$
s_0^+=\left (\begin{array}{cc} 1 & 0\\
0 & 0
\end{array}
\right )
$$
 and 
$$
\mu_+(A)\cdot s=A \,s\, {}^t\negthinspace A,
$$
for all $A \in SL(2,{\bf C}),\ s\in \cal S$.

Since
$$
\left (\begin{array}{cc} 1 & 0\\
0 & 0
\end{array}
\right )=\left (\begin{array}{c} 1 \\
0 
\end{array}
\right )\left (\begin{array}{cc} 1 & 0
\end{array}
\right )  $$
the    orbit of
$$
\left (\begin{array}{cc} 1 & 0\\
0 & 0
\end{array}
\right )$$ by $\mu_+$\ is composed by the elements of ${\cal S}^2$ of the 
form $$ 
 (A  \left (\begin{array}{c} 1 \\
0 
\end{array}
\right )  )\ {}^t \negthinspace (A  \left (\begin{array}{c} 1 \\
0 
\end{array}
\right )  ),
$$
for some $A \in SL(2,\bf C)$.
Then, one sees that the orbit is contained in
$$
{\cal B}=\{ z \ {}^t \negthinspace z: z\in {\bf C}^2-\{0\}\}.
$$
But every $z\in {\bf C}^2-\{0\}$\ has the form 
$$ A\, \left (\begin{array}{c} 1 \\
0 
\end{array}
\right ), 
$$
for some $A \in SL(2,{\bf C})$,\ 
so that the orbit coincides with ${\cal B}$.

For each $ s\in {\cal S}$ such that $s\neq 0$ and $Det\, s=0,$\ there exist 
exactly two $z\in {\bf C}^2-{0}$\ such that $s=z \ {}^t \negthinspace z.$\ 
In fact, if $$
s=\left (\begin{array}{cc} a & b\\
b & d
\end{array}
\right )\in B$$
the $z$ \ such that $s=z \ {}^t \negthinspace z$ \ are $$
\pm \left (\begin{array}{c} \alpha \\
\sigma \delta 
\end{array}
\right )$$
where $\alpha $ is a square root of $a$,\ $\delta$ is a square root of $d$,\ 
and $\sigma\in\{\pm 1\}$\ such that $b=\sigma \alpha \delta.$

As a consequence, we also have
$$
{\cal B}=\{ s\in {\cal S}: s\neq 0,Det\, s=0\}.
$$

The homogeneous space $SL/Ker(C_\gamma)_{SL}$ is identified		to ${\cal B}$\ by 
means of		$$ A\,Ker(C_\gamma)_{SL} \in SL/Ker(C_\gamma)_{SL}\longrightarrow	
\mu_+(A)\cdot s_0^+ \in {\cal B}
$$ 

Then,  the canonical map $$ 
SL/Ker(C_\gamma)_{SL}	\longrightarrow SL/(G_\gamma)_{SL},
$$
denoted in the following by $r_+,$\ becomes $$
r_+:(A \left (\begin{array}{cc} 1 & 0\\
0 & 0
\end{array}
\right ){}^t\negthinspace  A)\in {\cal B}	\longrightarrow A \left (\begin{array}{cc} 1 & 0\\
0 & 0
\end{array}
\right )A^*\in C^+,
$$
for all $A\in SL,$\  so that
$$
r_+(z{}^t \negthinspace z)= zz^*
$$
for all $z \in {\bf C}^2-{0}$.
																																							
\par		By  elementary operations with matrices, one can prove that 
the relation $$
r_+(s)=C $$is equivalent to
$$C=(+(Tr\,s\overline{s})^{-1/2})\,s\overline{s}$$
 and also to 
$$s=(+(Tr\,C {}^t \negthinspace C)^{-1/2})e^{i\phi }(C {}^t \negthinspace C)
$$
for some $\phi \in {\bf R}.$

To obtain Wave Functions, we need functions on ${\cal B}$\ homogeneous of degree -1 
under product by modulus one complex numbers. If $f$\ is one of such functions
the corresponding Prewave Function is
\begin{equation} \label{prefunciononda+1}
	\psi_f^{(-\eta,-\eta)}(H,K)=f(s)\,s\,e^{2 \pi i\eta \langle \, h^{-1}(K),\,h^{-1}(H)\rangle_m},
	\end{equation}
	where $s$\ is arbitrary in $r_+^{-1}(K).$\ In ${(-\eta,-\eta)}$ the first $-\eta$ stands for the sign of energy and the second by $\ell,$  helicity.

	Now, let us consider the cases $\chi=-1$ $i.e.$ $\ell=\eta.$ 
	
	In these cases, a                                                                
trivialisation  is
$$
(\mu_{-}, { \cal{S}},s_0^-),
$$
where $\cal S$\ is as in the $\chi=+1$ case, 
$$
s_0^-=\left (\begin{array}{cc} 0 & 0\\
0 & 1
\end{array}
\right )
$$
 and 
$$
\mu_-(A)\cdot s=(A^*)^{-1} \,s\, (\overline{A})^{-1},
$$
for all $A \in SL(2,{\bf C}),\ s\in \cal S$.

The    orbit of
$$
\left (\begin{array}{cc} 0 & 0\\
0 & 1
\end{array}
\right )$$ by $\mu_-$\ is, as in case $\chi=+1$, 
\  ${\cal B}$.
	
The canonical map $$
r_-: SL/Ker(C_{\gamma})_{SL}	\longrightarrow   SL/(G_{\gamma})_{SL}
$$
becomes a map from ${\cal B}$\ onto ${\bf C}^+,$\  given by
$$
r_-(	\mu_-(A)\cdot s_0^-)=
A \left (\begin{array}{cc} 1 & 0\\
0 & 0
\end{array}
\right )A^*.
$$

	The maps $r_+$\ and $r_-$\ are geometrically related as follows.
																									
	Let us consider the map
\begin{equation}\label{jota}
J:s\in {\cal S} \longrightarrow -\epsilon\overline{s}\epsilon  \in {\cal S}				.
\end{equation}
This is an antilinear map with
$$
J^2=I.
$$

Since 
$$
J(z{}^t\negthinspace z)=(-\epsilon \overline{z}){}^t\negthinspace (-\epsilon \overline{z}),
$$
we have $J({\cal B})={\cal B}.$

On the other hand
\begin{eqnarray*}
r_-\circ J(\mu_+(A)\cdot s_0^+)=r_-(-\epsilon\, \overline{A}\,\overline{s_0^+}\,A^*\,\epsilon)=r_-(-(A^*)^{-1}\,\epsilon \,\overline{s_0^+} \,\epsilon \, (\overline A)^{-1})=\\=r_-(\mu_-(A)\cdot s_0^-)=A\  ({\mbox{$G_\gamma$}} )_{SL}=r_+(\mu_+(A)\cdot s_0^+)	,																
\end{eqnarray*}				
so that
$$
				r_+=r_-\circ J.
$$
Since $J^2=I,$\  we also have
$$
				r_-=r_+\circ J.
$$

As a consequence, $J$ stablishes a principal fibre bundle isomorphism over the identical map of $C^+$,\ the isomorphism of structural groups being the map defined by sending  each element of $S^1$\ to its inverse.

Another consequence is
$$
r_-(z {}^t\negthinspace z)=-\epsilon \overline{zz^*}\epsilon.
$$

If $f$\ is a function on ${\cal B}$\ homogeneous of degree -1 
under product by modulus one complex numbers, it also defines   a Prewave Function for this kind of photon by means of 
\begin{equation} \label{prefunciononda-1}
	\psi_f^{(-\eta,\eta)}(H,K)=f(s)\,s\,e^{2 \pi i\eta \langle \, h^{-1}(K),\,h^{-1}(H)\rangle_m},
	\end{equation}
	where $s,$\ now, is arbitrary in $r_-^{-1}(K).$ Here, in ${(-\eta,\eta)},$  $-\eta$ stands for the sign of energy and the second $\eta,$ which in the present case coincides with $\ell,$  stands for helicity.

Thus, the Prewave Functions  corresponding to the four kinds of Photon are different. The sign of energy, $-\eta,$ and the sign of helicity, $\ell,$ determines  the type of Prewave Functions  to be used,  $\psi_f^{(-\eta, \ell)},$ that are given by    \eqref{prefunciononda+1} or  \eqref{prefunciononda-1}.

When $f$\ is continuous with compact support, the corresponding Wave Function is
\begin{equation}\label{WFsimetrica}
\widetilde \psi_f^{(-\eta, \ell)}(x)=\int_{C^+} \psi_f^{(-\eta, \ell)}(h(x),K) \omega_K,
\end{equation}
where $\omega $ is the invariant volume element on $C^+$ defined in \eqref{omegaC+}.

These Wave Functions represent states whose \bf energy \rm has sign $-\eta$\ and are \bf eigenvectors of  helicity \rm ($c.f. $ section \ref{helicidadgen}) corresponding to the eigenvalues
$$
\frac{\ell}{2\pi}.
$$

\subsubsection{Prehilbert Space estructure.}\label{prodherfot}

Let us denote by 	${\cal{F}}$\  the vector subspace composed by the  functions on ${\cal B}$\ homogeneous of degree -1 
under product by modulus one complex numbers, and by ${\cal{F}}_c$\  the subspace of ${\cal{F}}$\  composed by the continuous elements with compact support.

The space ${\cal{F}}_c$\  can be provided with a prehilbert space structure, by means of the general method described in section \ref{wave}, for each kind of Photon. In more detail we proceed as follows.

We separate the cases ${-\eta\ell}=\pm 1.$ 

If $\ f , f^\prime \in {\cal F}_c,$\ its hermitian product is in each case
$$
\langle f,\, f^\prime \rangle_{(-\eta, \ell)}=  \int_{C^+}  (\overline{f}  f^\prime )_{(-\eta, \ell)}\
{\mbox{$\omega$}} , $$
where  $ (\overline{f}  f^\prime)_{(-\eta, \ell)}$\  is the function defined on $C^+$\  by $$
(\overline{f}  f^\prime )_{(-\eta, \ell)} (K)=\overline{f(s)}  f^\prime (s)
$$
for all $K\in  C^+,$ where $\ s\in r_{-\eta\ell}^{-1}(K).$

With each of these inner products, ${\cal F}_c$\ becomes a prehilbert space.
	
For prewave functions we define
$$
\langle \psi_f^{(-\eta, \ell)},\, \psi_{f^\prime}^{(-\eta, \ell)} \rangle_{(-\eta, \ell)}=	
\langle f,\, f^\prime \rangle_{(-\eta, \ell)}.
$$

A sexquilinear form on {\cal S} is given by
$$
\Phi(s,s^\prime)=Tr(\overline{s}\,{s^\prime}).
$$
Thus ($c.f.$ section \ref{wave}), the hermitian product of Prewave Functions can be given in terms of the Prewave Functions themselves instead of the functions $f,$  by
\begin{equation}\label{prher1}
\langle \psi^{(-\eta, \ell)}_f,\, \psi^{(-\eta, \ell)}_{f^\prime} \rangle_{(-\eta, \ell)}=\int_{C^+} \psi_f^{(-\eta, \ell)}\Phi_{(-\eta, \ell)} \psi_{f^\prime}^{(-\eta, \ell)} \  \omega,
\end{equation}
where $\psi_f^{(-\eta, \ell)} \Phi_{(-\eta, \ell)} \psi_{f^\prime}^{(-\eta, \ell)}$\ is the function on $C^+$\ given by
\begin{equation}\label{prher2}
\psi_f^{(-\eta, \ell)} \Phi_{(-\eta, \ell)} \psi_{f^\prime}^{(-\eta, \ell)} (K)=\frac{Tr(\overline{\psi_f^{(-\eta, \ell)}(H,K)}{ \psi_{f^\prime}^{(-\eta, \ell)}(H,K))}}{Tr(\overline{s}\,{s})}
\end{equation}
for all $K\in C^+,$\ $H \in H(2)$\ and $s\in r_{-\eta\ell}^{-1}(K).$\

The hermitian product of Wave Functions is defined as being the  hermitian product of the corresponding Prewave Functions.

\subsubsection{Electromagnetic Potential. }\label{elpot}

If $K\in C^+$,\ the tangent space to $C^+$\ at $K$ can be identified to the subspace of $H(2)$\ given by 
 $$
T_KC^+=\{ M\in H(2): Tr\,M\epsilon \overline{K}\epsilon =0 \}.
$$

Then, the  complexified tangent space can be identified to
$${}^{\bf C}\negthinspace T_KC^+=\{ M\in gl(2,\mathbb{ C}\rm): Tr\,M\epsilon \overline{
K}\epsilon =0 \}.
$$

A real vector field on $C^+$ is thus a map
$$
A:C^+ \longrightarrow H(2),
$$
such that
\begin{equation}\label{simreal}
Tr\,A(K)\epsilon \overline{K}\epsilon =0,
\end{equation}
for all $K\in C^+$.\ 			

A  complex vector field  on $C^+$\ is thus given by a function
$$
A:C^+ \longrightarrow gl(2,\bf C),
$$
 whose real and imaginary hermitian parts are real vectorfields on $C^+$. 

Since $Tr M\epsilon \overline{K}\epsilon$\ is real for $M$ and $K$ hermitian, we see that $A$ is a complex vector field on $C^+$ if and only if 
\begin{equation}\label{simcom}
Tr\,A(K)\epsilon \overline{K}\epsilon =0,
\end{equation}
for all $K\in C^+$.\ 												

Equation \eqref{simcom}	is equivalent to say that $A(K)\epsilon \overline{K}$ is a symmetric matrix.	

Let $A$\ be a complex		vector field on $C^+.$\ We denote by $A_R$ and $A_I$ the real and imaginary hermitian parts of $A$ and
\begin{eqnarray*}
A_R(K)&=&h(A_R^1(K),.., A_R^4(K))\\
A_I(K)&=&h(A_I^1(K),.., A_I^4(K))\\
A^\mu (K)&=&A_R^\mu+i\, A_I^\mu,\ \ \mu=1,..,4\\
\overrightarrow {A_R}(K)&=&(A_R^1(K),.., A_R^3(K))\\
\overrightarrow {A_I}(K)&=&(A_I^1(K),.., A_I^3(K))\\
\overrightarrow {A}(K)&=&(A^1(K),.., A^3(K)).
\end{eqnarray*}
If $A_i^j(K)$ is the element of $A(K)$ in the row $i$ column $j,$ we also have
\begin{eqnarray*}
A^1 (K)&=&\frac12(A_1^2(K)+A_2^1(K))\\
A^2 (K)&=&\frac{i}{2}(A_1^2(K)-A_2^1(K))\\
A^3 (K)&=&\frac12(A_1^1(K)-A_2^2(K))\\
A^4 (K)&=&\frac12(A_1^1(K)+A_2^2(K))
\end{eqnarray*}

If $A(K)$ is , for exemple, continuous with compact support, for $\mu=1,..,4,\ x\in \mathbb{R}^4$ we define
$$
\widetilde{ A^\mu}(x)=\int_{C^+} A^\mu(K) Exp(2\pi i \eta \langle h^{-1}( K),x \rangle ) \omega.
$$

Then
$$
\square \widetilde{A^\mu}=0, 
$$
for all $\mu,$ and
$$
\sum_{\mu=1}^4 \frac{\partial \widetilde{A^\mu}}{\partial x^\mu}=0.
$$

We thus see that the $\widetilde{A^\mu}(x)$\ define a complex electromagnetic potential in the Lorenz gauje.

The corresponding electric field is  given by
$$
\overrightarrow{E}(x)=-\nabla \widetilde{A^4}(x)-\frac{\partial \overrightarrow {\widetilde A}(x)}{\partial x^4 },
$$
where
$$
\overrightarrow {\widetilde A}(x)=(\widetilde{ A^1}(x),\widetilde{ A^2}(x),\widetilde{ A^3}(x)),
$$
that can be written
$$
\overrightarrow{E}(x)=2\pi i \eta \int_{C^+}(A^4(K)\overrightarrow{K}-K^4 \overrightarrow{A}(K)) Exp(2\pi i \eta  \langle h^{-1}( K),x \rangle )  \omega  .
$$
The magnetic field is given by
$$
\overrightarrow{B}(x)=\nabla \times \overrightarrow {\widetilde A}(x),
$$
that  can be written
$$
\overrightarrow{B}(x)=2\pi i \eta  \int_{C^+}(\overrightarrow{A}(K)\times \overrightarrow{K} ) Exp(2\pi i \eta  \langle h^{-1}( K),x \rangle )  \omega.
$$

This electromagnetic field take in general complex values, the real parts of $\overrightarrow{E}$ and $\overrightarrow{B}$ are real electric and magnetic fields, corresponding to the electromagnetic potential given by the real parts of the $A^\mu(x).$

\subsubsection{Wave Functions and the Electromagnetic Potential. }

	In section \ref{elpot} we have seen that a  complex vector field on $C^+$,	 define 
	a  complex			electromagnetic potential	in the Lorenz gauje.
	
	Let $A$\ be a complex			vector field on $C^+$. Then, the matrix $A(K)\epsilon \overline{K}$ is a symmetric matrix, for all $K\in C^+$.

If $ s\in r_+^{-1}(K)$\ there exist an unique complex number, $f_A(s),$\ such that 	
\begin{equation}\label{palapre}
A(K)\epsilon \overline{K}=f_A(s)\, s	.
\end{equation}	

In fact, if 	$z\in {\bf C}^2-{0}$,\ is such that $s= z {}^t \negthinspace z$,\ we have $K=zz^*$,\ and, since 
$$
0=Tr A(K)\epsilon \overline {z}{}^t \negthinspace z\epsilon={}^t\negthinspace z\epsilon A(K)\epsilon \overline {z},
$$
 there exist a number, $\lambda$,\ such that
$$
A(K)\epsilon \overline {z}=\lambda	\, z,
$$
so that
$$
A(K)\epsilon \overline{K}=\lambda	\, s	.
$$	
Since $s\neq 0$\ such a $\lambda$ is unique and is denoted by 
	$f_A(s).$
	
Thus the complex vector field $A$\ defines a 
 function, $f_A,$\  on ${\cal B}.$\

If we take $e^{i\phi}\,s$	\ instead of $s$ in $  r_+^{-1}(K)$\ , we can take in the preceeding reasoning  $e^{i\phi/2}z$ \ instead of $z$\ , and thus
$$
A(K)\epsilon \overline {e^{i\phi/2}z}=\lambda^\prime	\, e^{i\phi/2}z
$$
leads to
$$
\lambda^\prime=	e^{-i\phi}\,\lambda.
$$
We thus see that
$$
f_A(e^{i\phi}\,s)=e^{-i\phi}\,f_A(s)
$$
which proves that $f_A $\ is $S^1$-homogeneous of degree -1.

The Prewave Function corresponding to $f_A$ for $-\eta\ell=1$ can be written as
$$
\psi_A^{(-\eta, -\eta)}(H,K)=f_A(s)\,s\,e^{2 \pi i\eta \langle \, h^{-1}(K),\,h^{-1}(H)\rangle_m},
	$$
	where $s$\ is arbitrary in $r_+^{-1}(K).$\ 
	
	As a consecuence of \eqref{palapre} we have
	\begin{equation}\label{prefunciononda+2} 
\psi_A^{(-\eta, -\eta)}(H,K)=A(K)\epsilon \overline{K}e^{2 \pi i\eta \langle \, h^{-1}(K),\,h^{-1}(H)\rangle_m},
\end{equation}
that gives Prewave  Functions directly in terms of vectorfields.

If $A(K)$ is continuous with compact support, the corresponding Wave Function is
\begin{equation} \label{funciononda+2}
	\widetilde{\psi}_A^{(-\eta, -\eta)}(x)=\int_{C^+}A(K)\epsilon \overline{K}\,e^{2 \pi i\eta \langle \, h^{-1}(K),\,x\rangle_m}\omega_K.
	\end{equation}

Now we define
$$
\widehat{f}_A=\overline{f_A\circ J},
$$
where $J$ is given by \eqref{jota}.

I shall prove that $\widehat{f}_A$ is $S^1$-homogeneous of degree -1 on ${\cal B}$ and that, for all $ s\in r_-^{-1}(K),$ we have  
\begin{equation}\label{palapre2}
-\epsilon \overline{A(K)}\epsilon K \epsilon=\widehat{f}_A(s)\, s	.
\end{equation}	
This will enable us to give a Prewave Function of Photons with $\ell=\eta,$ directly in terms of $A$.

For all $ s\in B,\ a\in S^1$\ we have
$$
\widehat{f}_A(as)=\overline{f_A( J(as))}=\overline{f_A(\overline{a} J(s)}=\overline{af_A( J(s)}=\overline{a}\,\widehat{f}_A(s),
$$
so that $\widehat{f}_A$ has the appropiate homogeneity to define  a Prewave Function.

On the other hand, if $s \in r_-^{-1}(K)$,\  we have $$
\widehat{f}_A(s)\,s=\overline{f_A(J(s))}\,J(J(s))=J(f_A(J(s))\,J({s}))=$$$$
=
J(A(r_+(J(s)))\epsilon \overline{r_+(J(s))})=J(A(r_-(s))\epsilon \overline{r_-(s)}=$$$$
=
J(A(K)\epsilon \overline{K}))=-\epsilon \overline{A(K)}\epsilon K \epsilon .$$

Then, the Prewave Function for Photons with $\ell=\eta$ corresponding  to $\widehat{f}_A$ is 
 \begin{equation}\label{prefunciononda-2}
\psi_A^{(-\eta, \eta)}(H,K)=-\epsilon \overline{A(K)}\epsilon K \epsilon \,e^{2 \pi i\eta \langle \, h^{-1}(K),\,h^{-1}(H)\rangle_m}.
\end{equation}

Obviously
$$
\psi_A^{(-\eta, \eta)}=-\epsilon \overline{\psi_A^{(\eta, \eta)}}\epsilon, \ \ \psi_A^{(\eta, \eta)}=-\epsilon \overline{\psi_A^{(-\eta, \eta)}}\epsilon.
$$

If $A(K)$ is continuous with compact support, the corresponding Wave Function is
\begin{equation} \label{funciononda-2}
	\widetilde{\psi}_A^{(-\eta, \eta)}(x)=-\int_{C^+}\epsilon \overline{A(K)}\epsilon K \epsilon \,\,e^{2 \pi i\eta \langle \, h^{-1}(K),\,x\rangle_m}\omega_K.
	\end{equation}
	
	We have
$$
\widetilde{\psi}_A^{(-\eta, \eta)}=-\epsilon \overline{\widetilde{\psi }_A^{(\eta, \eta)}}\epsilon,\ \ \widetilde{\psi}_A^{(\eta, \eta)}=-\epsilon \overline{\widetilde{\psi }_A^{(-\eta, \eta)}}\epsilon.
$$

With equations \eqref{funciononda+2} and \eqref{funciononda-2} we see that a single complex vectorfied on $C^+$ gives rise to Wave Functions of the four kinds of Photons: those with positive and negative energy and helicity.

The inner products defined in section \ref{prodherfot} lead to the following results.

If $A$\ and $A^\prime$ are vector fields on $C^+$\ we have 
$$
\langle \psi^{(-\eta, -\eta)}_A,\, \psi^{(-\eta, -\eta)}_{A^\prime} \rangle_{(-\eta, -\eta)}= \langle f_A,\, f_{A^\prime} \rangle_{(-\eta, -\eta)},
$$
and
$$
\langle \psi^{(-\eta, \eta)}_A,\, \psi^{(-\eta, \eta)}_{A^\prime} \rangle_{(-\eta, \eta)}=\langle \widehat{f}_A,\, \widehat{f}_{A^\prime} \rangle_{(-\eta, \eta)}.
$$

Then, from \eqref{prher1} and \eqref{prher2} one can obtain
$$
\langle \psi^{(-\eta, -\eta)}_A,\, \psi^{(-\eta, -\eta)}_{A^\prime} \rangle_{(-\eta, -\eta)}=\int_{C^+} A\Phi A^\prime=
$$$$=\langle \psi^{(-\eta, \eta)}_{A^\prime},\, \psi^{(-\eta, \eta)}_{A} \rangle_{(-\eta, \eta)}
$$
where
$$
 A\Phi A^\prime(K)=  \frac{Tr(\overline{A(K)}\epsilon  K A^\prime(K)\epsilon\overline{K})}{Tr(\overline{s}{s})}
$$
for all $K\in C^+,$\  and $s\in r_{\pm}^{-1}(K).$

\subsubsection {Wave Functions in terms of the Electromagnetic Field.}

Let $A$ a complex vectorfield on $C^+$ and denote
\begin{eqnarray}
\overrightarrow{E}(K)&=& A^4(K)\overrightarrow{K}-K^4 \overrightarrow{A}(K) \\
\overrightarrow{B}(K)&=&  \overrightarrow{A}(K)\times \overrightarrow{K}
\end{eqnarray}

Thus, the Electric and Magnetic Fields can be given by
$$
\overrightarrow{E}(x)=2\pi i \eta \int_{C^+}\overrightarrow{E}(K) Exp(2\pi i \eta  \langle h^{-1}( K),x \rangle )  \omega_K  
$$

$$
\overrightarrow{B}(x)=2\pi i \eta  \int_{C^+}\overrightarrow{B}(K) Exp(2\pi i \eta  \langle h^{-1}( K),x \rangle )  \omega_K.
$$

Let us prove that
\begin{eqnarray}\label{campos1}
A(K) \epsilon \overline{K} \epsilon =\left(\overrightarrow{E}(K)+i\overrightarrow{B}(K)\right)\cdot \overrightarrow \sigma,º
\end{eqnarray}
which means
\begin{eqnarray}\label{campos2}
A(K) \epsilon \overline{K} \epsilon =\left (\begin{array}{cc}
(\overrightarrow{E}+i\overrightarrow{B})^3  & (\overrightarrow{E}+i\overrightarrow{B})^1-i(\overrightarrow{E}+i\overrightarrow{B})^2\\
(\overrightarrow{E}+i\overrightarrow{B})^1+i(\overrightarrow{E}+i\overrightarrow{B})^2  &-(\overrightarrow{E}+i\overrightarrow{B})^3
\end{array}
\right )
\end{eqnarray}

In fact, we have
$$
A(K) \epsilon \overline{K} \epsilon=\frac12 \left( A(K) \epsilon \overline{K}+ {}^{t}\negthinspace (A(K) \epsilon \overline{K}) \right)\epsilon=
$$
$$
=\frac12\left((A_R+iA_I)\epsilon \overline{K}\epsilon -K \epsilon ({}^{t}\negthinspace A_R+i {}^{t}\negthinspace A_I)\epsilon\right)=
$$
$$
=\frac12(A_R\epsilon \overline{K} \epsilon-K\epsilon \overline{A_R} \epsilon)+
\frac{i}{2}(A_I\epsilon \overline{K} \epsilon-K\epsilon \overline{A_I} \epsilon)
$$
and thus \eqref{campos2}  follows from \eqref{casicom}.

As a consequence, the Prewave Functions corresponding to $A$ are
\begin{equation}\label{prefuncfield+2} 
\psi_A^{(-\eta, -\eta)}(H,K)=-\left(\left(\overrightarrow{E}(K)+i\overrightarrow{B}(K)\right)\cdot \overrightarrow{ \sigma} \right)\epsilon \ e^{2 \pi i\eta \langle \, h^{-1}(K),\,h^{-1}(H)\rangle_m},
\end{equation}
\begin{equation}\label{prefuncfield-2}
\psi_A^{(-\eta, \eta)}(H,K)=-\epsilon \ \left(\overline{\left(\overrightarrow{E}(K)+i\overrightarrow{B}(K)\right)\cdot \overrightarrow { \sigma}}\right)   \ e^{2 \pi i\eta \langle \, h^{-1}(K),\,h^{-1}(H)\rangle_m}.
\end{equation}

If $A$ is continuous with compact support, we have
\begin{equation}\label{funcfield+2.1} 
\widetilde{\psi}_A^{(-\eta, -\eta)}(x)=-\int_{C^+}((\overrightarrow{E}(K)+i\overrightarrow{B}(K))\cdot \overrightarrow \sigma)\ \epsilon \ e^{2 \pi i\eta \langle \, h^{-1}(K),\,h^{-1}(H)\rangle_m}\omega_K,
\end{equation}
\begin{equation}\label{funcfield-2.1}
\widetilde{\psi}_A^{(-\eta, \eta)}(x)=-\int_{C^+}\ \epsilon \ (\overline{(\overrightarrow{E}(K)+i\overrightarrow{B}(K))\cdot \overrightarrow \sigma)}   \,e^{2 \pi i\eta \langle \, h^{-1}(K),\,h^{-1}(H)\rangle_m}\omega_K.
\end{equation}
so that the Wave Functions are
\begin{equation}\label{funcfield+2} 
\widetilde{\psi}_A^{(-\eta, -\eta)}(x)=\frac{i\eta}{2\pi}((\overrightarrow{E}(x)+i\overrightarrow{B}(x))\cdot \overrightarrow \sigma)\ \epsilon ,
\end{equation}
or
\begin{equation}\label{funcfield-2}
\widetilde{\psi}_A^{(-\eta, \eta)}(x)=\frac{i\eta}{2\pi}\epsilon \  (\overline{(\overrightarrow{E}(x)+i\overrightarrow{B}(x))\cdot \overrightarrow \sigma)}   .
\end{equation}

These Wave Functions are given in terms of the components of $\overrightarrow{E}(x)+i\overrightarrow{B}(x)$ as (up to constants) the Wave Functions of Bialynicki-Birula \cite{Bialynicki-Birula}.

\paragraph{Gauje invariance for  Photons.}

Let us denote by $\cal{D}$\ the complex vector space of complex vector fields  on $C^+.$

The map
	$$
	A\in {\cal D} \longrightarrow f_A \in {\cal F}
	$$	
	 is linear and its kernel is composed by the vector fields on $C^+$\ having the form
	$$
	A(K)=\left( \begin{array}{c}\lambda(z)\\ \mu(z) \end{array}\right)\,z^*
	$$
	where $z\in {\bf C}^2-{0}$ \ is such that $K=zz^*$\ and $\lambda,\ \mu$\ are funtions on ${\bf C}^2-{0}$\ $S^1$-homogeneous of degree +1.
	
	The kernel of the map
	$$
	A\in {\cal D} \longrightarrow \widehat{f}_A \in {\cal F}
	$$	
	is the same and will be denoted by $\cal{N}.$

	In particular, for each  map 
	$$
	L:K\in C^+ \rightarrow L(K)\in gl(2,{\bf C}),$$
	 the vectorfield 
	\begin{equation}\label{invargauje}
	A_L(K)=L(K)K
	\end{equation}
	is in  $\cal{N}.$
	
	If $A\in \cal{D}$,\ $N \in \cal{N}$\ we have 
$$f_{A+N}=f_A$$
$$\widehat{f}_{A+N}= \widehat{f}_{A},$$
so that  the Prewave and Wave Functions are invariant under the changes $A\rightarrow A+N$:
$$
\widetilde{\psi}_{A+N}^{(-\eta, \ell)}=\widetilde{\psi}_{A}^{(-\eta, \ell)}.
$$

Because of \eqref{funcfield+2} and \eqref{funcfield-2} we see that also the Electric and Magnetic Fiels corresponding to Photon are invariant under the changes $A\rightarrow A+N$.
 
In the particular case $N=A_L$\ with  $L(K)=g(K)\,I,$ where $g$ is a complex valued function on $C^+,$ the change $A\rightarrow A+N$ lead to the following change in the Electromagnetic Potential
$$
\widetilde{(A+N)}^\mu(x)=\widetilde{A}^\mu(x)+\partial^\mu \phi(x),
$$
where
$$
\phi(x)=-\frac{i\eta}{2\pi}\int_{C^+}g(K) e^{2 \pi i\eta \langle \, h^{-1}(K),\,x\rangle_m}\omega_K
$$
and
$$
\partial^\mu \phi=g^{\mu \nu}\frac{\partial}{\partial x^\nu}\,\phi
$$
where the $g^{\mu \nu}$ are the components of Minkowski metric.

Thus, the invariance of the Electromagnetic Field of Photons under the change $A(K) \rightarrow A(K)+g(K)K$ is a particular case of the well known gauje invariance of general Electromagnetic Fields.

\section{The Classical State Space}\label{SS}

\subsection{General remarks}\label{General view}

In this paper, State Space for a particle defined by $\alpha \in \underline G^*$  has been stated as the homogeneous space that correspond to the orbit of $(0,\alpha)$ in $H(2)\times \underline G^*.$
Thus, as we have seen in section \ref{wave}, it is identified to
$$
\frac{G}{G_{(0,\alpha)}}
$$
where
$$
G_{(0,\alpha)}=G_\alpha \cap (SL \oplus \{0\}),
$$
or, equivalently, to
$$
H(2) \times \frac{SL}{(G_\alpha)_{SL}\cap SL_1} ,
$$
that is the form adopted  in Figure \ref{diagr2} of that section.

 The equivariant maps
 $$
\iota_4:H(2) \times \frac{SL}{(G_\alpha)_{SL}\cap SL_1} \rightarrow H(2) \times \frac{SL}{(G_\alpha)_{SL}}
$$
and
$$
\nu_2:H(2) \times \frac{SL}{(G_\alpha)_{SL}\cap SL_1} \rightarrow \frac{G}{G_\alpha}.
$$
in that Figure, will be used in this section.

When we characterize 
$$
H(2) \times \frac{SL}{(G_\alpha)_{SL}\cap SL_1}
$$
in terms of other concrete manifolds or of some parametrization, we need to physically interpret the geometric objects that appear, and, in  order to do that,  the more effective  way is , in general,  to know the values of the Canonical  Dynamical Variables in terms of these objects.

But the value of these Dynamical Variables is well known on the coadjoint orbit \emph{i.e.} on $G/G_\alpha.$ Thus, its values in State Space can be obtained by composition with $\nu_2.$

Since Prewave Functions are defined on
$$
H(2) \times \frac{SL}{(G_\alpha)_{SL}}
$$
its composition with $\iota_4$ gives another version of Quantum States, now defined on the Classical State Space.

In the definition of Prewave Functions \ref{prefunc},  space-time only appears in the exponential, whose exponent is 
$$
-2\pi i\langle h^{-1}(P),h^{-1}(H)\rangle_m.
$$
where $P$ is the ``traslation" of momentum-energy to $SL/(G_\alpha)_{SL}.$ The exponent in the composition of Prewave Functions with $\iota_4$
 must have the same expression, but changing $P$ by $P\circ \iota_4.$ The conmutativity of the diagramm in Figure \ref{diagr2} imply that   $P\circ \iota_4$  can be  obtained as the impulsion-energy in $G/G_\alpha$ composite with $\nu_2.$

In the next sections I give the expressions of the canonical dynamical variables on State Space, for different kind of particles.

\subsection{Klein-Gordon particles}\label{SSKG}

We have seen in section \ref{WFKG} that State Space for Klein-Gordon particles can be identified to
$$
H(2) \oplus \frac{SL}{SU(2)},
$$
and this homogeneous space to ${\cal H}^m\times H(2).$

If $(K,H)\in {\cal H}^m\times H(2) $ and $A\in SL$ is such that $K=mAA^*,$ we denote 
\begin{eqnarray*}
H&=&h(\overrightarrow x,x^4)\\
\overrightarrow x&=& (x^1,x^2,x^3)\\
K&=&m h(\overrightarrow k, k_4)  \\
\overrightarrow k&=& (k^1,k^2,k^3)\\
k^4&=&\sqrt{1+\Vert \overrightarrow k \Vert^2}.
\end{eqnarray*}

Thus, the map \eqref{nuuno}  becomes, 
$$
\nu_1(K,H)=(A,H)*(mI,\overrightarrow 0,1)=
$$$$=(K,\overrightarrow x-\frac{x^4}{k^4} \overrightarrow {k},e^{-2\pi i \eta m x^4/k^4})
$$
and the map \eqref{nudos},
\begin{equation}\label{KGnudos}
\nu_2(K,H))=(A,H)*(mI,\overrightarrow 0)=(K,\overrightarrow x-\frac{x^4}{k^4}  \overrightarrow {k}).
\end{equation}

The Canonical Dynamical Variables on State Space are obtained as composition of \eqref{KGvardin} with $\nu_2.$  If we denote $V_{KG}=V\circ \nu_2$ for each dynamical variable, $V,$ we have
$$
P_{KG}(mh(\overrightarrow k,k_4),h(\overrightarrow x,x^4))=- \eta\,mh(\overrightarrow k,k_4),
$$$$
\overrightarrow l_{KG}(mh(\overrightarrow k,k_4),h(\overrightarrow x,x^4))=\overrightarrow x \times \left(\overrightarrow  P_{KG}(mh(\overrightarrow k,k_4),h(\overrightarrow x,x^4)\right) ,$$$$
\overrightarrow g_{KG}(mh(\overrightarrow k,k_4),h(\overrightarrow x,x^4))=$$$$ = P^4_{KG}(mh(\overrightarrow k,k_4),h(\overrightarrow x,x^4)) \overrightarrow x-x^4 \overrightarrow P_{KG}(mh(\overrightarrow k,k_4),h(\overrightarrow x,x^4)).
$$
where $ k_4=\sqrt{1+\Vert \overrightarrow k \Vert^2}.$

The Prewave functions have been  defined on a manifold that,  in this case,  coincides with State Space. The map $\iota_4$ is the identical map.

Another fact is that, in this case, State Space is the Universal Covering of the homogeneous  contact manifold. 

Let us prove that $\nu_1$ is a covering map.

The local expresion of $\nu_1$ in the charts corresponding to $\phi_\tau$ and 
$$
\phi^\prime:(k_1,k_2,k_3,x^1,x^2,x^3,x^4)\in {\mathbb R}^7 \longrightarrow $$$$ \longrightarrow  (h(m\overrightarrow k,m k_4),h(x^1,x^2,x^3,x^4))\in {\cal H}^m\times H(2) ,
$$
is given by
$$
(\phi_\tau)^{-1}\circ \nu_1 \circ \phi^{\prime}(k_1,\dots,x^4)=$$$$=(k_1,k_2,k_3,x^1-\frac{x^4}{k_4}k_1,
x^2-\frac{x^4}{k_4}k_2,x^3-\frac{x^4}{k_4}k_3,-\tau-\eta m \frac{x^4}{k_4}+N)
$$
where the domain of definition is defined by the condition that 
$$\tau+\eta m \frac{x^4}{k_4} \notin {\mathbb Z}+\frac12 $$ 
and $N$ is defined by the condition that
$$
-\tau-\eta m \frac{x^4}{k_4}+N\in (-\frac12,\frac12).
$$

We have
$$
((\phi_\tau)^{-1}\circ \nu_1 \circ \phi^{\prime})^{-1}\{(\overrightarrow n,\overrightarrow y,t)\}=$$$$=
\{(\overrightarrow n,\overrightarrow y-\frac{\eta}{m}(t+\tau+N)\overrightarrow n, -\frac{\eta}{m}(t+\tau+N)\sqrt{1+\Vert \overrightarrow n \Vert^2}): N\in {\mathbb Z}\}.
$$

Let $d\in {\cal H}^m \times {\mathbb R}^3\times {\mathbb S}^1,$ and let $\tau$ be such that $d$ is in the image of $(\phi_\tau)^{-1}.$ This image, $U_o,$ is an open neighborhood of $d$, and we shall prove that its antiimage by $\nu_1$ is union of disjoint open sets, each of them diffeomorphic to $U_o$ under the restriction of $\nu_1.$ 

We denote by ${\cal U}$ the domain of $\phi_\tau$. 

The domain of  $(\phi_\tau)^{-1}\circ \nu_1 \circ \phi^{\prime},$
$$
{\cal W}=((\phi_\tau)^{-1}\circ \nu_1 \circ \phi^{\prime})^{-1}({\cal U})
$$
is the union of the open sets
\begin{eqnarray*}
{\cal W}_N=\{&(&\overrightarrow n,\overrightarrow y-\frac{\eta}{m}(t+\tau+N)\overrightarrow n, -\frac{\eta}{m}(t+\tau+N))\sqrt{1+\Vert \overrightarrow n \Vert^2}: \\&(&\overrightarrow n,\overrightarrow y,t)\in {\cal U}\}.
\end{eqnarray*}
where $N\in \mathbb Z.$

But the ${\cal W}_N$ are disjoint. In fact, if
$$
(\overrightarrow n,\overrightarrow y-\frac{\eta}{m}(t+\tau+N)\overrightarrow n, -\frac{\eta}{m}(t+\tau+N)\sqrt{1+\Vert \overrightarrow n \Vert^2})=$$$$=(\overrightarrow n',\overrightarrow y'-\frac{\eta}{m}(t'+\tau+N')\overrightarrow n', -\frac{\eta}{m}(t'+\tau+N')\sqrt{1+\Vert \overrightarrow n' \Vert^2})
$$
where $(\overrightarrow n,\overrightarrow y,t),\ (\overrightarrow n',\overrightarrow y',t')\in \cal U,$ we obtain first $\overrightarrow n=\overrightarrow n'$ and then $t+N=t'+N'.$ Since $-1/2<t, t'<1/2,$ it follows that  $N=N'.$

The restriction of $\nu_1$ to each of the $W_N$ is a diffeomorphism onto $U.$

 Since ${\cal H}^m\times H(2) $ is symply connected, we see that  State Space is the universal covering space of the contact manifold $G/Ker\,C_\alpha.$  By reciprocal image it inherits a contact form that must also be homogeneous: it is the contact structure used in section \ref{clasical}. This homogeneous contact form can also be introduced directly by means of
$C^\prime_{\alpha_o},$ in the same way that the homogeneous contact structure has been introduced in $G/Ker\,C_{\alpha_o}.$

The covering  can be also considered in the following way.

We define an action of $\mathbb Z$ on ${\cal H}^m\times H(2) $  by
$$
N*(K,H)=(K,H-N\frac{\eta }{m^2} K).
$$

With the parametrization $\phi^\prime$ we have
$$
(\phi^\prime)^{-1}\circ (N*) \circ \phi^\prime(\overrightarrow k,\overrightarrow x,x^4)=$$$$=
(\overrightarrow k, \overrightarrow x-N\frac{\eta }{m}\overrightarrow k,x^4-N\frac{\eta }{m}k^4).
$$

We thus have a properly discontinuous action without fixed point  of $\mathbb Z$ on State Space and the quotient space is $G/Ker\,C_\alpha.$

\subsection{Massive particles with  $T\neq 0$.}

State Space is $H(2)\times (SL/[S^1])$ or, equivalently, $G/([S^1]\oplus \{0\}).$

In section \ref{WFDIRAC} we have identified $SL/[S^1]$ with ${\cal H}^m \times P_1(C).$

Contrarily to the Klein-Gordon case, if $T\neq 0,$ there exist  no equivariant map from  State Space onto the contact manifold $G/[R].$

The canonical map $\nu_2$  from State Space onto Movement Space, is such that
$$
\nu_2(A*(mI,\left[ \left(\begin{array}{c} 1\\0 \end{array}\right)\right]),H)=$$$$= (A,H)*\left(mI,\left[\left(\begin{array}{c}1\\0 \end{array}\right)\right],
 \overrightarrow 0 \right)$$
so that
$$
\nu_2(mAA^*,\left[A \left(\begin{array}{c} 1\\0 \end{array}\right)\right],H)=$$$$= \left(mAA^*,\left[A \left(\begin{array}{c}1\\0 \end{array}\right)\right],
 (H^1,H^2,H^3)-\frac{H^4}{(AA^*)^4}((AA^*)^1,(AA^*)^2,(AA^*)^3)\right)$$$$\in {\cal H}^m \times P_1(C)\times {\mathbb R}^3.
$$
where I have denoted, for each $L\in H(2)$, $h^{-1}(L)\stackrel{\mathrm {def}}{=}(L^1,L^2,L^3,L^4)).$
Then
$$
\nu_2(mh(\overrightarrow k,k_4),\left[ u\right],h(\overrightarrow x,x^4))=$$$$=
\left(mh(\overrightarrow k,k_4),\left[ u\right],\overrightarrow x-\frac{x^4}{k_4}\overrightarrow k\right)),
$$
where $k_4=\sqrt{1+\Vert \overrightarrow k \Vert^2}.$ 

Thus, the portrait in State Space of the movement $(mh(\overrightarrow k,k_4),\left[ z\right],\overrightarrow y)$ is
$$
(\nu_2)^{-1}\{ (mh(\overrightarrow k,k_4),\left[ z\right],\overrightarrow y)\}=$$$$=
\{ (mh(\overrightarrow k,k_4),\left[ z\right],h(\overrightarrow y,0)+\lambda\ h(\overrightarrow k, k_4)):
 \lambda\in {\mathbb R} \}.
$$

These results on Canonical Dynamical Variables are better expressed when we use the identification of $P_1(C)$ with $S^2$ given by \eqref{otradefbeta}.

If $(mh(\overrightarrow k,k_4),\overrightarrow v,h(\overrightarrow x,x_4))\in {\cal H}^m \times S^2 \times H(2),$ is interpreted as a state of a  particle with mass $m>0$, the values of the canonical dynamical variables in this state are obtained by composition of 
\eqref{momentosdiracabajo} with the map $\nu_2$.

When for each dynamical variable $V$ we denote $V\circ \nu_2$ by $V_{massive}$ we have
\begin{eqnarray*}
P_{massive}(mh(\overrightarrow k,k_4),\overrightarrow v,h(\overrightarrow x,x^4))&=&-\eta m h(\vec{k},k_4),\\
\overrightarrow l_{massive}(mh(\overrightarrow k,k_4),\overrightarrow v,h(\overrightarrow x,x^4))&=&\frac{T}{4\pi}\frac{k_4\overrightarrow v-\vec{k}}{k_4-\langle \vec{ k},\overrightarrow v \rangle }+\overrightarrow x\times\overrightarrow P_{massive},\\
\overrightarrow g_{massive}(mh(\overrightarrow k,k_4),\overrightarrow v,h(\overrightarrow x,x^4))&=&\frac{T}{4\pi}\frac{\vec{k}\times\overrightarrow v}{k_4-\langle \vec{ k},\overrightarrow v \rangle}+  \\ &+&  P^4_{massive}\overrightarrow x-x^4\overrightarrow P_{massive},
\end{eqnarray*}

In what concerns the Pauli-Lubanski fourvector, we have 
\begin{equation}   
\begin{split}
\overrightarrow W_{massive}&=P^4_{massive}\frac{T}{4\pi}     \frac{k_4\overrightarrow v-\vec{k}}{k_4-\langle \vec{ k},\overrightarrow v \rangle }   \\
W^4_{massive}&= -\eta m
\frac{T}{4\pi}     \frac{\langle k_4\overrightarrow v-\vec{k},\vec k \rangle}{k_4-\langle \vec{ k},\overrightarrow v \rangle},
\end{split}
\end{equation}

As in $T=0$ case, the prewave functions have been defined directly on State Space, the map $\iota_4$ being the identical map. Thus they are exactly as in section \ref{generconmasa}.

\subsection{Massless particles of type 4}\label{ssnula}

Since $(G_\alpha)_{SL}\cap SL_1=[S^1],$ State Space for massless particles, $H(2)\times (SL/((G_\alpha)_{SL}\cap SL_1),$ can be identified to State Space for massive particles, ${\cal H}^m \times P_1(C) \times H(2)$. Here $m$ is an arbitrary  positive number, with no physical significance.

On the other hand, Movement Space, $G/G_\alpha,$ is identified to $P^{sign(\nu)}_3$ by means of the action \eqref{accionproy}.

The canonical map $\nu_2$ from State Space onto Movement Space thus becomes for this kind of particle
$$
\nu_2((A,H)*\left(mI,\left[ \left(\begin{array}{c} 1\\0 \end{array}\right) \right],0 \right )=
(A,H)*[q]
$$
where
$$
q=\left( \begin{array}{c} 0\\2\nu \\0\\1\end{array}  \right).
$$

Then
$$
\nu_2\left(mAA^*,\left[A \left(\begin{array}{c} 1\\0 \end{array}\right) \right],H\right )=
\left[\begin{array}{c} 2\nu A \left(\begin{array}{c} 0\\1 \end{array}\right)-iH(A^*)^{-1}
\left(\begin{array}{c} 0\\1 \end{array}\right)\\ (A^*)^{-1}
\left(\begin{array}{c} 0\\1 \end{array}\right)
\end{array} \right].
$$

But $(A^*)^{-1}=-\varepsilon \overline A \varepsilon,$ so that
$$
\nu_2\left(mAA^*,\left[A \left(\begin{array}{c} 1\\0 \end{array}\right) \right],H\right )=
\left[\begin{array}{c} (2\nu AA^* -iH)(A^*)^{-1}
\left(\begin{array}{c} 0\\1 \end{array}\right)\\ (A^*)^{-1}
\left(\begin{array}{c} 0\\1 \end{array}\right)
\end{array} \right]=$$$$=
\left[\begin{array}{c} (2\nu AA^* -iH)\varepsilon \overline A
\left(\begin{array}{c} 1\\0 \end{array}\right)\\\varepsilon \overline A
\left(\begin{array}{c} 1\\0 \end{array}\right)
\end{array} \right].
$$

Then
\begin{equation}\label{nu2nula}
\nu_2:(K,[u],H)\in {\cal H}^m \times P_1(C) \times H(2) \rightarrow 
\left[\begin{array}{c} \left(\frac{2\nu}{m} K -iH \right)\varepsilon \overline u
\\    \varepsilon \overline u
\end{array} \right] \in P_3^{sign(\nu)}.
\end{equation}

If we fix, for exemple, $m=1,$ then for all
$$
\left[\left(\begin{array}{c} \omega\\\pi \end{array}\right)  \right]\in  P_3^{sign(\nu)},
$$
we have
$$
(\nu_2)^{-1}\left\{ \left[\left(\begin{array}{c} \omega\\\pi \end{array}\right)  \right]\right\}=$$$$=
\{ (K,[\varepsilon \overline \pi],H)\in {\cal H}^1  \times P_1(C) \times H(2):  (\frac{2\nu}{m} K-iH)\pi=\omega\}.
$$

This set represent the ``portrait" of the movement
$$
\left[\left(\begin{array}{c} \omega\\\pi \end{array}\right)\right]
$$
in ${\cal H}^1  \times P_1(C) \times H(2)$ considered as State Space of massless particles.

The values of  dynamical variables $P,\ \overrightarrow l ,\ \overrightarrow g,$ on ${\cal H}^m \times P_1(C) \times H(2)$ for the massless Type 4 particles can be obtained by composition of its expressions \eqref{Pproy}, \eqref{lproy}, \eqref{gproy}, with the map $\nu_2$ given in \eqref{nu2nula}.

If $(mh(\overrightarrow k,k_4),\overrightarrow v,h(\overrightarrow x,x_4))\in {\cal H}^m \times S^2 \times H(2),$ is interpreted as a state of a massless Type 4 particle, where now $m$ is an arbitrary positive number with no physical significance, the values of the canonical dynamical variables in this state are
\begin{eqnarray*}
\overrightarrow P_{massless}(mh(\overrightarrow k,k_4),\overrightarrow v,h(\overrightarrow x,x^4))&=&P_{massless}^4 \overrightarrow v,\\
P_{massless}^4(mh(\overrightarrow k,k_4),\overrightarrow v,h(\overrightarrow x,x^4))&=&\frac{-\eta}{ 2 (k_4-\langle \vec{ k},\overrightarrow v \rangle) }  \\
\overrightarrow l_{massless}(mh(\overrightarrow k,k_4),\overrightarrow v,h(\overrightarrow x,x^4))&=&\frac{\chi T}{4\pi}\frac{k_4\overrightarrow v-\vec{k}}{k_4-\langle \vec{ k},\overrightarrow v \rangle}+  \\ &+&\overrightarrow x\times\overrightarrow P_{massless},\\
\overrightarrow g_{massless}(mh(\overrightarrow k,k_4),\overrightarrow v,h(\overrightarrow x,x^4))&=&\frac{\chi T}{4\pi}\frac{\vec{k}\times\overrightarrow v}{k_4-\langle \vec{ k},\overrightarrow v \rangle}+  \\ &+&  P^4_{massless}\overrightarrow x-x^4\overrightarrow P_{massless}.
\end{eqnarray*}

The Pauly-Lubanski four vector for type 4 particles is (\emph{c.f.} section \ref{concreteqforms})
$$
W_{massless}=\frac{\chi  T}{4\pi} P_{massless}.
$$

We thus see that the expression of $
\overline l$ and $\overline  g$ are \bf formally \rm almost identical for massive and massless particles. But this is not the case for $P.$

The point $(mh(\overrightarrow k,k_4),\overrightarrow v,h(\overrightarrow x,x^4))$ can be interpreted as being the  state of a massive particle or the state of a massless particle, but the value in this state of linear momentum, energy and angular momentum is different in the massive or in the massless case.

The map $\iota_4,$ what in the case of massive particles whas the identical map, in the case of  massless particles can be found to be
$$
\iota_4:(mh(\overrightarrow k,k_4),\overrightarrow v,h(\overrightarrow x,x^4))\in {\cal H}^m \times S^2 \times H(2) \rightarrow $$$$ \rightarrow  \left( \frac{h(\overrightarrow v,1)}{2(k_4-\langle \vec{ k},\overrightarrow v \rangle)}, h(\overrightarrow x,x^4))\right)\in C^+ \times H(2)
$$

The $P_{KG},\ P_{massive}$ and $P_{massless}$ appears in the exponent of the Prewave Functions in State Space of the corresponding particles.

\bibliography{bibliopr}

\bibliographystyle{alpha}

\end{document}